\documentclass[         %
aps,                    
prd,                    
showpacs,               
superscriptaddress,     
nofootinbib,            
eqsecnum,               %
twocolumn,              %
showkeys,               %
preprintnumbers,        %
amsmath,                %
amssymb,                %
floatfix]               
{revtex4-1}             
\usepackage{graphicx, amsmath}
\graphicspath{{plots/}}
\usepackage{mathrsfs, bm} 
\usepackage{mathtools} 
\usepackage{sublabel}
\usepackage{amsmath}
\usepackage{amssymb}
\usepackage[utf8]{inputenc}
\usepackage{tikz}
\usepackage{multirow}
\usepackage{color}
\usepackage{hyperref}
\usepackage{caption}
\usepackage{subcaption}
\usepackage{standalone}
\usepackage{nameref}
\usepackage{nicefrac}
\usepackage{feynmp}
\usepackage{soul}
\usepackage{textcomp} 

\newcommand{\msp}{\hspace{0.5pt}} 

\newcommand{\s}{$\mkern5mu$} 

\newcommand{\f}{\uparrow\downarrow\mkern-23mu\textrm{---}}
\newcommand{\e}{\textrm{---}}

\begin{document}
\title{Neutrino spectral split in the exact many-body formalism}
\author{Savas Birol}
\email{savas.birol@istanbul.edu.tr}
\affiliation{Institude of Science, Istanbul University, Istanbul 34116, Turkey}
\author{Y.~Pehlivan}
\email{yamac.pehlivan@msgsu.edu.tr}
\affiliation{Mimar Sinan Fine Arts University, Sisli, Istanbul, 34380, Turkey}
\affiliation{National Astronomical Observatory of Japan, Mitaka, Tokyo 181-8588,
Japan}
\author{A.~B.~Balantekin}
\email{baha@physics.wisc.edu}
\affiliation{Department of Physics, University of Wisconsin, Madison, WI 53706, USA}
\affiliation{National Astronomical Observatory of Japan, Mitaka, Tokyo 181-8588,
Japan}
\author{T. Kajino}
\email{kajino@nao.ac.jp}
\affiliation{School of Physics and Nuclear Energy Engineering and IRCBBC,
Beihang University, Beijing 100083, People's Republic of China}
\affiliation{National Astronomical Observatory of Japan, Mitaka, Tokyo 181-8588,
Japan}
\affiliation{Graduate School of Science, The University of Tokyo, Bunkyo-ku,
Tokyo 113-0033, Japan}
\date{\today}
\begin{abstract}
We consider the many-body system of neutrinos interacting with each other
through neutral current weak force. Emerging many-body effects in such a system
could play important roles in some astrophysical sites such as the core collapse
supernovae. In the literature this many-body system is usually treated within
the mean field approximation which is an effective one-body description based on
omitting entangled neutrino states. In this paper, we consider the original
many-body system in an effective two flavor mixing scenario under the single
angle approximation and present a solution without using the mean field
approximation. Our solution is formulated around a special class of many-body
eigenstates which do not undergo any level crossings as the neutrino 
self-interaction rate decreases while the neutrinos radiate from the supernova. 
In particular, an initial state which consists of electron neutrinos and
antineutrinos of an orthogonal flavor can be entirely decomposed in terms of
those eigenstates. Assuming that the conditions are perfectly adiabatic so that
the evolution of these eigenstates follow their variation with the interaction
rate,  we show that this initial state develops a spectral split at exactly the
same energy predicted by the mean field formulation.
\end{abstract}
\medskip
\pacs{14.60.Pq, 
95.85.Ry, 
97.60.Bw. 
02.30.Ik, 
03.65.Fd 
67.85.-d, 
74.20.Fg, 
}
\keywords{
Collective neutrino oscillations,
spectral swaps,
supernova neutrinos,
many-body effects,
exact many-body methods,
Richardson-Gaudin method,
Bethe ansatz.
}
\preprint{}
\maketitle

\section{Introduction}
\label{section:introduction}

Neutrinos and photons are the most abundant particle species in the Universe
\cite{Patrignani:2016xqp}. In some astrophysical sites, neutrinos can reach
sufficiently high densities to form a many-body system through mutual neutral
current weak interactions.  A prominent example of such a site is a
core-collapse supernova explosion \cite{Burrows:1990ts, Kotake:2005zn,
Beacom:2010kk, Raffelt:1996}: When the inert iron core of a massive star reaches
the Chandrasekar mass limit, it collapses until a dense
proto-neutron star forms at the center.  The proto-neutron star is initially
very hot because it carries a large amount of gravitational potential energy
which is converted into heat during the collapse.  In about $10$ s, it
cools down by emitting of the order of $10^{58}$ neutrinos \cite{Colgate:1966ax,
Woosley:1986ta, 1989ARA&A..27..629A}. Since the neutrinos can easily pass
through the outer layers of the star as the matter is pushed into the space by
the shock wave, they can quickly carry the energy and the entropy away from the
proto-neutron star. In fact, neutrino emission is a very fast and efficient
cooling mechanism not only for a proto-neutron star, but also for black hole
accretion disks \cite{Ruffert:1998qg, Popham:1998ab, Narayan:2001qi,
Matteo:2002ck, Chen:2006rra, Malkus:2012ts} and binary neutron star mergers
\cite{Tian:2017xbr}.

In this paper we explore the impact of many-body effects due to
neutrino-neutrino interactions on the flavor evolution of neutrinos. We use the
core-collapse supernova as the backdrop of our discussion on many-body effects.
However, our intention is not to give a sophisticated analysis of neutrino
flavor evolution in a realistic supernova environment, which has already been
the subject of intense research for the past few decades due to the importance
of neutrinos in several aspects of supernova physics(see, e.g.,
\cite{Fowler:1964zz, barkat, Epstein:1988gt, Woosley:1989bd, Woosley:1994ux,
Woosley:2002zz}) and the possibility of observing them with the current
detectors \cite{Totani:1997vj, Katz:2011ke, Scholberg:2012id}. The interested
reader is referred to the excellent review articles in the literature
\cite{Duan:2010bg, Chakraborty:2016yeg}.

Here our focus is on the many-body system formed by the neutrinos. In 
particular, we report on an exact solution of this many-body system in a 
simplified case and its comparison with the results obtained with the mean field
approximation, which is used to treat the neutrino-neutrino interactions in an
overwhelming majority of studies in the literature. The mean field approximation
basically amounts to replacing the mutual interactions between individual
particles by a description in which each particle interacts with an average field
formed by all other particles. Such a treatment greatly simplifies an
interacting  many-body system by effectively reducing it to the study of
independent particles moving in an ``external'' field. This field is determined
from a simple self-consistency requirement: Since the mean field is collectively
created by all the particles in the system, it should change in line with the
evolution of individual particles. At the mathematical level, the mean field
approximation works by blocking all entangled states in the Hilbert space
because particles moving independently in a field cannot develop entanglements
if they are not entangled in the first place. In practice, one starts with an
unentangled one-body initial state, i.e., a state which consists of
multiplication of one-body states. Then the evolution of the system is restricted
to such states at later times. For a system consisting of $n$ particles, each of
which can occupy $k$ different states, this amounts to replacing the original $k^n$ 
dimensional Hilbert space with $n$ individual $k$-dimensional Hilbert spaces.

The mean field approximation was applied to self-interacting neutrinos quite
early on \cite{Samuel:1993uw, Raffelt:1992uj, Sigl:1992fn, McKellar:1992ja} and
widely adopted in subsequent studies. It is also extensively used in various
areas such as nuclear physics, condensed matter physics, and the physics of cold
atom systems. Unlike the case for neutrinos, in those other fields one has the
advantage of experimental access to the system under consideration. In
particular, it is usually possible to measure the fluctuations of various
quantities around their mean field values in order to assess the accuracy of the
approximation.  Although the general consensus is that the mean field
approximation becomes more and more accurate with an increasing number of
particles, there are some cases which do not agree with this simple
expectation \cite{2013PhRvA..88c3608G}. In fact the latter situations are
usually sought after and actively studied by theorists
\cite{2001Natur.409...63S, 2009PhRvL.102j0401P} and experimentalists
\cite{2008Natur.455.1216E} in cold atom systems in connection to such
applications as cryptography and quantum computing. Those studies
intentionally create conditions under which an initially unentangled system
develops into a macroscopically entangled state through time evolution and, in
doing so, significantly deviates from its mean field description.

In the context of neutrino astrophysics, one is naturally interested in the
opposite question, i.e., if the mean field approximation provides an accurate
description of self-interacting neutrinos. References
\cite{Friedland:2003dv, Bell:2003mg} were the first papers to tackle this
difficult question. Reference \cite{Friedland:2003dv} was concerned with the
microphysics and worked with small neutrino wave packages which undergo distinct
scatterings from one another. The authors  were interested in whether the
entanglement can build up on the system which starts from a one-body state.  In
particular, they considered two intersecting beams of neutrinos, each of which
consisted only of a particular flavor. Using a physically very transparent
argument, they showed that the buildup of entangled states occurs at the 
timescale of an incoherent effect which is much longer than the timescale of
a coherent effect relevant for self-interacting neutrinos.%
\footnote{As the neutrinos scatter from background particles, including other
neutrinos, some diagrams with definite relative phases add up at the amplitude
level whereas others with random relative phases add up at the probability
level. In a dense environment, the former (coherent) addition give rise
to a much faster flavor evolution than the latter (incoherent) addition since it
is proportional to the square of the background density.} Therefore Ref.
\cite{Friedland:2003dv} concluded that the mean field picture provides an
accurate description of the problem. Reference \cite{Bell:2003mg} used a 
different picture in which neutrinos are represented by plane waves in a box so 
that they all interact with each other at the same time. The authors approached 
the problem numerically by simulating the exact\footnote{Here, and throughout 
this article, we use the word \emph{exact} to indicate that we are avoiding the 
mean-field approximation. It does not imply a precise treatment of 
self-interacting neutrinos, as one usually has to employ several other 
simplifying assumptions.} many-body behavior of $14$ neutrinos and comparing it 
with the mean field prediction. In particular if the vacuum oscillations are 
ignored, which was the case in both Refs. \cite{Friedland:2003dv} and 
\cite{Bell:2003mg}, then the mean field picture predicts that no flavor 
evolution would occur for a one-body initial state in which all neutrinos 
occupy flavor eigenstates. However, Ref. \cite{Bell:2003mg} has found some 
flavor conversion for such an initial state in the exact many-body picture 
which indicated a possible breakdown of the mean field approximation. This 
apparent contradiction was resolved in a later study \cite{Friedland:2003eh} 
which established the following two results: (1) As far as the coherent effects 
are concerned, the study of the problem is independent of the size of the 
neutrino wave packages, so the two descriptions of the problem in Refs.  
\cite{Friedland:2003dv, Bell:2003mg} are equivalent. (2) The time required for 
the flavor conversion observed in Ref. \cite{Bell:2003mg} scales as expected 
from an incoherent effect with an increasing number of particles; i.e., it 
develops more slowly until it becomes irrelevant in comparison to the much 
faster coherent effects.

Here, our treatment of the exact many-body effects is, in some ways, similar to
that of Ref. \cite{Bell:2003mg}: We use the plane wave picture, and we compare
the exact many-body behavior of the system with the mean field prediction.
However, there are important differences between our work and Refs.
\cite{Friedland:2003dv, Bell:2003mg, Friedland:2003eh}. First of all, our
formalism includes the vacuum oscillations which were ignored in those studies.
This allows us to obtain a spectral split in the exact many-body
formalism for the first time. Spectral splits (or swaps) are  a particular kind
of emergent behavior in which neutrinos of different flavors totally or
partially exchange their energy spectra \cite{Duan:2006an, Duan:2010bg}. Since
they are caused by an interplay between the one-particle terms (vacuum
oscillations) and two-particle terms (self-interactions), one needs to
incorporate both effects in the calculations in order to unfold such a
behavior \cite{Raffelt:2007xt, Raffelt:2007cb, Pehlivan:2011hp, Raffelt:2011yb,
Galais:2011gh}. For this reason, spectral splits are so far observed only in
the mean field calculations.  Another difference between our work and the
previous studies is that we use a semianalytical technique which allows us to
numerically work with as many as $10^8$ neutrinos. Our technique is based on the
algebraic Bethe ansatz formalism
\cite{Bethe:406121, Richardson:1963, Gaudin:1976, Gaudin:1983, Dukelsky:2004re}
coupled with the relatively new Bethe ansatz solver method
\cite{Faribault:2011rv} which enables us to convert the problem of diagonalizing
a many-body Hamiltonian into a problem of finding the roots of a polynomial.

The most important caveat of our study is that we do not carry out a dynamical
evolution calculation of the neutrino many-body system. With $10^8$ neutrinos,
this would indeed be a very difficult task. Instead we consider perfectly
adiabatic evolution conditions, and assume that the dynamical evolution of
eigenstates follows their slow transformation as the external conditions change.
In fact our study is mostly concerned with how the eigenstates of the exact
many-body Hamiltonian change with the decreasing neutrino density, e.g., as the
neutrinos occupying a comoving volume element move away from the center
of the supernova.  In general, exact many-body eigenvalues of the neutrino
Hamiltonian cross each other at several points, in which case the adiabatic
theorem is not necessarily applicable.\footnote{Under some simplifying
assumptions, the Hamiltonian of self-interacting neutrinos have several
conserved quantities \cite{Pehlivan:2011hp, Raffelt:2011yb} which may be useful
in examining the behavior of the system at those crossings. For example, in
some cases conserved quantities allow us to map the dynamics near a crossing
point to the adiabatic dynamics of another system which has no such crossings
\cite{2013AAMOP..62..117T}. Such a scheme may be the subject of another paper.}
However, we are able to identify a certain class of many-body eigenstates whose
eigenvalues do not undergo any crossings. We also find that a state consisting
only of electron neutrinos and antineutrinos of an orthogonal flavor can be
decomposed entirely in terms of those eigenstates. This makes it possible to 
follow the adiabatic transformation of the neutrino ensemble if it starts from 
such an initial configuration. The fact that we can presently apply this 
technique to only a certain initial flavor composition is the second caveat of 
our study. In our Conclusions, we briefly comment on the possibility of 
extending the range of applicability of this method. We also note that we work 
in the two flavor mixing scheme, and ignore the angular dependence of the 
neutrino-neutrino interactions.

Between the straightforward application of the mean field approximation, and the
challenging study of the exact many-body system lies a middle ground where one
tries to calculate corrections to the mean field results in an order-by-order
fashion. Such a systematical approach was first adopted in Ref.
\cite{Balantekin:2006tg} where the authors developed a path integral
representation for the evolution operator of the exact many-body system and
showed that the application of the saddle-point approximation to this path
integral yielded identical flavor evolution equations with the mean field 
approximation. Having established this result, the authors wrote down a 
Gaussian integral which captures the next order correction. However, the 
numerical evaluation of this Gaussian integral proved to be very difficult.

More recently, a different approach was adopted in Refs. \cite{Volpe:2013uxl,
Volpe:2015rla} based on the Bogoliubov-Born-Green-Kirkwood-Yvon (BBGKY)
hierarchy \cite{Bogoliubov1, Born:1946vqa, Kirkwood, Yvon}. In this method,
instead of treating an $n$-body system with an $n$-body density operator (which
would be an exact treatment), one constructs a hierarchy of $m$-body density
operators for $m=1,2,\dots,n-1$. The mean field approximation corresponds to the
lowest order in this scheme and one can investigate the domain of validity of
the mean field approximation by calculating the next order terms in the
hierarchy.  BBGKY hierarchy is the most systematical way to go beyond the mean
field approximation. Unlike our study in this paper, it can be applied to any
initial state, and with enough computational power, one can look into the actual
dynamics of the system. Such systematical studies of the self-interacting 
neutrinos beyond the mean field approximation using  order-by-order methods, 
and the study of exact many-body solutions where they are available can nicely 
complement one another. In particular, it would be interesting to apply the 
BBGKY hierarchy method to the initial state that we consider, under the 
particular circumstances that we work with: That the mean field result is 
identical to the exact many-body result in this case might be indicative of 
some symmetries of the equations describing the $m>1$ terms in the hierarchy.

This paper is organized as follows: In Sec. II, we introduce the isospin
formalism based on the $SU(2)$ flavor symmetry of mixing neutrinos and discuss
the exact many-body Hamiltonian describing their vacuum oscillations and 
self-interactions. The isospin formulation helps us to emphasize the analogy
between self-interacting neutrinos, interacting spin systems, and fermionic
systems with pairing interaction that we discuss in Sec. II. In Sec. III, 
we discuss the
eigenstates of the many-body Hamiltonian in the two limits where self
interactions are very strong and very weak in comparison to the vacuum
oscillations. Finding many-body eigenstates in these two limits is a
simple exercise in algebra. In Secs IV-VI we elaborate on the exact
many-body eigenstates away from these limits: We
discuss the classification of those eigenstates with respect to the
$z$-component of the total neutrino mass isospin (Sec. IV), and apply the
Richardson-Gaudin diagonalization scheme with one (Sec. V) or more (Sec.
VI) Bethe ansatz variables. Note that Richardson-Gaudin diagonalization was
applied to self-interacting neutrinos in a previous study
\cite{Pehlivan:2011hp}. We recapitulate the some details only for a
subset of eigenstates that we are interested in here. In Sec. VII we
present our main results and show that for an ensemble of electron neutrinos,
the initial state projects only to those many-body eigenstates which do not
undergo any crossings as the neutrino density decreases, and that following the
transformation of those eigenstates with the assumption of a perfect adiabatic
evolution leads to a spectral split. We present our results for both a simple
box distribution and a thermal distribution of the initial neutrino ensemble. In
order to simplify our notation, we exclude antineutrinos from our main
discussion.  In the main text, we work only with neutrinos for simplicity, and
adopt the normal mass hierarchy. We convert our results into inverted
mass hierarchy in Sec. VIII, and show that antineutrinos of nonelectron
flavor can be included into the formalism without changing the main results of
the paper in Sec. IX. Sec. X is devoted to discussion and conclusions.

\section{The Hamiltonian}
\label{section:Hamiltonian and Isospin Formulation of the Problem}

\subsection{Isospin operators}
\label{subsection:Isospin Operators}

In this paper, we consider an (effective) two flavor mixing scenario between an
electron neutrino $\nu_e$ and an orthogonal flavor that we denote by $\nu_x$,
which can be a muon neutrino, a tau neutrino, or a normalized linear combination
of the two. The flavor eigenstates $\nu_e$ and $\nu_x$ are mixtures of two mass 
eigenstates $\nu_1$ and $\nu_2$. We denote a state in which we have a 
$\nu_\alpha$ (for $\alpha=1,2,e,x$) with momentum $\mathbf{p}$ by
$|\nu_\alpha,\mathbf{p}\rangle$. The corresponding annihilation operator is 
denoted by $a_\alpha(\mathbf{p})$. In principle there are other quantum numbers 
which distinguish the neutrinos with the same momentum from each other but we 
suppress them in our notation for simplicity. Neutrino operators in the flavor 
and mass bases are related by
\begin{equation}
\label{eq:fermion relations}
\begin{split}
a_e(\mathbf{p})&=\cos\theta \; a_1(\mathbf{p})+\sin\theta \; a_2(\mathbf{p}) \\
a_x(\mathbf{p})&=-\sin\theta \; a_1(\mathbf{p})+\cos\theta \; a_2(\mathbf{p})
\end{split}
\end{equation}
where $\theta$ is the mixing angle.

The $SU(2)$ symmetry of the two-dimensional flavor space gives rise to the
concept of neutrino isospin whereby one of these states is designated as
isospin up and the other as isospin down. The isospin assignment
is arbitrary, and in this paper we use the following isospin doublets in the 
mass
and flavor bases, respectively:
\begin{equation}
\label{doublets}
\begin{pmatrix}
|\nu_1,\mathbf{p}\rangle \\ |\nu_2,\mathbf{p}\rangle
\end{pmatrix}
\quad
\mbox{and}
\quad
\begin{pmatrix}
|\nu_e,\mathbf{p}\rangle \\ |\nu_x,\mathbf{p}\rangle
\end{pmatrix}.
\end{equation}
We emphasize that neutrino isospin is an entirely abstract concept which greatly
simplifies the calculations, and has nothing to do with the actual neutrino
spin. In this paper, neutrino spin does not play a role as we  completely ignore
the \emph{wrong} helicity states; i.e., we assume that all neutrinos have
negative helicity, whereas all antineutrinos have positive 
helicity.\footnote{The spin components which are ignored here, i.e., positive 
helicity neutrinos, and negative helicity antineutrinos come into play in a 
number of situations. For example, a strong magnetic field may flip the 
neutrino helicity and the resulting effect may be amplified by the nonlinear 
nature of collective oscillations \cite{deGouvea:2012hg, deGouvea:2013zp, 
Pehlivan:2014zua} Even without any magnetic fields, many-body correlations may 
develop between \emph{right} and \emph{wrong} helicity states in the presence of 
a net flow in the matter background as is the case in an exploding supernova 
\cite{Volpe:2013uxl, Serreau:2014cfa}. However, these effects are outside of the 
scope of this paper.}. For a neutrino with negative chirality, these components 
are suppressed by the ratio of neutrino mass to its energy.

The doublet structures given in Eq. (\ref{doublets}) lead to the definition of
neutrino isospin operators $\vec{J}_\mathbf{p}$ whose components are denoted by
\begin{equation}
\label{isospin operator}
\vec{J}_\mathbf{p}
=\left(
J^+_{{\tiny \mathbf{p}},{\mbox{\tiny mass}}},
J^-_{{\tiny \mathbf{p}},{\mbox{\tiny mass}}}
J^0_{{\tiny \mathbf{p}},{\mbox{\tiny mass}}}
\right)
=\left(
J^+_{{\tiny \mathbf{p}},{\mbox{\tiny flavor}}},
J^-_{{\tiny \mathbf{p}},{\mbox{\tiny flavor}}},
J^0_{{\tiny \mathbf{p}},{\mbox{\tiny flavor}}}
\right)
\end{equation}
in the mass and flavor bases, respectively. These components are given by
\begin{equation}
\begin{split}
\label{eq:isospin mass}
&{J^+_{{\tiny \mathbf{p}},{\mbox{\tiny 
mass}}}}=a^\dagger_1(\mathbf{p})a_2(\mathbf{p}),\quad
{J^{-}_{{\tiny \mathbf{p}},{\mbox{\tiny 
mass}}}}=a^\dagger_2(\mathbf{p})a_1(\mathbf{p}) \\
&{J^z_{{\tiny \mathbf{p}},{\mbox{\tiny 
mass}}}}={\frac{1}{2}}{\big(a^\dagger_1(\mathbf{p})a_1(\mathbf{p}
)-a^\dagger_2(\mathbf{p})a_2(\mathbf{p})\big)}
 \end{split}
\end{equation}
and
\begin{equation}
\label{eq:isospin flavor}
\begin{split}
&{J^+_{{\tiny \mathbf{p}},{\mbox{\tiny 
flavor}}}}=a^\dagger_e(\mathbf{p})a_x(\mathbf{p}),\quad
{J^{-}_{{\tiny \mathbf{p}},{\mbox{\tiny 
flavor}}}}=a^\dagger_x(\mathbf{p})a_e(\mathbf{p}) \\
&{J^z_{{\tiny \mathbf{p}},{\mbox{\tiny 
flavor}}}}={\frac{1}{2}}{\big(a^\dagger_e(\mathbf{p})a_e(\mathbf{p}
)-a^\dagger_x(\mathbf{p})a_x(\mathbf{p})\big)}.
\end{split}
\end{equation}
Note that we use bold letters to indicate vectors in configuration space 
and the arrows to indicate vectors in isospin space. The components 
$\vec{J}_\mathbf{p}$ satisfy the $SU(2)$ commutation relations
\begin{equation}
\label{su2}
[J^+_{{\tiny \mathbf{p}}}, J^-_{{\tiny \mathbf{q}}}]= 
2\delta_{\mathbf{p,q}}J^0_{{\tiny \mathbf{p}}}
\qquad
[J^0_{{\tiny \mathbf{p}}}, J^\pm_{{\tiny \mathbf{q}}}]=\pm 
\delta_{\mathbf{p,q}}J^\pm_{{\tiny \mathbf{p}}}.
\end{equation}
These relations hold in both bases and tell us that, if we have $n$ neutrinos
in the ensemble, the dynamics of the system takes place in the group space of
$SU(2)_1\otimes SU(2)_2\otimes \cdots SU(2)_n$.

In this paper, we mostly work in the mass basis. For simplicity we
drop the ``mass'' index from the isospin components in the mass basis, i.e.,
\begin{equation}
J^{\pm,0}_{{\tiny \mathbf{p}},{\mbox{\tiny mass}}}\to
J^{\pm,0}_{{\tiny \mathbf{p}}}.
\end{equation}
However, we keep the ``flavor'' index to distinguish it from the mass basis.
In the two flavor mixing scheme, a neutrino with momentum $\mathbf{p}$
oscillates with angular frequency
\begin{equation}
\label{w}
\omega=\frac{m_2^2-m_1^2}{2E}
\end{equation}
in vacuum. Here $E=|\mathbf{p}|$ is the energy of the neutrino and $m_i$ is the
mass of mass eigenstate $\nu_i$ for $i=1,2$. Since all neutrinos with the same
energy oscillate with the same frequency in vacuum, it is convenient to
define the total isospin operator
\begin{equation}
\label{eq:total isospin operators for each mode}
{\vec{J}_\omega}=\sum_{|\mathbf{p}|=E}{\vec{J}_{\mathbf{p}}}.
\end{equation}
This sum runs over all neutrinos with the same energy $E$ and is referred to as
the total isospin of the oscillation mode $\omega$. If there are
additional quantum numbers besides the momentum which distinguishes the
neutrinos, they are also summed over here. It is also useful to introduce the
total isospin operator $\vec{J}$ of the whole ensemble by summing over all
oscillation modes, i.e.,
\begin{equation}
\label{eq:global isospin operator}
\vec{J}=\sum_{\omega}{\vec{J}_{\omega}}.
\end{equation}

The relation between the mass and flavor bases given in
Eq. (\ref{eq:fermion relations}) can also be written in an operator form as
\begin{equation}
\label{eq:global SU(2) transformation}
\begin{split}
a_e(\mathbf{p})&=U^\dagger a_1(\mathbf{p})U \\
a_x(\mathbf{p})&=U^\dagger a_2(\mathbf{p})U
\end{split}
\end{equation}
where $U$ is given by
\begin{equation}
\label{eq:transformation operator}
U=e^{z{J^+}}e^{\ln(1+|z|^2){J^z}}e^{-z{J^-}}
\end{equation}
with $z=\tan\theta$. The operator $U$ represents a rotation by $\theta$ in the
flavor space. It is a global transformation in the sense that it involves the
total isospin operator $\vec{J}$ and acts in the same way on all neutrinos with
different momenta. Clearly, the isospin operators in mass and flavor bases are
also related by
\begin{equation}
\label{eq:transformation of isospin operators}
{\vec{J}_{\mathbf{p},{\mbox{\tiny flavor}}}}=U^\dagger \vec{J}_{\mathbf{p}} U
\end{equation}
With the repeated use of the Baker-Campbell-Hausdorff formulas, this leads
to
\begin{eqnarray}
\label{BCH formula}
J^z_{{\tiny \mathbf{p}},{\mbox{\tiny flavor}}} &=&\cos2\theta{J}^z_\mathbf{p}+
\frac{1}{2}\sin2\theta({J^+_\mathbf{p}}+{J^-_\mathbf{p}}) \nonumber \\
J^+_{{\tiny \mathbf{p}},{\mbox{\tiny flavor}}} 
&=&\cos^2\theta{J^+_\mathbf{p}}-\sin^2\theta{J^-_\mathbf{p}}
-\sin2\theta{J}^z_\mathbf{p} \\
J^-_{{\tiny \mathbf{p}},{\mbox{\tiny flavor}}} 
&=&\cos^2\theta{J^-_\mathbf{p}}-\sin^2\theta{J^+_\mathbf{p}}
-\sin2\theta{J}^z_\mathbf{p}\nonumber.
\end{eqnarray}
These formulas tell us that ${\vec{J}_{\mathbf{p},{\mbox{\tiny flavor}}}}$ is
obtained by rotating $\vec{J}_{\mathbf{p}}$ by $2\theta$ around the $y$-axis.

\subsection{Summation convention}
\label{subsection: summation convention}

Equations (\ref{eq:total isospin operators for each mode}) and (\ref{eq:global
isospin operator}) are particular examples of the general summation convention
that we use in this paper.  Any operator which is labeled by $\omega$ indicates
that it is summed over all neutrinos with the same energy corresponding to
$\omega$. The same quantity with no indices means that it is summed over all
neutrinos:
\begin{equation}
\label{eq: general summation convention}
Q_\omega=\sum_{|\mathbf{p}|=E}Q_{\mathbf{p}}
\qquad
Q=\sum_{\omega}Q_{\omega}.
\end{equation}
If an operator refers to only those neutrinos
in a particular flavor or mass eigenstate, we denote this with an upper index as
$Q^{(a)}_{\mathbf{p}}$, $Q^{(a)}_\omega$ and $Q^{(a)}$
where $a=1,2,e,x$. In that case, we have
\begin{equation}
\label{number summation convention}
Q_\mathbf{p}
=Q^{(1)}_\mathbf{p}+Q^{(2)}_\mathbf{p}
=Q^{(e)}_\mathbf{p}+Q^{(x)}_\mathbf{p}
\end{equation}
due to the completeness of both mass and flavor bases.

As an example, if we denote the number operator for those neutrinos in mass
eigenstate $\nu_i$ with momentum $\mathbf{p}$ as
\begin{equation}
\label{number operators}
N^{(i)}_\mathbf{p}= a_i^\dagger({\mathbf{p}})a_i({\mathbf{p}})
\end{equation}
then, $N^{(i)}_\omega$ and $N^{(i)}$ represent the number operator for
all $\nu_i$ neutrinos in the oscillation mode $\omega$, and in the entire
ensemble, respectively. In this case, Eqs. (\ref{eq: general summation
convention}) and (\ref{number summation convention}) tell us that
\begin{equation}
N_\omega =N^{(1)}_\omega+N^{(2)}_\omega
\quad\mbox{and}\quad
N =N^{(1)}+N^{(2)}
\end{equation}
denote the number operators for all neutrinos in the oscillation mode $\omega$
and in the entire ensemble, respectively.

We denote the eigenvalue relevant for the operator $Q_\mathbf{p}$ with the
corresponding lowercase letter $q_\mathbf{p}$. For example, the eigenvalue of
the operator in Eq. (\ref{number operators}) is denoted by $n^{(i)}_\mathbf{p}$.
It is important to note that $q_\omega$ and $q$ do not denote the sum of the
eigenvalues but the eigenvalues of the total operators $Q_\omega$ and $Q$,
respectively. While this distinction does not make a difference in some cases
(for example, for the number operators), it is important in the case of isospin.
If the isospin algebra $(J^+_\mathbf{p}, J^z_\mathbf{p},J^-_\mathbf{p})$ is
realized in the representation with quantum numbers $j_\mathbf{p}$, then
$j_\omega$ is the quantum number corresponding to the total isospin algebra
$(J^+_\omega, J^z_\omega,J^-_\omega)$.  Therefore, in principle, it can take all
values starting from $0$ or $1/2$ up to the literal sum
$\sum_{|\mathbf{p}|=E}j_{\mathbf{p}}$. The same is also true for the total
isospin quantum number $j$ of the whole neutrino ensemble.

Finally, we note that the only exception to our convention of denoting
eigenvalues of the operators with the corresponding lowercase letters are the
$z$ components of the isospins. The eigenvalues of $J^z_\mathbf{p}$,
$J^z_\omega$, and $J^z$ are denoted by $m_\mathbf{p}$, $m_\omega$, and $m$,
respectively.

\subsection{The Hamiltonian}
\label{subsection:Hamiltonian}

The Hamiltonian describing the propagation of neutrinos in vacuum is
given by
\begin{equation}
\label{eq:vacuum mixing mb H}
H_\nu=\sum_{\mathbf{p}}
\left(E_1(p)N^{(1)}_\mathbf{p}+E_2(p)N^{(2)}_\mathbf{p}\right)
\end{equation}
where $E_i(p)=\sqrt{\mathbf{p}^2+m_i^2}$ is the energy of the neutrino with mass
$m_i$ and momentum $\mathbf{p}$.  The Hamiltonian in Eq. (\ref{eq:vacuum mixing
mb H}) can also be written as
\begin{equation}
\label{eq:vacuum mixing mb H revised 0}
\begin{split}
H_\nu=\frac{1}{2}\sum_{\mathbf{p}}&\left\{(E_1(p)+E_2(p))
(N^{(1)}_\mathbf{p}+N^{(2)}_\mathbf{p})\right.
\\ &\left.
+(E_1(p)-E_2(p))(N^{(1)}_\mathbf{p}-N^{(2)}_\mathbf{p})\right\}.
\end{split}
\end{equation}
Here we consider neutrinos in the freely streaming regime, i.e., after
they decouple from the proto-neutron star and the processes which annihilate or
create them can be ignored. Therefore,  the total number of neutrinos
$N^{(1)}_\mathbf{p}+N^{(2)}_\mathbf{p}$ in any momentum mode $\mathbf{p}$ is a
constant. In treating such a problem, it is natural to start with an initial
state which is an eigenstate of the total number operator $N_{\mathbf{p}}$. In
this case the first term in Eq.  (\ref{eq:vacuum mixing mb H revised 0}) is only
a number and can be dropped from the Hamiltonian. In the second term,
the ultrarelativistic approximation $E_i(p)\approx p + \frac{m_i^2}{2p}$ can be
applied which leads to
\begin{equation}
\label{eq:vacuum mixing mb H revised}
H_\nu=\sum_{\omega}\;\omega\;\hat{B}\cdot \vec{J}_\omega
\end{equation}
where $\omega$ is the vacuum oscillation frequency given by Eq. (\ref{w}) and
the unit vector $\hat{B}$ is defined as
\begin{equation}
\label{eq:magnetic field}
\hat{B}=(0,0,-1)_{\mbox{\tiny mass}}
\end{equation}
in the mass basis. Here the appearance of the minus sign in the third component
is due to our adoption of the normal mass hierarchy (i.e., $m_1<m_2$).  In
writing Eq. (\ref{eq:vacuum mixing mb H revised}), we also used the definition
of neutrino isospin in the mass basis given in Eq. (\ref{eq:isospin mass}). 
Since the Hamiltonian is a scalar, it has the same form as in Eq. 
(\ref{eq:vacuum mixing mb H revised}) in the flavor basis as well.  When 
creation and annihilation operators are rotated by $\theta$ as given in Eq.  
(\ref{eq:fermion relations}), the isospin operators which are quadratic in them 
are rotated by $2\theta$. As a result, the components of $\hat{B}$ are given by
\begin{equation}
\label{eq:magnetic field in flavor}
\hat{B}=(\sin2\theta,0,-\cos2\theta)_{\mbox{\tiny flavor}}
\end{equation}
in the flavor basis.

\begin{figure}
\includegraphics[scale=0.9]{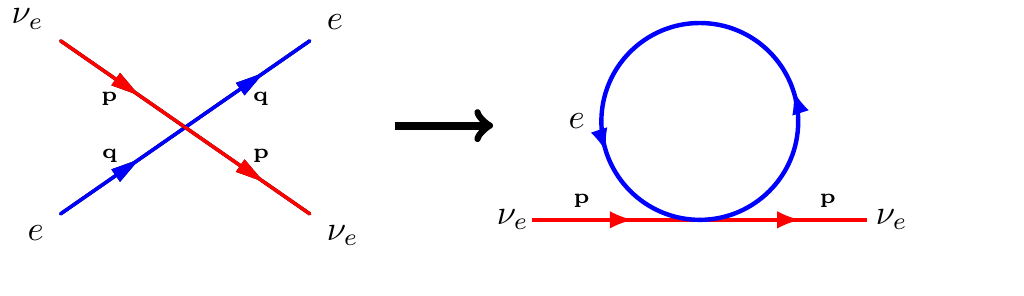}
\caption{\footnotesize \raggedright Forward scattering of 
neutrinos from the background particles involves no momentum exchange. These 
diagrams always add up coherently (i.e., at the amplitude level) and dominate 
over the other diagrams which add up incoherently (i.e., at the probability 
level). They manifest themselves as an effective mass which is well defined in 
the weak interaction basis.}
\label{neutrino e forward}
\end{figure}

In the free-streaming regime outside of the proto-neutron star the only
important effect is scattering, which can significantly modify the flavor
evolution of neutrinos.  The scattering of neutrinos from each other and from
other background particles should be discussed separately because in the former
case identical particle effects play an important role. In a general 
astrophysical environment (as opposed to, say, a periodic crystal), scattering 
amplitudes from different background particles into a given direction generally 
add up incoherently, so their combined effect increases only linearly with the 
density of background particles. However, in the forward direction, scattering 
amplitudes from different background particles always add up coherently, in 
which case their combined effect increases quadratically with the background 
density. Therefore, in a dense environment, it is enough to work with an 
effective Hamiltonian which only includes the forward scattering terms in which 
there is no momentum transfer between neutrinos and the background particles
\cite{Wolfenstein:1977ue, PhysRevD.22.2718} (also see Ref.
\cite{PhysRevD.27.1228}). When averaged over the background, such terms manifest
themselves in the form of an effective mass (Fig. \ref{neutrino e forward}). The
total mass includes both the effective mass which is diagonal in the weak
interaction basis ($\nu_e$, $\nu_x$) and the ordinary neutrino mass which is
diagonal in the mass basis ($\nu_1$, $\nu_2$), and it can be diagonalized in a
new basis, which is called the matter basis ($\tilde{\nu}_1$,
$\tilde{\nu}_2$). Therefore, the net effect of the background particles is to
modify the mixing angle into a corresponding matter effective value. (For
a review, see Ref. \cite{Kuo:1989qe}).  These effective mixing parameters depend
on the background density and they vary as the neutrinos move from the inner
dense regions of the supernova into to outer layers. However, if the density
does not change significantly in the region of a few hundred kilometers outside
of the proto-neutron star where the collective neutrino oscillations occur, one
can assume that the effective mixing parameters are approximately constant.
This would be a good approximation for the cooling period of the proto-neutron
star since the shock wave is far away from its surface at those later times.
(See Ref. \cite{Janka:2006fh} for a review.) However, at earlier times, the
changes in the density profile just outside of the proto-neutron star are more
dramatic. In this paper, our methods and conclusions are independent of the
actual values of the (effective) mixing parameters, as long as they can be
considered constant. In particular, the Richardson-Gaudin diagonalization method
depends on the constant density assumption in its present form. In what follows,
our notation will refer to the vacuum values of the mixing parameters, but they
can be easily exchanged with constant matter effective values.
\begin{figure}
\hspace*{5mm}
\begin{subfigure}[t]{0.5\columnwidth}
\hspace*{-10mm}
\vspace*{-5mm}
\includegraphics[scale=1]{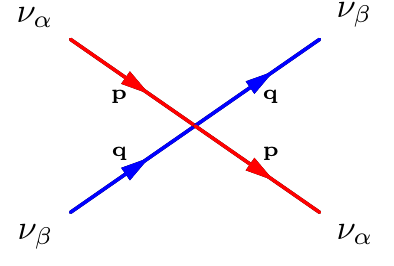}
\caption{\footnotesize \hspace*{1cm}}
\label{nu-nu forward}
\end{subfigure}%
\begin{subfigure}[t]{0.5\columnwidth}
\hspace*{-10mm}
\vspace{-5mm}
\includegraphics[scale=1]{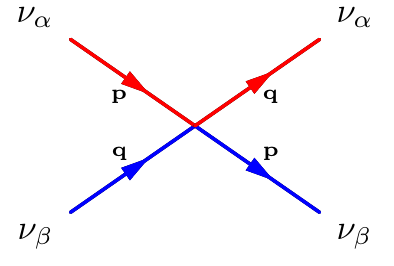}
\caption{\footnotesize \hspace{1cm}\label{nu-nu exchange}}
\end{subfigure}
\vspace*{-3mm}
\captionsetup{justification=RaggedRight}
\caption{\footnotesize Forward (a) and exchange (b) diagrams 
which add up coherently in $\nu\nu$ scattering.\label{nu-nu} }
\label{neutrino-neutrino scattering}
\vspace*{-5mm}
\end{figure}

When the scattering of neutrinos from each other is considered, in addition to
those diagrams which involve no momentum transfer (forward scattering diagrams 
discussed
above), those diagrams which involve a complete momentum exchange between
neutrinos also add up coherently (Fig. \ref{neutrino-neutrino scattering})
\cite{Pantaleone:1992eq,Pantaleone:1992xh}. This is due to
the fact that, these diagrams can also be viewed as forward scattering diagrams
in which neutrinos exchange their flavors. The effect of the exchange diagrams
cannot be included in the form of effective mixing parameters as in the case of
forward scattering. In fact, with the inclusion of the exchange diagrams, the
problem turns into a many-body phenomenon because the flavor transformation of
each neutrino is now affected by the flavor evolution of the entire neutrino
ensemble.

The effective Hamiltonian which describes the forward scattering and exchange
diagrams between neutrinos is given by \cite{Sawyer:2005jk,Pehlivan:2011hp}
\begin{eqnarray}
\label{eq:self-interaction H}
H_{\nu\nu}&=&\frac{G_F}{\sqrt2 V}\sum_{\mathbf{p}}\sum_{\mathbf{q}}
\left(1-\hat{\mathbf{p}}\cdot\hat{\mathbf{q}}\right)
\left\{
a^\dagger_e(\mathbf{p})a_e(\mathbf{p})
a^\dagger_e(\mathbf{q})a_e(\mathbf{q})\right.\nonumber \\
&& +a^\dagger_x(\mathbf{p})a_x(\mathbf{p})
a^\dagger_x(\mathbf{q})a_x(\mathbf{q})
+a^\dagger_x(\mathbf{p})a_e(\mathbf{p})
a^\dagger_e(\mathbf{q})a_x(\mathbf{q}) \nonumber \\
&&\left.+a^\dagger_e(\mathbf{p})a_x(\mathbf{p})
a^\dagger_x(\mathbf{q})a_e(\mathbf{q})\right\}
\end{eqnarray}
where the first two terms in the curly brackets correspond to the forward
scattering diagrams shown in Fig. \ref{nu-nu forward} whereas the last two
terms represent the exchange diagrams shown in Fig. \ref{nu-nu exchange}. In
writing this Hamiltonian, the space coordinates are integrated out with the
assumption of spacial uniformity, so one effectively works with neutrino
plane waves. Neutral current interactions between neutrinos are treated with
the Fermi 4-point interaction scheme and $G_F$ denotes the Fermi constant. This
description is accurate for the MeV scale energies relevant for the supernova
neutrinos. We assume that neutrinos are quantized in a box with volume $V$ so
that the momentum $\mathbf{p}$ and its direction $\hat{\mathbf{p}}$ can take
discrete values. As we follow the neutrinos in the comoving frame from the
surface of the proto-neutron star to the point where neutrino self-interactions
become negligible, this box expands corresponding to a decreasing neutrino
density. But since we ignore neutrino creation in the free-streaming regime, new
momentum modes do not appear.

The relativistic factor $1-\hat{\mathbf{p}}\cdot\hat{\mathbf{q}}$ in the
Hamiltonian of Eq. (\ref{eq:self-interaction H}) implies that
relativistic neutrinos traveling along parallel paths cannot scatter off each
other. This term turns the flavor evolution of a neutrino into a function of its
direction of travel and significantly complicates the problem. Replacing this
term with an average constant value results in the so called single angle
approximation in which neutrinos are assumed to undergo identical flavor
evolutions regardless of their direction. In this paper, we adopt the single
angle approximation together with the neutrino bulb model
\cite{Duan:2006an} which approximately accounts for the fact that the average
value of the angle between the neutrinos, and hence the factor
$1-\hat{\mathbf{p}}\cdot\hat{\mathbf{q}}$, decreases with the distance $r$ from
the center of the supernova by replacing the latter with
\begin{equation}
\label{D}
1-\hat{\mathbf{p}}\cdot\hat{\mathbf{q}}\approx
D(r)=\frac{1}{2}\left(1-\sqrt{1-\frac{r^2}{R_\nu^2}}\right)^2.
\end{equation}
Here $R_\nu$ denotes the radius of the neutrino-sphere which is an imaginary
sphere just inside the surface of the proto-neutron star from which neutrinos
thermally decouple and start free streaming. In principle, $R_\nu$ is a
function of time, and it decreases from almost $100$ km at the time of the
bounce to about $10$ km at late times \cite{Roberts:2016rsf}. However, the
character of collective neutrino oscillations does not depend strongly on the
value of $R_\nu$ \cite{Duan:2010bf}.

Using this approximation scheme, and the neutrino isospin operators given in Eq.
(\ref{eq:isospin flavor}) together with the adopted summation conventions, one
can write the Hamiltonian in Eq. (\ref{eq:self-interaction H}) as
\begin{equation}
\label{eq:self-interaction H revised}
H_{\nu\nu}=\mu(r)\vec{J}\cdot\vec{J}
\end{equation}
where we discard some terms which are proportional to the total number of 
neutrinos in accordance with the discussion following Eq. (\ref{eq:vacuum
mixing mb H revised 0}). We also use the fact that scalar quantities have the 
same form in both mass and flavor bases, i.e., 
$\vec{J}\cdot\vec{J}=\vec{J}{\mbox{\tiny flavor}}\cdot\vec{J}{\mbox{\tiny
flavor}}$. In Eq. (\ref{eq:self-interaction H revised}),
\begin{equation}
\mu(r)=\frac{\sqrt{2}G_F}VD(r)
\end{equation}
plays the role of an effective interaction constant. Since the normalization
volume $V$ is inversely proportional to the neutrino density, it increases as
$r^2$ with distance from the proto-neutron star. Together with the change of the
average angle between the neutrinos in accordance with Eq. (\ref{D}), $\mu(r)$
decreases roughly as ${1}/{r^4}$.  The total Hamiltonian of an ensemble of
neutrinos undergoing vacuum oscillations and self-interactions is found by
adding Eqs. (\ref{eq:vacuum mixing mb H revised}) and (\ref{eq:self-interaction
H revised}):
\begin{equation}
\label{eq:H Total}
H=\sum_\omega\omega\hat{B}\cdot\vec{J}_\omega+\mu(r){\vec{J}}\cdot{\vec{J}}.
\end{equation}
In the rest of this paper, we work with this Hamiltonian.

It was already pointed out by several authors that Eq. (\ref{eq:H Total}) is
analogous to the Hamiltonian of a hypothetical one-dimensional spin system with
long-range interactions in the presence of a position-dependent external
magnetic field, given by
\begin{equation}
\label{eq:H spin}
H_{\mbox{\tiny spin}}=\sum_i H_i \;
\hat{B}\cdot\vec{S}_i+G(t){\vec{S}}\cdot{\vec{S}}.
\end{equation}
Here, $i$ is a discrete position parameter in one dimension, and it is assumed
that a (real) spin $\vec{S}_i$ is located at that point. The external magnetic
field everywhere points in the direction of $\hat{B}$, while its magnitude $H_i$
at $i$ may be position dependent.\footnote{Magnetic permeability and 
gyromagnetic ratio at the cite $i$ are inserted into the definition of the 
magnetic field so that the latter is in units of energy.} In Eq. (\ref{eq:H 
spin}), $\vec{S}=\sum_i \vec{S_i}$ denotes the total spin of the system. It 
appears in the Hamiltonian because the range of the spin-spin interactions is 
assumed to be infinite, so every spin in the system interacts with every 
other one with the same strength. The analogy of this problem with the 
self-interacting neutrinos is clear once one identifies the spin-up 
($|\uparrow\rangle$) and spin-down ($|\downarrow\rangle$) states of the former 
with the isospin states of the latter as
\begin{equation}
\label{analogy 1}
|\uparrow\rangle\leftrightarrow|\nu_1\rangle
\qquad
|\downarrow\rangle\leftrightarrow|\nu_2\rangle.
\end{equation}
The interaction strength $G(t)$ in Eq. (\ref{eq:H spin}) may depend on time $t$.
In particular, in the context of the analogy with self-interacting neutrinos
$G(t)$ is assumed to decrease with time. In this case the spins are strongly
correlated at the beginning when $G$ is large, but their dynamics are dominated
by the external field at later times when $G$ is small. This analogy gives us a
nice picture of spectral splits because, under the adiabatic evolution
conditions, the spins eventually align or antialign themselves with the
magnetic field as their mutual interactions slowly cease. Since $\hat{B}$ points
in the direction of $J_z$ in the isospin space (see Eqs. (\ref{eq:isospin mass})
and (\ref{eq:magnetic field})), the neutrinos end up in one of the mass (or
matter) eigenstates after the collective oscillations cease 
\cite{Raffelt:2007xt}.

Correlated spin systems like the one described in Eq. (\ref{eq:H spin}) is a
popular subject in many-body physics because several problems with internal
$su(2)$ symmetries can be described in analogy with them. Besides the self
interacting neutrinos considered here, another example is a system of fermions
with pairing interactions. While neutrino isospin plays the role of the spins in
the former case, this role is played by the so-called pair quasispin in
the latter. As a result, an analogy also exists between self-interacting
neutrinos, and fermions with pairing interaction whereby the neutrino isospin
and the pair quasispin are the analogous quantities. However, this analogy does
not imply that supernova neutrinos form pairs. As is explained below, it is a
more subtle analogy which should only be thought of as a mathematical
similarity.

Pairing interaction appears in several fermionic many-body systems. It was
originally suggested by Bardeen, Cooper, and Schrieffer (BCS) in connection with
their theory of superconductivity \cite{Bardeen:1957mv} as an effective
interaction between electrons in the presence of a lattice. Soon it was realized
that pairing also plays an important role in the nuclear shell model as the 
residual interaction between nucleons \cite{1950PhRv...78...16M, 
1950PhRv...78...22M}. In trapped ultracold atomic systems, a pairing 
interaction can be created between fermionic atoms \cite{RevModPhys.80.885, 
RevModPhys.80.1215, Randeria:2013kda}, and its strength can be controlled via 
Feshbach resonances by changing the applied magnetic field \cite{Inouye1998}. 
This is particularly important as it allows direct experimental access to the 
behavior of the many-body system as the interaction constant changes with time.

The pairing model is described by the Hamiltonian
\begin{equation}
\label{eq:pairing Hamiltonian}
H_{\mbox{\tiny pair}}=\sum_k \sum_i \epsilon_k c^\dagger_{ki}
c_{ki}^{\phantom{\dagger}}-g(t)\sum_{kk^\prime} \sum_{ii^\prime}c^\dagger_{ki} 
c^\dagger_{\bar{k}i}
c_{\bar{k}^\prime i^\prime}^{\phantom{\dagger}}c_{k^\prime 
i^\prime}^{\phantom{\dagger}}\msp.
\end{equation}
where $\epsilon_k$ denotes a group of possibly degenerate energy levels, with
the index $i$ running over such degeneracies. These levels can be either empty
or occupied by a pair of spin-up and spin-down fermions,
\footnote{Singly occupied energy levels decouple from the pairing dynamics in
these kinds of models because pairs cannot scatter into these levels due to the
Pauli exclusion principle. Although such levels can be important for other
characteristics of the system under consideration, they are irrelevant for our
purposes.}
which are created by the operators $c^\dagger_{ki}$ and
$c^\dagger_{\bar{k}i}$, respectively. The interaction strength $g(t)$ is
a function of time in general. The pair quasispin operators mentioned above are
defined by
\begin{equation}
\begin{split}
\label{eq:quasispin operators}
&K^+_{ki}=c^\dagger_{ki} c^\dagger_{\bar{k}i} \quad K^-_{ki}=
c_{\bar{k}i}^{\phantom{\dagger}} c_{ki}^{\phantom{\dagger}}\\
&K^z_{ki}=\frac{1}{2}\left(c^\dagger_{ki} c_{ki}^{\phantom{\dagger}}
-c_{\bar{k}i}^{\phantom{\dagger}}
c^\dagger_{\bar{k}i}\right)\msp,
\end{split}
\end{equation}
and satisfy the $SU(2)$ commutation relations given in Eq. (\ref{su2}), with
$\vec{J_\mathbf{p}}$ replaced with $\vec{K}_{ki}$.
These definitions reflect the use of the quasispin doublet
\begin{equation}
\label{quasispin}
\left( \begin{array}{c}
\f \\ \e
\end{array}\right)
\end{equation}
in which an empty level $|\e\rangle$ is defined to have quasispin down while an
occupied level $|\f\rangle$ is defined to have quasispin up.
Defining a summation convention analogous to the ones introduced for
neutrinos whereby $\vec{K}_k$ denotes the isospin
operator which is summed over the degeneracy index $i$, and $\vec{K}$ denotes
the total isospin operator summed over the index $k$, we can write the
Hamiltonian in Eq.\s(\ref{eq:pairing Hamiltonian}) as
\begin{equation}
\label{eq:pairing Hamiltonian in quasi-spin formalism}
 H_{\mbox{\tiny pair}}=\sum_k 2\epsilon_k K^z_k-g(t) \vec{K}\cdot\vec{K}
\msp.
\end{equation}
In writing this Hamiltonian, we assume that the system contains a definite
number of pairs, and discard a constant term related to this number.  The
similarity with the previous models becomes apparent with the identification
\begin{equation}
\label{analogy 2}
|\f\rangle\leftrightarrow|\uparrow\rangle\leftrightarrow|\nu_1\rangle
\qquad
|\e\rangle\leftrightarrow|\downarrow\rangle\leftrightarrow|\nu_2\rangle
\end{equation}

The pairing problem was solved in the 1960s by Richardson 
\cite{Richardson:1963, RICHARDSON1964253, Richardson:1966zza} who was able to 
analytically write down the exact eigenstates and eigenvalues of the 
Hamiltonian given in Eq. (\ref{eq:pairing Hamiltonian in quasi-spin 
formalism}). His solution was based on the Bethe ansatz technique 
\cite{Bethe:406121}.  In this method, one first forms a trial eigenstate 
depending on some unknown parameters and then tries to determine the values of 
these parameters by substituting the state into the eigenvalue-eigenstate 
equation. This process yields a coupled set of algebraic equations in the 
unknown parameters which are known as the Bethe ansatz equations. In 
1976, also using the Bethe ansatz technique, Gaudin \cite{Gaudin:1976} solved a 
family of interacting spin model Hamiltonian's which are today known as the 
(rational) Gaudin magnet Hamiltonians. These Hamiltonians were, at first 
glance, unrelated to the spin Hamiltonian given in Eq. (\ref{eq:H spin}), but 
Gaudin found the same Bethe ansatz equations as Richardson. In 1997, 
unaware of both Richardson's and Gaudin's work, Cambiaggio, Rivas, and Saraceno 
\cite{Cambiaggio:1997vz} showed that the Gaudin magnet Hamiltonians are in 
fact constants of motion of the pairing Hamiltonian given in Eq. 
(\ref{eq:pairing Hamiltonian}) corresponding to its dynamical symmetries. 
Today, we have a complete picture in which (a larger class of) Gaudin magnet
Hamiltonians, and all models which are related or analogous to them, can be
solved exactly by using the Bethe ansatz technique. For a review, see Ref.
\cite{Dukelsky:2004re}.

The analogy between self-interacting neutrinos, and the fermions with pairing
interaction was first pointed out in Ref. \cite{Pehlivan:2011hp}, where the
Richardson-Gaudin solution was used to obtain the exact many-body eigenstates
and eigenvalues of the neutrino Hamiltonian given in Eq. (\ref{eq:H Total}). It
was also pointed out that the Gaudin magnet Hamiltonians mentioned above form a
set of invariants for the collective neutrino oscillations. (These invariants
were also mentioned in Ref. \cite{Raffelt:2011yb} at the mean field level.)
Reference \cite{Pehlivan:2011hp} also showed that, at the mean field level, 
and for an initial box distribution of electron neutrinos, the formation of a 
spectral split can be viewed as the evolution of relevant fermionic degrees of 
freedom from quasiparticle to particle degrees of freedom in the pairing model. 
A more recent study \cite{Pehlivan:2016lxx} found that, again at the mean 
field level, but for a more generic initial energy distribution of electron 
neutrinos, formation of a spectral split is analogous to the BCS-Bose Einstein
condensation (BEC) crossover which was experimentally observed in cold atom
systems \cite{PhysRevLett.92.120403, RevModPhys.80.885, RevModPhys.80.1215,
Randeria:2013kda}. Here we will further exploit the analogy between self
interacting neutrinos and the fermions with pairing interaction to show, for the
first time, the emergence of a spectral split in the exact many-body picture.

\section{Eigenstates in special cases}
\label{section:Eigenstates in special cases}

The Bethe ansatz method can be applied to the neutrino Hamiltonian for any 
value of
the interaction constant $\mu$. However, in the limits of strongly and weakly
interacting systems, one can find the eigenstates and eigenvalues by
conventional methods as well. For the sake of an intuitive understanding, it is
useful to study these limits first before we apply the Richardson-Gaudin
diagonalization for the arbitrary values of the interaction constant.

\subsection{Eigenstates at $\mu\to 0$ limit}
\label{subsection:Eigenstates at mu=0 limit}

When the neutrino density is very low, as would be the case when neutrinos are 
far from the center of the supernova, the self-interaction term of the 
Hamiltonian in Eq. (\ref{eq:H Total}) can be ignored. In this case, the 
Hamiltonian only consists of the vacuum oscillation terms:
\begin{equation}
\label{eq:H low mu limit}
\lim_{\mu\to 0} H=\sum_\omega\omega\hat{B}\cdot\vec{J}_\omega
=-\sum_\omega\omega J_\omega^z.
\end{equation}
Since there is no coupling between different oscillation modes in this limit,
the eigenstates of the Hamiltonian are tensor products of the eigenstates of
individual isospin components $J_\omega^z$. In other words, they can be written
as
\begin{align}
\label{eq: mu=0 eigenstate}
\prod_\omega |j_\omega,m_\omega\rangle.
\end{align}
The eigenvalue corresponding to the state in Eq. (\ref{eq: mu=0
eigenstate}) is given by
\begin{align}
\label{energy for mu=0}
E=-\sum_\omega \omega m_\omega.
\end{align}
Note that since the total isospin quantum number $j_\omega$ can take several
(degenerate) values ranging from $0$ or $1/2$ to $n_\omega/2$, the states
$|j_\omega,m_\omega\rangle$ form a reducible representation. However, in what
follows, we will assume that $\vec{J}_\omega$ lives in the highest weight
representation, i.e., $j_\omega=n_\omega/2$ for every $\omega$. This choice is 
dictated by the symmetries of our simplified model and the initial state. Our 
simplified Hamiltonian given in Eq. (\ref{eq:H Total}) does not include any 
dependence on the position, or propagation direction of the neutrinos. In fact, 
it remains unchanged if we exchange any two neutrinos with the same energy. 
Moreover, in this paper we restrict ourselves to the study of a neutrino 
ensemble which initially consists only of electron neutrinos; i.e., our 
initial state is completely symmetric under the exchange of any two 
neutrinos including those with different energies.%
\footnote{Since neutrinos are fermions, their total many-body state, which
consists of spin, isospin and space parts, is antisymmetric under the
exchange of any two neutrinos. Here, our symmetry assumption applies only to the
isospin part.}%
Naturally the symmetry of the initial state between different energy modes
will be broken by the vacuum oscillations once they start. But when the state
evolves according to the Hamiltonian in Eq. (\ref{eq:H Total}), it will continue
to be symmetric under the exchange of any two neutrinos with the same energy.
Among the possible $|j_\omega,m_\omega\rangle$ states, only those that live in 
the
highest weight ($j_\omega=n_\omega/2$) representation satisfy this requirement.
That is to say, only those states are invariant under the exchange of any two
neutrinos with the same $\omega$. For this reason, although there are many more
eigenstates of the Hamiltonian, the dynamics of our initial state is restricted
to those which involve only the highest weight representation for each $\omega$.

Note that the total isospin quantum number $j$ of all neutrinos
is \emph{not} restricted by this symmetry. In
accordance with Eq. (\ref{eq:global isospin operator}), $j$ can take any value
from $0$ up to $n/2$.

Instead of using the basis in which the isospins are summed for each $\omega$,
one can also use the basis of individual neutrino isospins. Since the isospin 
being up or down corresponds to the neutrino being in the first or second 
mass eigenstate, respectively, the many-body eigenstates in Eq. (\ref{eq: mu=0
eigenstate}) can also be written as
\begin{equation}
\label{eq: mu=0 eigenstate 2}
|\nu_{i_1},\nu_{i_2}, \dotsc \nu_{i_n}\rangle
\end{equation}
in the $\mu\to 0$ limit, where $i_k=1,2$. As per our discussion above, those
neutrinos in the same oscillation modes should be symmetrized. The eigenstates
in Eq. (\ref{eq: mu=0 eigenstate 2}) are in a form that one would intuitively
expect because neutrinos in mass eigenstates do not oscillate in vacuum and
therefore form the stationary states of the Hamiltonian when there are no
interactions. However, those eigenstates in Eq. (\ref{eq: mu=0 eigenstate}) are
more suitable for the Bethe ansatz scheme. These two sets of eigenstates are
related by a set of complicated Clebsh-Gordon coefficients which will not be
reproduced here.

\subsection{Eigenstates at $\mu\to \infty$ limit}
\label{subsection:Eigenstates at mu=infty limit}

When neutrino density is sufficiently high so that $\mu$ is much larger than
the relevant $\omega$ values in the system, one can ignore the vacuum
oscillations in the Hamiltonian in Eq. (\ref{eq:H Total}):
\begin{equation}
\label{eq:H hign mu limit}
\lim_{\mu\to \infty} H=\mu(r){\vec{J}}\cdot{\vec{J}}
\end{equation}
This limit would be realized when the neutrinos are close to the proto-neutron
star at the center. The eigenstates of the Hamiltonian in this limit are the
$|j,m\rangle$ states of the total (mass) isospin with the eigenvalue $\mu
j(j+1)$:
\begin{equation}
\label{eq:eigenstates at mu=infinite}
\lim_{\mu\to\infty} H |j,m\rangle=\mu j(j+1)|j,m\rangle
\end{equation}
Here the total isospin quantum number $j$ can take any value from $0$ or $1/2$
to $n/2$. We emphasize once again that our assumption of highest weight
representation applies only to the total isospin of individual oscillation modes
$\vec{J}_\omega$, not to the total isospin $\vec{J}$.

The operator $U$ given in Eq. (\ref{eq:transformation operator}) converts the
mass isospin states to flavor isospin states:
\begin{equation}
\label{eq:eigenstates at mu=infinite, flavor}
|j,m\rangle_{\mbox{\tiny flavor}}=U|j,m\rangle \end{equation}
This can be seen by summing Eq. (\ref{eq:transformation of isospin operators})
over all neutrinos, and using it on both sides of Eq. (\ref{eq:eigenstates at
mu=infinite, flavor}). The operator $U$ commutes with the Hamiltonian in Eq.
(\ref{eq:H hign mu limit}), which tells us that the total flavor isospin states
given in Eq. (\ref{eq:eigenstates at mu=infinite, flavor}) are also eigenstates
of the Hamiltonian with the same eigenvalue in the $\mu\to\infty$ limit. This 
can also be seen by noting that, when $U$ acts on $|j,m\rangle$, it cannot 
change the value of $j$. In other words, the right-hand side of Eq. 
(\ref{eq:eigenstates at mu=infinite, flavor}) yields
\begin{equation}
\label{eq:eigenstates at mu=infinite, flavor 2}
|j,m\rangle_{\mbox{\tiny flavor}}
=\sum_{m=-j}^j \alpha_m^{(j)}(z) |j,m\rangle
\end{equation}
where $\alpha_m^{(j)}(z)$ are some coefficients that can be calculated from Eq.
(\ref{eq:transformation operator}). Since all the states on the right-hand side 
of Eq. (\ref{eq:eigenstates at mu=infinite, flavor 2}) are degenerate with 
energy $\mu j(j+1)$ in the $\mu\to\infty$ limit, the state on the left-hand 
side is also an eigenstate with the same energy in this limit.

\section{Number Conservation and Classification of Eigenstates}
\label{subsection:Eigenstates for any mu}

Richardson-Gaudin diagonalization \cite{Richardson:1963, RICHARDSON1964253,
Richardson:1966zza, Gaudin:1976} was applied to self-interacting neutrinos in
Ref. \cite{Pehlivan:2011hp}, and the resulting eigenvalues were presented in
their most generic form. In this section, we reproduce the relevant results of
Ref. \cite{Pehlivan:2011hp} both for the convenience of the reader and to set 
our notation. Here we restrict ourselves to only those eigenstates which meet 
our symmetry criteria, i.e., only those involving the highest weight 
representations $j_\omega=n_\omega/2$ for each $\omega$. (See the discussion 
following Eq. (\ref{energy for mu=0}).)

The Hamiltonian in Eq. (\ref{eq:H Total}) commutes with the operator
$\vec{B}\cdot\vec{J}= J^z$. In the interacting spin model analogy, this
corresponds to the fact that the problem is unchanged if we rotate the spin
system around the magnetic field. For neutrinos, it reflects the fact that we
can multiply mass eigenstates with arbitrary phases without changing the
Hamiltonian.  Since $J_z$ commutes with the Hamiltonian,\footnote{This is only
one of the many conserved quantities related to the dynamical symmetries of the
exact many-body Hamiltonian \cite{Pehlivan:2011hp}.} it can be diagonalized
together with it; i.e., for an energy eigenket $|\psi_E\rangle$, we can always
write
\begin{equation}
\label{m}
J_z|\psi_E\rangle=m|\psi_E\rangle.
\end{equation}
In what follows we classify eigenstates of the Hamiltonian according to Eq.
(\ref{m}).

Using Eqs. (\ref{eq:isospin mass}) and (\ref{number operators}), together with
the related summation conventions, we can write $J_z$ as
\begin{equation}
\label{eq: number conservation}
J^z=\frac{N^{(1)}-N^{(2)}}{2}.
\end{equation}
Since we are in the free-streaming regime, the total number of neutrinos
$N=N^{(1)}+N^{(2)}$ is also conserved. Together with Eq. (\ref{eq: number
conservation}), this tells us that $N^{(1)}$ and $N^{(2)}$ are separately
conserved, and they can also be diagonalized together with the Hamiltonian. For
the energy eigenket in Eq. (\ref{m}) we can write
\begin{equation}
\label{n1 and n2}
n^{(1)}=\frac{n}{2}+m
\quad \mbox{and} \quad
n^{(2)}=\frac{n}{2}-m.
\end{equation}
For example, for those states with $(n^{(1)},n^{(2)})=(n,0)$ and 
$(n^{(1)},n^{(2)})=(0,n)$ the action of $J_z$ yields
\begin{equation}
\label{m explicit}
m=\frac{n^{(1)}-n^{(2)}}{2}=\pm \frac{n}{2}
\end{equation}
respectively. Therefore, these states can only belong to the $j=n/2$ 
representation. In fact, they are, respectively, the highest and lowest weight 
states of the total isospin given by
\begin{equation}
\begin{split}
\label{eq:highest-lowest weight states}%
|n/2,+ n/2\rangle&=\prod_\omega |\frac{n_\omega}{2}, \; +
\frac{n_\omega}{2}\rangle=|\nu_1,\nu_1,\dots,\nu_1\rangle\\
|n/2,- n/2\rangle&=\prod_\omega |\frac{n_\omega}{2}, \; - 
\frac{n_\omega}{2}\rangle
=|\nu_2,\nu_2,\dots,\nu_2\rangle.
\end{split}
\end{equation}%
These are also the simultaneous eigenstates of ${\vec{J}}\cdot{\vec{J}}$ and
$J^z_{\omega}$ with the respective eigenvalues $n/2(n/2+1)$ and
$\pm\frac{n_\omega}{2}$. As a result, they are eigenstates of the total
neutrino Hamiltonian for any value of $\mu$, i.e.,
\begin{equation}
\label{eq:Energy eigenvalues of highest/lowest weight states}
H\;|n/2,\pm n/2\rangle=E_{\pm n/2} \;|n/2,\pm
n/2\rangle \end{equation}
with
\begin{equation}
\label{eq:E_max}
E_{\pm n/2}=\mp \sum_{\omega} \omega \frac{n_\omega}{2}
+ \mu \; \frac{n}{2} (\frac{n}{2} +1).
\end{equation}
It is easy to understand intuitively why the two states in Eq.
(\ref{eq:highest-lowest weight states}) are eigenstates of the Hamiltonian. A
hypothetical ensemble of neutrinos which are all in mass eigenstates would not
undergo vacuum oscillations. If, in addition, all of these neutrinos occupy the
\emph{same} mass eigenstate (i.e., all $\nu_1$ or all $\nu_2$), then the
many-body state would also remain unchanged under the neutrino-neutrino
interactions because both the forward and exchange diagrams would take the 
state onto
itself.

\section{One Bethe ansatz variable}
\label{sec: one BAV}

Other eigenstates of the Hamiltonian can be obtained with the Bethe ansatz
technique. As mentioned earlier, this method is based on a trial state
depending on some unknown parameters which are known as the Bethe ansatz
variables. For our particular Hamiltonian, Bethe ansatz states are formed with 
the help of the so-called Gaudin algebra operators
\begin{equation}
\label{eq:Gaudin operators}
\vec{Q}(\xi)=\sum_\omega \frac{\vec{J}_\omega}{\omega-\xi}.
\end{equation}
Here $\xi$ is a generic complex number which will later turn into a Bethe ansatz
variable and its value will be determined from the requirement that the trial
state is an eigenstate.

Before we consider the most general case, it is instructive to study the
simplest nontrivial application of the formalism in detail. For this purpose, we
consider those eigenstates with $(n^{(1)},n^{(2)})=(1,n-1)$.  These eigenstates 
yield
\begin{equation}
\label{m for one BAV}
m=\frac{n^{(1)}-n^{(2)}}{2}=-\frac{n}{2}+1,
\end{equation}
under the action of $J_z$. Therefore, they live in $j=n/2, n/2-1$
representations. In order to find their explicit form, one starts from
the Bethe ansatz state
\begin{equation}
\label{eq: one Q+ state}
|\xi_1\rangle=Q^+(\xi_1)|n/2,-n/2\rangle
\end{equation}
where $Q^+(\xi_1)$ is given by the + component of Eq. (\ref{eq:Gaudin 
operators}).
Note that this state is not normalized, but
the corresponding normalized state can be easily found as
\begin{equation}
\label{eq:BA state with norm}
{|\xi_1\rangle}^\prime=\frac{1}{\sqrt{G}}Q^+(\xi_1) |n/2,-n/2\rangle\msp,
\end{equation}
where
\begin{equation}
\label{eq:normalization}
G = \sum_\omega \frac{1}{(\xi_1-\omega)^2}.
\end{equation}
The non-normalized Bethe ansatz states in the form
of Eq. (\ref{eq: one Q+ state}) are usually more convenient to work with.
In this paper, we mostly work with unnormalized eigenstates unless we
specifically state otherwise. We denote the normalized eigenstates with a prime
as in Eq. (\ref{eq:BA state with norm}).

\begin{figure*}
\raggedright
\begin{subfigure}[t]{0.32\textwidth}
\raggedright
\includegraphics{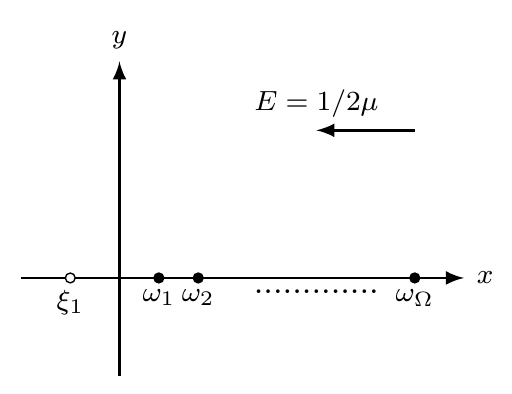}
    \caption{\footnotesize \raggedright Electrostatic analogy for the Bethe 
ansatz equation given in Eq. (\ref{eq:BA eq with one parameter}).}
    \label{fig:Electrostatic analogy with one BA parameter-highest}
    \end{subfigure}
\hspace{0mm}
  \begin{subfigure}[t]{0.32\textwidth}
  \raggedright
  \includegraphics{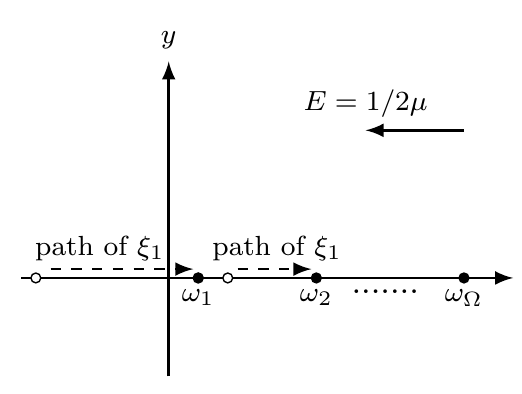}
\caption{\footnotesize \raggedright Paths of two representative solutions of the
Bethe ansatz equations as $\mu$ decreases from very large to very small values.}
\label{fig:Electrostatic analogy with one BA parameter-second highest}
  \end{subfigure}
\hspace{3mm}
\begin{subfigure}[t]{0.3\textwidth}
\includegraphics[width=\columnwidth]{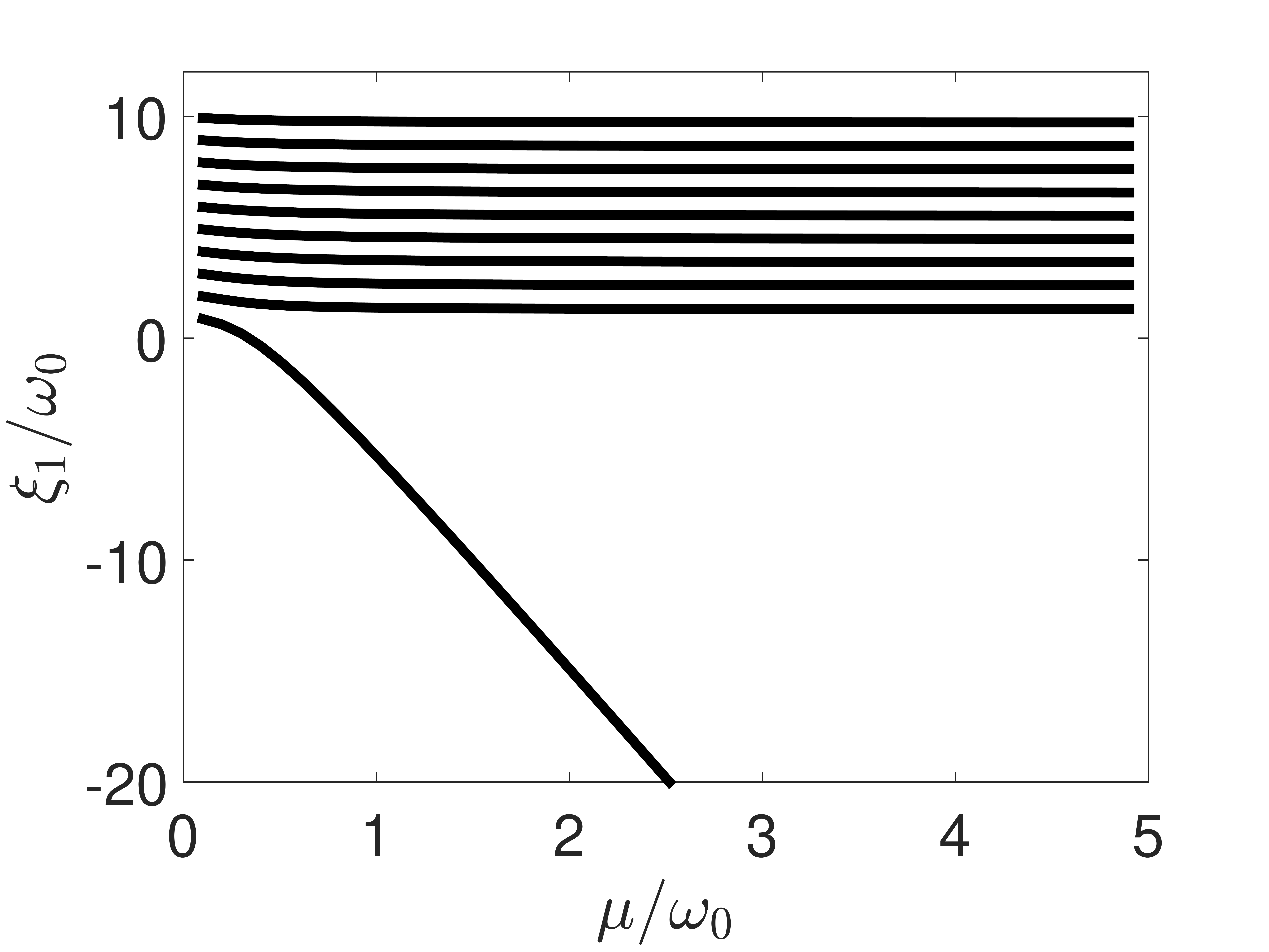}
\caption{\footnotesize \raggedright Numerical solutions of the Bethe ansatz
equation given in Eq. (\ref{eq:BA eq with one parameter}) for the toy model
introduced in Eq. (\ref{toy modes}).}
\label{one BAV numerical solution}
\end{subfigure}
\caption{\footnotesize \raggedright (a) The Bethe ansatz variable $\xi_1$ is
interpreted as the position of a point particle with one unit of positive
electric charge in the complex plane, while the neutrino oscillation frequencies
$\{\omega_1,\cdots,\omega_n\}$ are interpreted as the positions of fixed
particles with negative charges $\{-j_{\omega_1},\cdots,-j_{\omega_n}\}$,
respectively. The self-interaction constant $\mu$ enters the electrostatic 
picture as an external electric field. The equilibrium position of $\xi_1$ in 
this setup is the solution of Eq.  (\ref{eq:BA eq with one parameter}). 
Although Bethe ansatz variables can be complex, in general, in the particular 
case of a single Bethe ansatz variable the solution is always real due to the 
symmetry of the problem around the $x$-axis.
(b) If the equilibrium position of the free charge is at $-\infty$ in 
the $\mu\to\infty$ limit, then it approaches $\omega_1$ as $\mu\to0$. If 
its equilibrium position is in between $\omega_k$ and $\omega_{k+1}$ for $k\geq 
1$ in the $\mu\to\infty$ limit, then it approaches $\omega_{k+1}$ as 
$\mu\to0$. We assume that $\omega_1<\omega_2<\dots<\omega_\Omega$.
(c)
Numerical solutions for the toy model introduced in Eq. (\ref{toy modes}) agree
with these expectations. As $\mu$ decreases (from right to left in the figure),
one of the solutions starts from $-\infty$ and approaches the lowest
oscillation frequency, while the others start in between oscillation 
frequencies and approach the larger frequencies in their respective intervals.
}
\label{fig:Electrostatic analogy with one BA parameter}
\end{figure*}

In order to understand the trial state $|\xi_1\rangle$, first note that in
$|n/2,-n/2\rangle$, all neutrinos are $\nu_2$. (See Eq. (\ref{eq:highest-lowest
weight states}).) The operator $Q^+(\xi_1)$ given in Eq.  (\ref{eq:BA state with
norm}) turns one of these $\nu_2$ into a $\nu_1$ in such a way that the
probability amplitude of finding this single $\nu_1$ in the oscillation mode
$\omega$ is $1/(\sqrt{G}(\omega-\xi_1))$. We want to choose $\xi_1$ so that this
state satisfies
\begin{equation}
\label{eq:eigenvalue eqs for BA states with 1 parameter}
H|\xi_1\rangle=E(\xi_1)|\xi_1\rangle
\end{equation}
for some energy $E(\xi_1)$. A direct substitution of the state in Eq.
(\ref{eq: one Q+ state}) into the left-hand side of Eq. (\ref{eq:eigenvalue eqs
for BA states with 1 parameter}) yields\footnote{In deriving Eq.
(\ref{eq:commutation [H,Q] on |j,-j>}), it is helpful to first calculate the
commutator
\begin{align}
\nonumber
 [H,Q^+(\xi_1)]=Q^+(\xi_1)(-\xi_1+ 
2\mu{J}^z)-J^+\left(1+2\mu\sum_\omega\frac{J_\omega^z}{\omega-\xi_1}\right)\msp
\end{align}
by using the isospin commutation relations given in Eq. (\ref{su2}).}
\begin{equation}
\label{eq:commutation [H,Q] on |j,-j>}
\begin{split}
 H|\xi_1\rangle\;=\;\;&(E_{-n/2}-\mu n -\xi_1)|\xi_1\rangle\\
&\mkern-25mu-\left(1+2\mu\sum_\omega\frac{{-n_\omega}/2}{\omega-\xi_1}
\right)J^+|n/2,-n/2\rangle\msp.
\end{split}
\end{equation}
where $E_{-n/2}$ is the energy of the lowest weight state given in Eq.
(\ref{eq:E_max}). A comparison between Eqs.\s(\ref{eq:commutation [H,Q] on
|j,-j>}) and (\ref{eq:eigenvalue eqs for BA states with 1 parameter}) leads to
the conclusion that $|\xi_1\rangle$ is an eigenstate of the Hamiltonian if one
chooses $\xi_1$ so as to satisfy
\begin{equation}
\label{eq:BA eq with one parameter}
-\frac{1}{2\mu}+ \sum_\omega \frac{n_\omega/2}{\omega-\xi_1}=0.
\end{equation}
Equation \s(\ref{eq:BA eq with one parameter}) is a Bethe ansatz equation.
Solving this equation for $\xi_1$ and substituting the solution retrospectively 
in Eq. (\ref{eq: one Q+ state}) yields an exact eigenstate whose energy 
eigenvalue is given by
\begin{equation}
\label{eq:energy eigenvalue of BA state with one parameter}
E(\xi_1)=E_{-n/2}-\mu n -\xi_1\msp.
\end{equation}
In general, Bethe ansatz variables can take complex values. But in the
particular case of a single Bethe ansatz variable, Eq. (\ref{eq:BA eq with one
parameter}) admits only real solutions. Therefore, $\xi_1$ and the resulting
energy given in Eq.  (\ref{eq:energy eigenvalue of BA state with one parameter})
are always real.

In constructing the Bethe ansatz state given in Eq. (\ref{eq: one Q+ state}), we
started from the lowest weight (all $\nu_2$) state and converted one $\nu_2$
into a $\nu_1$. This is what we call a raising operator formulation.
It is also possible to use the opposite lowering operator formalism by
starting from the highest weight (all $\nu_1$) state and converting one $\nu_1$ 
into a
$\nu_2$, i.e.,
\begin{equation}
\label{eq: one Q- state}
|\zeta_1\rangle=Q^-(\zeta_1)|n/2,n/2\rangle.
\end{equation}
These eigenstates have $(n^{(1)},n^{(2)})=(n-1,1)$ and yield
\begin{equation}
\label{m for one BAV -}
m=\frac{n^{(1)}-n^{(2)}}{2}=\frac{n}{2}-1,
\end{equation}
under the action of $J_z$. Therefore, they
also live in $j=n/2, n/2-1$ representations of total isospin.
Through direct substitution, one can show that
the state in Eq. (\ref{eq: one Q- state}) is an eigenstate of the Hamiltonian
with the energy
\begin{equation}
\label{eq:energy eigenvalue of BA state with one parameter -}
E(\zeta_1)=E_{+n/2}-\mu n +\zeta_1\msp,
\end{equation}
if $\zeta_1$ obeys the Bethe ansatz equation
\begin{equation}
\label{eq:BA eq with one parameter -}
\frac{1}{2\mu}+ \sum_\omega \frac{n_\omega/2}{\omega-\zeta_1}=0.
\end{equation}
In Eq. (\ref{eq:energy eigenvalue of BA state with one parameter -}), $E_{+n/2}$
denotes the energy of $|n/2,n/2\rangle$ and is given in Eq. (\ref{eq:E_max}).

Note that the Bethe ansatz equations for the raising and lowering formalisms
(Eq.  (\ref{eq:BA eq with one parameter}) and (\ref{eq:BA eq with one parameter
-}))  are the same except for the sign of the $1/2\mu$ term.  In what follows,
we discuss the solutions of the Bethe ansatz equations in the context of the
raising operator formalism. Our conclusions also apply to the lowering operator
formalism with appropriate sign changes.

In general, Bethe ansatz equations admit several solutions, each one yielding 
an eigenstate. In particular, for an $n$-particle system there should 
be $n$ linearly independent eigenstates with $(n^{(1)},n^{(2)})=(1,n-1)$. 
However, we restrict ourselves to those states which are completely symmetric 
under the exchange of any two neutrinos in the same oscillation mode. The 
number of such symmetric states with $(n^{(1)},n^{(2)})=(1,n-1)$ is equal to 
$\Omega$, i.e., the number of energy modes in the system. Therefore, we expect 
to find $\Omega$ eigenstates in the form of Eq. (\ref{eq: one Q+ state}). 
Indeed it is easy to see that the number of solutions of Eq. (\ref{eq:BA eq 
with one parameter}) is $\Omega$. As discussed below, each one of these 
solutions yields a linearly independent eigenstate when substituted in Eq. 
(\ref{eq: one Q+ state}). That these states satisfy the required symmetry 
condition is guaranteed by the fact that each $J_\omega^+$ in $Q^+(\xi_1)$ lives 
in the highest weight representation $j_\omega=n_\omega/2$.

Although one can work out the solutions of the Bethe ansatz equations directly,
it is often useful to refer to the so-called electrostatic analogy which
was first suggested by Gaudin \cite{Gaudin_Electrostatic_Analogy} and elaborated
by Richardson \cite{richardson:1802}. This analogy is based on the observation
that the Bethe ansatz equations can be interpreted as the stability conditions
for an electrostatic system on a complex plane.  Let us denote the real and
imaginary axes of the complex plane by $x$ and $y$, respectively. In the
electrostatic analogy, the Bethe ansatz variable $\xi_1=x_1+iy_1$ is interpreted
as the position of a point particle which carries one unit of positive electric
charge in the complex plane (see Fig \ref{fig:Electrostatic analogy with one BA 
parameter-highest}). The neutrino oscillation frequencies
$\{\omega_1,\cdots,\omega_n\}$ are interpreted as the positions of some
\emph{fixed} point electric charges with magnitudes
$\{-j_{\omega_1},\cdots,-j_{\omega_n}\}$, respectively. Since the oscillation
frequencies are real and positive in our case, the fixed charges are positioned
along the positive $x$ axis.\footnote{Note that, due to our box quantized
treatment, neutrino energies and therefore the oscillation frequencies have
discrete values, leading to an electrostatic picture with point charges.
However, one can go to the continuum limit and work with a continuous
distribution of fixed charges, and a (piecewise) continuous distribution of free
charges \cite{richardson:1802, doi:10.1080/01630560008816948, McMillen2009862}.
Also note that the inclusion of antineutrinos introduces negative oscillation
frequencies as discussed in Section \ref{section:Antineutrinos}.} The whole
system is placed in a uniform electric field in the $-x$ direction with a
strength of $1/{2\mu}$. It can be shown that the electrostatic potential energy
of such a configuration is proportional to
\begin{eqnarray}
\label{Electrostatic Energy}
V&\propto&\frac{1}{2\mu} Re(\xi_1)
-\frac{1}{2\mu} \sum_{\omega} j_{\omega} \omega
-\frac{1}{2}\sum_{\substack{\omega,\omega^\prime \\
\omega\neq \omega^\prime}} j_{\omega} 
j_{\omega^\prime}\ln{|\omega-\omega^\prime|}
\nonumber
\\
&+&
\sum_{\omega} j_{\omega} \ln{|\xi_1-\omega|}~.
\end{eqnarray}
The free charge comes to an equilibrium when the electrostatic potential energy
reaches a local minimum, i.e., when $\partial V/\partial \xi_1=0$. It is easy to
show that this equilibrium condition yields the Bethe ansatz equation given in
Eq. (\ref{eq:BA eq with one parameter}).  Note that the positions of the fixed
charges and the external field are such that the total electric field in which
the free charge moves is symmetric with respect to the $x$ axis. For this
reason, the equilibrium position of the free charge lies on the $x$ axis for any
value of $\mu$. This is another way of saying that the Bethe ansatz equations
given in Eq. (\ref{eq:BA eq with one parameter}) can only have real solutions.

We find the electrostatic analogy particularly helpful in visualizing the
transformation of eigenstates with changing $\mu$. In what follows, we first
consider the solutions of the Bethe ansatz equations and the corresponding
eigenstates in the $\mu\to\infty$ and $\mu\to 0$ limits, respectively. Our aim 
is to show that they indeed agree with those discussed in Sec.
\ref{section:Eigenstates in special cases}. Then we discuss how these
eigenstates transform into each other as $\mu$ changes between these two limits.

${\bm \mu\to\infty}$ {\bf limit:}
In the limit where $\mu$ is very large, the external electric field in the
analogy becomes very weak. In this limit, the stable configurations of the free
charge $\xi_1$ lie either at $-\infty$ or in between the fixed charges (Fig.
\ref{fig:Electrostatic analogy with one BA parameter-second highest}). Since
there is always an electric field in the $-x$ direction, even if it is
vanishingly small, we do not have a stable solution at $+\infty$. Considering
that there are $\Omega-1$ intervals between fixed charges, the total number of
solutions is $\Omega$, as was mentioned earlier.

In Sec. \ref{subsection:Eigenstates at mu=infty limit} we mentioned that
in the $\mu\to\infty$ limit the Hamiltonian is proportional to 
${\vec{J}}\cdot{\vec{J}}$ so the eigenstates must approach $|j,m\rangle$. Can 
we tell which $|j, m\rangle$ states these $\Omega$ solutions
correspond to? The hint lies in Eq. (\ref{m for one BAV})
which tells us that for the states with one Bethe ansatz variable, total isospin
quantum number $j$ can take only two values: $j=n/2$ or $j=n/2-1$.
Therefore, the states $|\xi_1\rangle$ can only be related to
\begin{equation}
\underbrace{|\frac{n}{2},\; -\frac{n}{2}+1\rangle}_{\mbox{\tiny unique}} \quad
\mbox{and}\quad
\underbrace{|\frac{n}{2}-1,\; -\frac{n}{2}+1\rangle}_{(\Omega-1)-
\mbox{\tiny fold degenerate}}
\quad
\end{equation}
and will go to one of these states in the $\mu\to\infty$ limit. Since $j$ is 
found by adding the individual $j_\omega$'s of $\Omega$ different oscillation 
modes, $j=n/2-1$ is $(\Omega-1)$-fold degenerate while $j=n/2$ is unique.  One
naturally suspects that in the $\mu\to\infty$ limit, the unique $\xi_1\to
-\infty$ solution yields the unique $|n/2, n/2-1\rangle$ state, while $\Omega-1$
finite solutions between fixed charges correspond to the $\Omega-1$ states in
the form of $|n/2-1, n/2-1\rangle$. This is indeed the case as can be shown very
easily. For the $\xi_1\to -\infty$ solution, we can ignore the finite $\omega$
values in Eq.\s(\ref{eq:BA eq with one parameter}). This tells us that $\xi_1$
approaches $-\infty$ as
\begin{equation}
\label{xi_1 to infty}
\xi_1\to -2\mu\sum_\omega \frac{n_\omega}{2}=-\mu n.
\end{equation}
Therefore, Eq.\s(\ref{eq:energy eigenvalue of BA state with one parameter}) 
gives the corresponding energy eigenvalue as
\begin{equation}
\label{eq:energy eigenvalue of BA state with one parameter (infinite) at high 
density}
\lim _{\mu\to\infty} E(\xi_1) =\mu \frac{n}{2}(\frac{n}{2}+1)
\end{equation}
in agreement with our guess that $\xi_1\to -\infty$ solution yields the
$|\frac{n}{2},\; -\frac{n}{2}+1\rangle$ state. For those solutions in which
$\xi_1$ remains finite, one can compute the energy from Eq.\s(\ref{eq:energy
eigenvalue of BA state with one parameter}) by ignoring $\xi_1$ and $\omega$
with respect to $\mu$. The result is
\begin{equation}
\label{eq:energy eigenvalue of BA state with one parameter (finite) at high 
density}
\lim _{\mu\to\infty} E(\xi_1)=\mu (\frac{n}{2} -1)\frac{n}{2}
\end{equation}
confirming our guess that these solutions yield the $|\frac{n}{2},\;
-\frac{n}{2}+1\rangle$ states. Technically, this only proves that finite $\xi_1$
solutions yield linear combinations of $|\frac{n}{2},\; -\frac{n}{2}+1\rangle$
states since they all have the same energy in the $\mu\to\infty$ limit. 
However, any linear combination of $j=n/2-1$ representations is a $j=n/2-1$ 
representation itself, and we can always choose the appropriate combinations of 
these representations so that each finite $\xi_1$ solution yields a single 
$|\frac{n}{2},\; -\frac{n}{2}+1\rangle$ state. See the Appendix for a further 
discussion of this point.

${\bm \mu\to 0}$ {\bf limit:}
In the limit where $\mu$ approaches zero, the external electric field in the
analogy becomes very large. In such a large external field, the free charge can
find a stable configuration only when it is practically on top of one of the
fixed charges. This can also be seen easily from Eq.\s(\ref{eq:BA eq with one
parameter}): When $\mu\to 0$, the Bethe ansatz equation can only be satisfied if
$\xi_1$ approaches one of the oscillation frequencies, say $\tilde{\omega}$.
In this limit, the Gaudin operator given in Eq. (\ref{eq:Gaudin operators})
diverges. In particular, we can write
\begin{equation}
\label{mu to 0 limit of xi_1 Gaudin}
(\tilde{\omega}-\xi_1)Q^+(\xi_1)
\underset{\xi_1\to \tilde{\omega}}{\longrightarrow}
{J^+_{\tilde{\omega}}}
\end{equation}
which tells us that
\begin{equation}
\label{mu to 0 limit of xi_1 state}
(\tilde{\omega}-\xi_1)|\xi_1\rangle
\underset{\xi_1\to \tilde{\omega}}{\longrightarrow}
{J^+_{\tilde{\omega}}}\prod_{{\omega^\prime}}
|\frac{n_{\omega^\prime}}{2},-\frac{n_{\omega^\prime}}{2}\rangle
\end{equation}
where we used Eqs. (\ref{eq:highest-lowest weight states}) and (\ref{eq: one Q+
state}). We can get rid of the coefficient on the left-hand side by normalizing
both sides of Eq. (\ref{mu to 0 limit of xi_1 state}), which yields
\begin{equation}
\label{mu to 0 limit of xi_1 prime}
|\xi_1\rangle^\prime
\underset{\mu\to 0}{\longrightarrow}
|\frac{n_{\tilde{\omega}}}{2},-\frac{n_{\tilde{\omega}}}{2}+1\rangle
\prod_{{\omega^\prime(\neq \tilde{\omega})}}
|\frac{n_{\omega^\prime}}{2},-\frac{n_{\omega^\prime}}{2}\rangle
\end{equation}
where the prime indicates the normalized state (see Eq. (\ref{eq:BA state with 
norm})). The resulting state in Eq. (\ref{mu to 0 limit of xi_1
prime}) is clearly in the form of Eq. (\ref{eq: mu=0 eigenstate}). Note that
Eq. (\ref{mu to 0 limit of xi_1 prime}) corresponds to a state in which all
oscillation modes contain only $\nu_2$ neutrinos, except for the mode
$\tilde{\omega}$, which contains a single neutrino in $\nu_1$ and
$n_{\tilde{\omega}}-1$ neutrinos in $\nu_2$.

\begin{figure*}[t!]
 \includegraphics[width=0.98\textwidth]{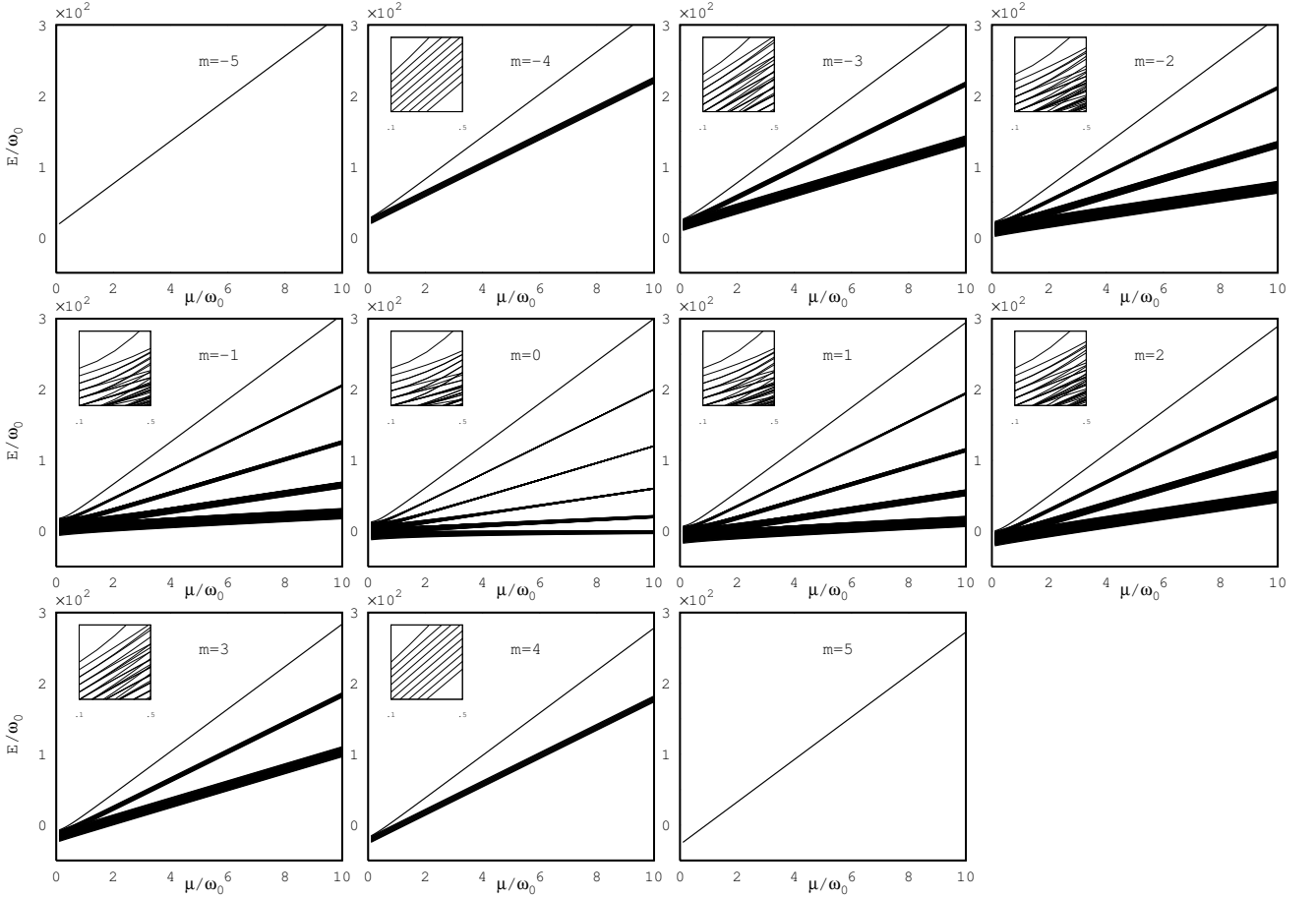}
 \caption{\footnotesize \raggedright Energy eigenvalues of the Hamiltonian as a 
function of $\mu/\omega_0$. Vacuum frequencies are taken as 
$\omega_i=i\omega_0$. Each graph represents the states with different numbers 
of Bethe ansatz parameters, so each figure has a different eigenvalue of $J^z$ 
from $m=-5$ to $m=5$. At high densities, one can see that energy eigenvalues 
have a slope of $j(j+1)$ and are grouped by the length of total isospin as 
$j=5,4,3,2,1,0$. One important result from the figures is that level crossings 
appear between all states except the highest level state in the low density 
region.}
 \label{fig:energies of BA states}
 \end{figure*}

{\bf Transformation of eigenstates:}
Which eigenstates in the $\mu\to\infty$ limit transform to which ones in the
$\mu\to 0$ limit as $\mu$ decreases? It was already mentioned that the 
equilibrium position of the free charge has to be on the $x$ axis for all $\mu$ 
values. Suppose that the free charge is in equilibrium at $\xi_1\to-\infty$ in 
the $\mu\to\infty$ limit. As $\mu$ decreases and the electric field becomes 
stronger, this equilibrium position has to shift until it is on top of one of 
the free charges. However, since $\xi_1$ can never be imaginary, by shifting on 
the $x$ axis it can only end up on top of the lowest oscillation frequency 
$\omega_1$ in the $\mu\to 0$ limit. (See Fig. \ref{fig:Electrostatic analogy 
with one BA parameter-second highest} and note that we 
take $\omega_1<\omega_2<\dots<\omega_\Omega$.) On the other hand, if $\xi_1$ is 
in equilibrium in between two fixed charges $\omega_k$ and $\omega_{k+1}$ for 
$k\geq 1$ in the $\mu\to\infty$ limit, then its equilibrium
position shifts towards the larger oscillation frequency $\omega_{k+1}$ as $\mu$
decreases. In the $\mu\to 0$ limit, this equilibrium position will be on top of
$\omega_{k+1}$.

These considerations tell us that the eigenstate $|n/2,\; n/2-1\rangle$ in the 
$\mu\to\infty$ limit transforms into the eigenstate given in Eq. (\ref{mu to 0
limit of xi_1 prime}) for $\tilde{\omega}=\omega_1$ as $\mu\to 0$.
The corresponding eigenvalue transforms as
\begin{equation}
E(\xi_1)=
\begin{cases}
\displaystyle{\mu\frac{n}{2}(\frac{n}{2}+1)}, & \mbox{as } \mu\to\infty\\
& \\
\displaystyle{\omega_1 (\frac{n_{\omega_1}}{2}-1)+\sum_{\omega(\neq\omega_1)} 
\omega
\frac{n_\omega}{2}}, & \mbox{as } \mu\to 0.
\end{cases}
\end{equation}
On the other hand, the degenerate eigenstates $|n/2-1,\; n/2-1\rangle$ in the 
$\mu\to\infty$ limit turn into the eigenstates given in Eq. (\ref{mu to 0 limit
of xi_1 prime}) for $\tilde{\omega}>\omega_1$ as $\mu\to 0$. The corresponding
eigenvalues transform as
\begin{equation}
E(\xi_1)=
\begin{cases}
\displaystyle{\mu(\frac{n}{2}-1)\frac{n}{2}}, & \mbox{as } \mu\to\infty\\
& \\
\displaystyle{\tilde{\omega}
(\frac{n_{\tilde{\omega}}}{2}-1)+\sum_{\omega(\neq\tilde{\omega})} \omega
\frac{n_\omega}{2}}, & \mbox{as } \mu\to 0
\end{cases}
\end{equation}

In order to illustrate these results, we consider a toy model with $10$ equally
spaced and nondegenerate oscillation modes,
\begin{equation}
\label{toy modes}
\omega_i=i\omega_0\qquad\mbox{for } i=1,2,\dots,10
\end{equation}
where $\omega_0$ is an arbitrary oscillation frequency. The nondegeneracy
assumption means that each mode contains only one neutrino so we have 
$n_{\omega_i}=1$ and $j_{\omega_i}=1/2$ for each $i$. The dimension of the 
corresponding Hilbert space is $2^{10}=1024$. In this particular example, we 
add $10$ isospin $1/2$'s, so the total isospin quantum number can take the 
values $j=5,4,3,2,1,0$ with the respective multiplicities of $1,9, 35, 75, 90, 
42$. As per our classification scheme, we can discuss the eigenstates of the 
Hamiltonian by grouping them in terms of their eigenvalues under $J_z$, which 
can take the values $m=\pm5, \pm4, \pm3, \pm2, \pm1, 0$. Those states with 
$m=\pm 5$ are the trivial eigenstates discussed in Eq. (\ref{eq:highest-lowest 
weight states}). With one Bethe ansatz variable, we can obtain those 
eigenstates with $m=\pm4$, depending on whether we use the raising or lowering 
formalism. Therefore, the only $j$ values which are relevant to us in this 
section are $j=5,4$. In this example, we only work with $m=-4$ states. 
Note that $m=+4$ states can be similarly studied with the lowering formalism.

Based on the above discussions we expect to find $10$ eigenstates with $m=-4$:
One of these eigenstates should approach $|5,-4\rangle$ with its energy
growing as $30\mu$ in the $\mu\to\infty$ limit. This state is expected to 
approach
$|\nu_1, \nu_2, \nu_2, \nu_2, \nu_2, \nu_2, \nu_2, \nu_2, \nu_2, \nu_2 \rangle$
in the $\mu\to 0$ limit, while its energy approaches $53\omega_0/2$. The 
other nine eigenstates should approach $|4,-4\rangle$ in the $\mu\to\infty$ 
limit with their energies becoming degenerate and growing as $20\mu$. In the
$\mu\to 0$ limit, we expect these states to be like the one given above, except 
that single $\nu_1$ will move to larger oscillation modes, i.e., $|\nu_2, 
\nu_1, \nu_2, \nu_2, \nu_2, \nu_2, \nu_2, \nu_2, \nu_2, \nu_2 \rangle$, 
$|\nu_2, \nu_2, \nu_1, \nu_2, \nu_2, \nu_2, \nu_2, \nu_2, \nu_2, \nu_2 
\rangle$, $\dots$. The energies of these states will be $51\omega_0/2, 
49\omega_0/2, 47\omega_0/2, \dots, 35\omega_0/2$, respectively.

In Fig. \ref{one BAV numerical solution} we show the numerical solutions of the
Bethe ansatz equations given in Eq. (\ref{eq:BA eq with one parameter}). The
behavior of the solution $\xi_1$ agrees with what we expect from the
electrostatic analogy described above: One solution starts from $-\infty$ in
the $\mu\to\infty$ limit and approaches the lowest oscillation frequency in the 
$\mu\to 0$ limit. The other solutions start in between the oscillation modes in
the $\mu\to\infty$ limit and move towards the larger oscillation modes in their
respective intervals. Corresponding energy eigenvalues calculated from Eq.
(\ref{eq:energy eigenvalue of BA state with one parameter}) are shown in Fig.
\ref{fig:energies of BA states} on the panel marked with $m=-4$.  The energy
eigenvalue corresponding to the first solution mentioned above is the top line
in this panel. As expected, it increases as $30\mu$ as $\mu\to\infty$, and
becomes $53\omega_0/2$ as $\mu\to 0$.  The energy eigenvalues corresponding to
the other nine solutions increase as $20\mu$ as $\mu\to\infty$ and approach the 
values $51\omega_0/2, 49\omega_0/2, 47\omega_0/2, \dots, 35\omega_0/2$ as
$\mu\to 0$. The lowering formalism also yields $10$ solutions for $m=+4$
eigenstates. The energy eigenvalues corresponding to these solutions are shown
in Fig. \ref{fig:energies of BA states}. Their behavior
is qualitatively similar to (although not exactly the same as) the $m=-4$ case.
The other panels of Fig. \ref{fig:energies of BA states}  show the energy
eigenvalues of the states with other $m$ values which are discussed in the next
section.

Note that our computation power in a standard desktop computer allows us to
solve Bethe ansatz equations for up to $16$ neutrinos. However, since the
resulting $2^{16}=65536$ eigenstates make our plots almost unreadable on paper,
we choose to present a simpler example with $10$ neutrinos. Also, our choice of
equally spaced and nondegenerate oscillation modes is due to its usefulness for
a simple discussion. In our numerical simulations involving more than $10$
neutrinos occupying nonequally spaced modes, we do not see any behavior which
is qualitatively different than this example.

\section{More Bethe ansatz variables}
\label{sec: More Bethe ansatz variables}

The method outlined in the previous section can be generalized to obtain the
eigenstates with generic occupation numbers.  For example, let us consider those
eigenstates with $(n^{(1)},n^{(2)})=(2,n-2)$. These eigenstates yield
\begin{equation}
m=\frac{n^{(1)}-n^{(2)}}{2}=-\frac{n}{2}+2,
\end{equation}
under the action of $J_z$ which tells us that they live in $j=n/2, n/2-1, n/2-2$
representations of the total isospin. They can be computed by starting from a
trial state with two Bethe ansatz parameters, i.e.,
\begin{equation}
\label{eq: BA state with two xi}
|\xi_1,\xi_2\rangle=Q^+(\xi_1)Q^+(\xi_2)|n/2,-n/2\rangle
\end{equation}
The state $|n/2,-n/2\rangle$ has only $\nu_2$ neutrinos, but each one of the
Gaudin operators turns one of them into a $\nu_1$.  Note that $\xi_1$ which
appears in this equation is not the same $\xi_1$ which appears in Eq. (\ref{eq:
one Q+ state}). Here, $\xi_1$ and $\xi_2$ are coupled with each other and
satisfy a set of equations which is different from Eq. (\ref{eq:BA eq with one
parameter}).  This method of denoting the Bethe ansatz variables may be
confusing at first.  But since this is the standard notation in the literature,
we adhere to it.  Also note that the state in Eq. (\ref{eq: BA state with two
xi}) is invariant under the exchange of the Bethe ansatz variables $\xi_1$ and
$\xi_2$, i.e.,
\begin{equation}
\label{eq: commutation of bethe ansatz variables}
|\xi_1,\xi_2\rangle=|\xi_2,\xi_1\rangle
\end{equation}
because the operators $Q^+(\xi_1)$ and $Q^+(\xi_2)$ commute with one another.

As in the previous case, we derive the equations satisfied by $\xi_1$ and 
$\xi_2$
by requiring that $|\xi_1,\xi_2\rangle$ is an eigenstate:
\begin{equation}
\label{eq:eigenvalue eq. for BA states with 2 parameter}
H|\xi_1,\xi_2\rangle = E(\xi_1,\xi_2) |\xi_1,\xi_2\rangle
\end{equation}
Direct substitution of Eq. (\ref{eq: BA state with two xi}) into the left-hand
side of Eq. (\ref{eq:eigenvalue eq. for BA states with 2 parameter})
yields
\begin{equation}
\label{eq:[H,QQ]|j,-j>}
\begin{split}
&H|\xi_1,\xi_2\rangle= -(\xi_1+\xi_2+2\mu n-2\mu+E_{-n/2})|\xi_1,\xi_2\rangle\\
&\mkern-3mu-\mkern-3mu\left(1\mkern-5mu+\mkern-3mu2\mu\sum_\omega\frac{
-n_\omega/2}{\omega-\xi_2}\mkern-3mu+\mkern-3mu\frac{2\mu}{\xi_1-\xi_2}\right)
\mkern-3mu Q^+(\xi_1)J^+|n/2,-n/2\rangle \\
&\mkern-3mu-\mkern-3mu\left(1\mkern-3mu+\mkern-3mu2\mu\sum_\omega\frac{
-n_\omega/2}{\omega-\xi_1}\mkern-3mu+\mkern-3mu\frac{2\mu}{\xi_2-\xi_1}\right)
\mkern-3muQ^+(\xi_2)J^+|n/2,-n/2\rangle
\end{split}
\end{equation}
For $|\xi_1,\xi_2\rangle$ to be an eigenstate, we need to choose
$\xi_1$ and $\xi_2$ in such a way that the last two terms on the right-hand side
of Eq.\s(\ref{eq:[H,QQ]|j,-j>}) vanish. This yields a coupled set of two Bethe
ansatz equations given by
\begin{equation}
\label{eq:BA eq with two parameters}
\begin{split}
&\sum_\omega\frac{-n_\omega/2}{\omega-\xi_1}=-\frac{1}{2\mu}+\frac{1}{
\xi_1-\xi_2}\\
&\sum_\omega\frac{-n_\omega/2}{\omega-\xi_2}=-\frac{1}{2\mu}+\frac{1}{
\xi_2-\xi_1}
\end{split}
\end{equation}
Solving these equations for $\xi_1$ and $\xi_2$, and substituting them
retrospectively in Eq. (\ref{eq: BA state with two xi}) gives us an eigenstate
with energy
\begin{equation}
\label{eq:BAE eq with two parameters}
E(\xi_1,\xi_2)=E_{-n/2}-\xi_1-\xi_2-2\mu(n-1)
\end{equation}

In general the eigenstates with $(n^{(1)},n^{(2)})=(\kappa,n-\kappa)$ can be 
obtained by
starting from a Bethe ansatz with $\kappa$ variables:
\begin{equation}
\label{eq:trial states from particle vacuum}
|\xi_1,\xi_2\cdots,\xi_\kappa\rangle=Q^+(\xi_1)\cdots Q^+ (\xi_\kappa)|n/2, 
-n/2\rangle
\end{equation}
These states have
\begin{equation}
\label{m for kappa BAV}
m=\frac{n^{(1)}-n^{(2)}}{2}=-\frac{n}{2}+\kappa,
\end{equation}
telling us that they can live in $j=n/2, n/2-1, \dots, n/2-\kappa$
representations.
It can be shown that the state in Eq. (\ref{eq:trial states from particle
vacuum}) is an eigenstate of the Hamiltonian with the energy
\begin{equation}
\label{eq:general energies of BA states}
E(\xi_1,\dotsc,\xi_\kappa)=E_{-n/2}-\sum_{\alpha}^\kappa \xi_\alpha - \kappa\mu 
(n-\kappa+1)
\end{equation}
if $\xi_1$,$\xi_2$,$\cdots$,$\xi_\kappa$ satisfy the following set of coupled 
Bethe ansatz equations:
\begin{equation}
\label{eq:general BA eq.}
\begin{split}
&\sum_\omega \frac{-n_\omega/2}{\omega-\xi_\alpha}=
-\frac{1}{2\mu}+\sum_{\beta\neq\alpha}^{\kappa} \frac{1}{\xi_\alpha-\xi_\beta}\\
& (\mbox{for every } \alpha=1,2,\dots,\kappa).
\end{split}
\end{equation}
When written out for every $\xi_\alpha$, Eq. (\ref{eq:general BA eq.})
represents a set of $\kappa$ equations in $\kappa$ unknowns.
As discussed below, these equations admit several solutions, each one
yielding a linearly independent eigenstate when substituted in Eq. 
(\ref{eq:trial states from particle vacuum}). Since the Bethe ansatz equations 
have real coefficients, each solution $(\xi_1,\xi_2,\dots,\xi_\kappa)$ involves 
either real numbers, or complex conjugate pairs. As a result, the energy given 
in Eq. (\ref{eq:general energies of BA states}) is always real.

As we increase $\kappa$, (which increases $n^{(1)}$ and decreases $n^{(2)}$ 
such that $n=n^{(1)}+n^{(2)}$ remains constant) we need to solve a larger and 
larger system of coupled algebraic equations in order to find the relevant 
eigenstates. When we go from the states with $\kappa$ variables ($\xi_1$, 
$\xi_2$, $\dots$, $\xi_\kappa$) to the states with $\kappa+1$ variables 
($\xi_1$, $\xi_2$, $\dots$, $\xi_\kappa$, $\xi_{\kappa+1}$), we need to solve 
the Bethe ansatz equations all over again because in the latter case the 
coupling to the variable $\xi_{\kappa+1}$ changes the values of the previous 
Bethe ansatz variables.\footnote{Some
approximation techniques exists in the literature \cite{2011PhyC..471..566P}
which relate the values of the Bethe ansatz variables from the step $\kappa$ to
those in the step $\kappa+1$. But we do not employ such approximations here.}

Bethe ansatz states presented above are not normalized. The norm of
the general Bethe ansatz state given in Eq. (\ref{eq:trial states from particle
vacuum}) is equal to
\begin{equation}
\langle\xi_1,\xi_2\cdots,\xi_\kappa|\xi_1,\xi_2\cdots,\xi_\kappa\rangle
=\det G(\xi_1,\xi_2\cdots,\xi_\kappa)
\end{equation}
where $G(\xi_1,\xi_2\cdots,\xi_\kappa)$ is a $\kappa\times\kappa$ matrix whose
elements are \cite{PhysRevB.65.060502}
\begin{equation}
\label{eq:normalization many}
G_{\alpha \beta} = \left\{
	 \begin{array}{l l}
	  \sum_\omega \frac{1}{(\xi_\alpha-\omega)^2}-2 \sum_{\alpha^\prime\neq
\alpha}\frac{1}{(\xi_\alpha-\xi_{\alpha^\prime})^2} & \quad
	  \text{if $\alpha=\beta$} \\
	  \frac{2}{(\xi_\alpha-\xi_\beta)^2} & \quad \text{if $\alpha\neq 
\beta$}.
	 \end{array} \right.
\end{equation}
Therefore, corresponding normalized states can be written as
\begin{equation}
\label{eq:general BA states with norm}
{|\xi_1,\cdots,\xi_\kappa\rangle}^\prime=\frac{1}{\sqrt{\det 
G}}Q^+(\xi_1)\dotsc Q^+(\xi_\kappa)
|n/2,-n/2\rangle\msp.
\end{equation}

In order to find those eigenstates for which $n^{(1)}>n^{(2)}$, it is more 
economical to use the Bethe ansatz states constructed with lowering operators, 
i.e.,
\begin{equation}
\label{eq:trial states from particle vacuum -}
|\zeta_1,\zeta_2\cdots,\zeta_\kappa\rangle=Q^-(\zeta_1)\cdots Q^- 
(\zeta_\kappa)|n/2, n/2\rangle
\end{equation}
which have
\begin{equation}
\label{m for kappa BAV -}
m=\frac{n^{(1)}-n^{(2)}}{2}=\frac{n}{2}-\kappa,
\end{equation}
telling us that they also live in $j=n/2, n/2-1, \dots, n/2-\kappa$
representations. They can similarly be shown to be eigenstates of the 
Hamiltonian with the energy
\begin{equation}
\label{eq:general energies of BA states -}
E(\zeta_1,\dotsc,\zeta_\kappa)=E_{+n/2}-\sum_{\alpha}^\kappa \zeta_\alpha - 
\kappa\mu (n-\kappa+1)
\end{equation}
if $\zeta_1$,$\zeta_2$,$\cdots$,$\zeta_\kappa$ satisfy
\begin{equation}
\label{eq:general BA eq. -}
\begin{split}
&\sum_\omega \frac{-n_\omega/2}{\omega-\zeta_\alpha}=
\frac{1}{2\mu}+\sum_{\beta\neq\alpha}^{\kappa} 
\frac{1}{\zeta_\alpha-\zeta_\beta}\\
& (\mbox{for every } \alpha=1,2,\dots,\kappa).
\end{split}
\end{equation}
As for the case with one Bethe ansatz variable, the Bethe ansatz equations for
the raising and lowering formalisms are identical except for a change in the
sign of the $1/2\mu$ term. In what follows, we only discuss the solutions of the
former, but our conclusions also apply to the latter with appropriate sign 
changes.

\begin{figure}[t]
\includegraphics{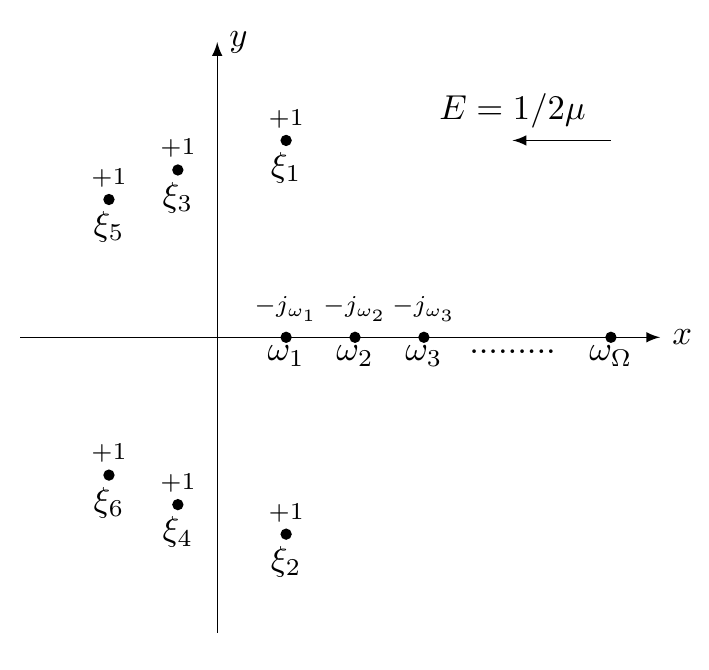}
\caption{\footnotesize \raggedright Electrostatic analogy for the Bethe ansatz
equations for more than one Bethe ansatz variables. Each Bethe ansatz variable
$\xi_\alpha$ is interpreted as the position of a free point particle with one
unit of positive electric charge. See the caption of Fig.
\ref{fig:Electrostatic analogy with one BA parameter} }
\label{fig:Electrostatic analogy}
\end{figure}

The electrostatic analogy introduced in Sec. \ref{subsection:Eigenstates for
any mu} can be generalized to any number of Bethe ansatz variables (See Fig.
\ref{fig:Electrostatic analogy}.) For $\kappa$ free particles carrying $+1$ unit
of electric charge at positions $\xi_\alpha=x_\alpha+iy_\alpha$, the
electrostatic potential energy is given by
\begin{eqnarray}
\label{Electrostatic Energy many}
V&\propto& \frac{1}{2\mu}\sum_{\alpha} Re(\xi_\alpha)
-\frac{1}{2\mu} \sum_{\omega} j_{\omega} \omega
-\frac{1}{2}\sum_{\substack{\alpha,\beta\\ (\alpha\neq\beta)}}
\ln{|\xi_\alpha-\xi_\beta|}\nonumber \\
&-&\frac{1}{2}\sum_{\substack{\omega,\omega^\prime \\
\omega\neq \omega^\prime}} j_{\omega} 
j_{\omega^\prime}\ln{|\omega-\omega^\prime|}+
\sum_{\alpha,\omega} j_{\omega} \ln{|\xi_\alpha-\omega|}~.
\end{eqnarray}
The free charges come to equilibrium when this electrostatic potential energy
reaches a local minimum, i.e., when $\partial V/\partial \xi_\alpha=0$ is
satisfied for every $\alpha$. It is easy to show that this equilibrium
condition yields the Bethe ansatz equations given in Eq. (\ref{eq:general BA
eq.}). It was already mentioned above that the complex solutions of the Bethe 
ansatz
equations always come as conjugate pairs. This is clearly visible in
the electrostatic analogy: The positions of the fixed charges and the external
electric field are such that the system can be in equilibrium only if the
free charges distribute themselves symmetrically with respect to the
$x$ axis.

In what follows, we employ the electrostatic analogy to show that the Bethe
ansatz states presented in this section agree with those presented in Secs
\ref{subsection:Eigenstates at mu=0 limit} and \ref{subsection:Eigenstates at
mu=infty limit} in the $\mu\to 0$ and $\mu\to\infty$ limits, respectively. After
that we discuss how the eigenstates and eigenvalues transform into each other as
$\mu$ decreases from very large to very small values.

${\bm \mu\to \infty}$ {\bf limit:} The external electric field in the
electrostatic analogy tends to zero in the $\mu\to\infty$ limit. Clearly there 
is a \emph{unique} equilibrium solution in which all $\kappa$ free charges are 
in the $x\to-\infty$ region.
There are also some equilibrium configurations in which $\kappa-1$ of the free
charges are in the $x\to-\infty$ region while one free charge is located in
between the fixed charges. Since interchanging Bethe ansatz variables does not
change the corresponding Bethe ansatz state (see Eq. (\ref{eq: commutation of
bethe ansatz variables})), it does not matter which free charge is in the finite
region. Therefore, the number of such configurations is $\Omega-1$ because there
are $\Omega-1$ intervals in which the single free charge in the finite region
can be located. We can continue in this manner to identify that equilibrium
configurations with $\kappa-k$ free charges are at $x\to-\infty$, while $k$ free
charges are located in the finite region near the free charges.

As mentioned in Sec \ref{subsection:Eigenstates at mu=infty limit}, the
eigenstates of the Hamiltonian must approach $|j, m\rangle$ states of the
total isospin in the $\mu\to\infty$ limit. From Eq. (\ref{m for kappa BAV}), we 
see that $j$ can only take the values $n/2\leq j \leq n/2-\kappa$. Therefore, 
the equilibrium configurations of the free charges mentioned above must yield 
the following states:
\begin{equation}
\label{jm states for kappa variables in infty}
|\frac{n}{2},\; -\frac{n}{2}+\kappa\rangle \quad
|\frac{n}{2}-1,\; -\frac{n}{2}+\kappa\rangle \quad
\dots\quad
|\frac{n}{2}-\kappa,\; -\frac{n}{2}+\kappa\rangle \quad
\end{equation}
The corresponding energy eigenvalues should also approach $\mu j(j+1)$ at the
same time. There is only one state with $j=n/2$ because the highest weight
representation is unique. Inspired by our results in the previous section, we
guess that this state is produced by the unique solution of the Bethe ansatz
equations in which all $\kappa$ free charges are in the $x\to-\infty$ region.
The number of states with $j=n/2-1$ is $\Omega-1$ which hints at the fact that
these states are produced by the solutions in which $\kappa-1$ of the free
charges are at $x\to-\infty$ while one of them is in between the fixed charges.
In fact, it is very easy to analytically show that the following is true: the
equilibrium configuration(s) in which $\kappa-k$ of the variables are located at
the $x\to-\infty$ region while $k$ variables are located near the fixed charges
in the $\mu\to\infty$ limit produce the states $|\frac{n}{2}-k,\;
-\frac{n}{2}+\kappa\rangle$ for $k=0,1,\dots,\kappa$. This proof can be found in
the Appendix.

${\bm \mu\to 0}$ {\bf limit:} As $\mu$ becomes vanishingly small, the electric
field in the electrostatic analogy becomes very strong. In that limit the free
charges can find their equilibrium positions only on top of the fixed charges.
Since free particles have $+1$ unit of electric charge while the fixed ones have
$-n_\omega/2$ unit of charge, several free particles can end up on the same
fixed particle. This can also be seen from the Bethe ansatz equations given in
Eq. (\ref{eq:general BA eq.}): As $\mu\to 0$, the divergence of the $1/2\mu$
term on the right-hand side can only be counteracted if $\xi_\alpha$ approaches 
one $\omega$, say $\tilde{\omega}^{(\alpha)}$. As a result, we can write
\begin{equation}
\label{mu to 0 limit of xi_alpha Gaudin}
(\tilde{\omega}^{(\alpha)}-\xi_\alpha)Q^+(\xi_\alpha)
\underset{\xi_\alpha\to \tilde{\omega}^{(\alpha)}}{\longrightarrow}
{J^+_{\tilde{\omega}^{(\alpha)}}}
\end{equation}
for the relevant Gaudin operator. Since Bethe ansatz equations in Eq.
(\ref{eq:general BA eq.}) have to be satisfied for every
$\alpha=1,2,\dots,\kappa$, Eq. (\ref{mu to 0 limit of xi_alpha Gaudin}) is true
for every $\xi_\alpha$. Therefore, we can write
\begin{eqnarray}
\label{mu to 0 limit of xi_alpha state}
&\displaystyle{(\tilde{\omega}^{(1)}-\xi_1) (\tilde{\omega}^{(2)}-\xi_2) \dots
(\tilde{\omega}^{(\kappa)}-\xi_\kappa) |\xi_1, \xi_2, \dots, \xi_\kappa\rangle}
\nonumber\\
&\displaystyle{\longrightarrow
{J^+_{\tilde{\omega}^{(1)}}} {J^+_{\tilde{\omega}^{(2)}}} \dots
{J^+_{\tilde{\omega}^{(\kappa)}}}
\prod_\omega|\frac{n_{\omega^\prime}}{2},-\frac{n_{\omega^\prime}}{2}\rangle
}
\end{eqnarray}
where we used Eqs. (\ref{eq:highest-lowest weight states}) and (\ref{eq:trial
states from particle vacuum}). The coefficients on the left-hand side
drop when we normalize both sides of Eq. (\ref{mu to 0 limit of xi_alpha
state}). The result is
\begin{equation}
\label{mu to 0 limit of xi_alpha prime}
|\xi_1,\xi_2,\dots,\xi_\kappa\rangle^\prime
\longrightarrow
\prod_\omega |\frac{n_{\omega}}{2},m_\omega\rangle
\end{equation}
where the values of $m_\omega$ depend on the particular equilibrium 
configuration
reached, i.e., the values of $\tilde{\omega}^{(\alpha)}$. This state is
in the form of Eq. (\ref{eq: mu=0 eigenstate}), as expected.

\begin{figure}[t!]
\includegraphics[width=1\columnwidth]{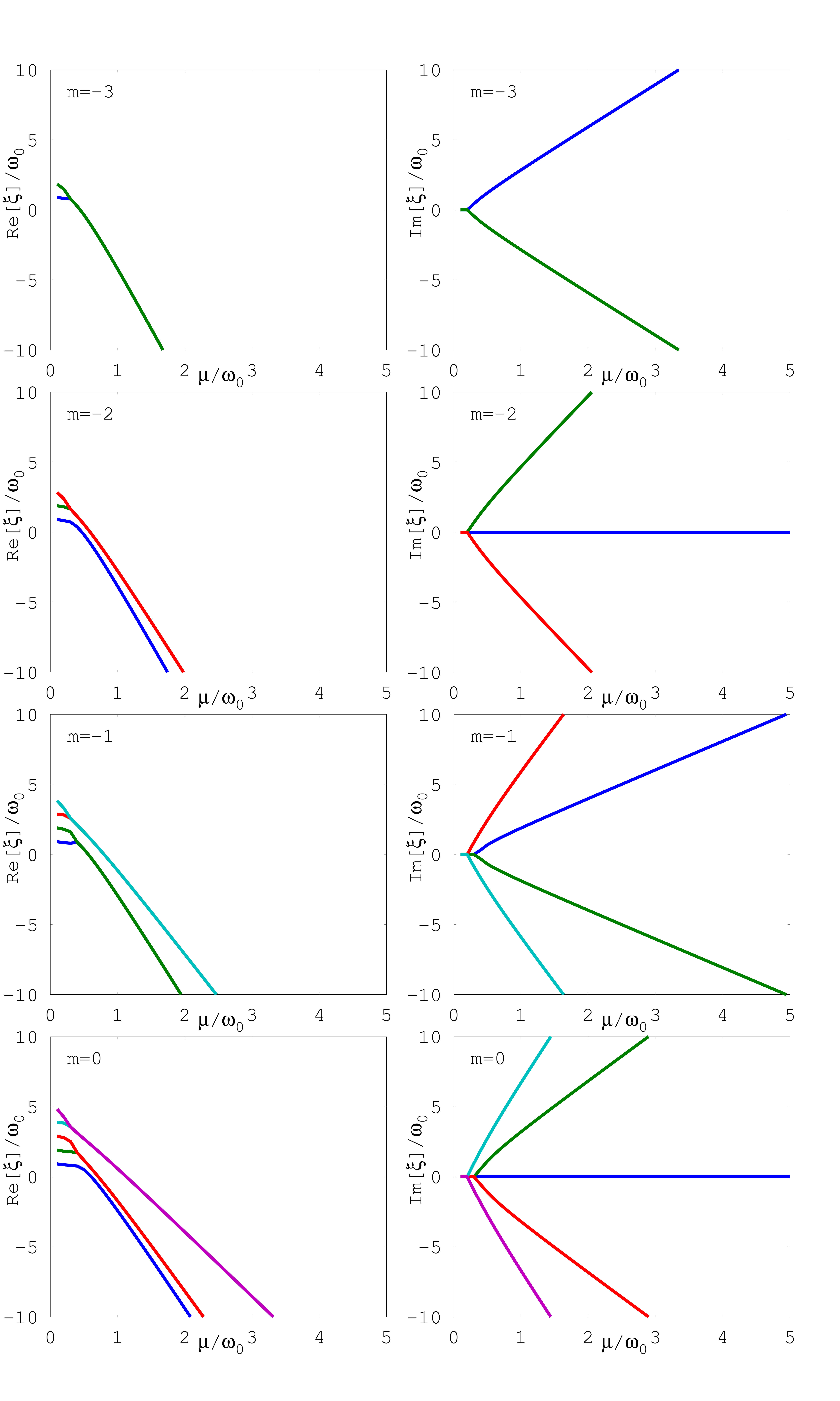}
\caption{\footnotesize \raggedright (Color online) Real and imaginary parts of 
the solutions of Bethe ansatz equations for the simple example given in Eq. 
(\ref{toy modes}). These are the solutions which correspond to the highest 
energy eigenstates. All Bethe ansatz variables approach infinity as 
$\mu\to\infty$. At the $\mu\to 0$ limit they settle onto the vacuum 
oscillation frequencies beginning from the lowest one. This configuration 
yields the highest energy eigenstate at the $\mu\to 0$ limit. [See Eq. 
(\ref{transformation 5}.)]. Note that for low values of $\mu$, complex 
conjugate Bethe ansatz variables approach each other and \emph{collide} on the 
real axis, forming two distinct real solutions.}
\label{fig:Solutions of BAEs-highest level}
\end{figure}

{\bf Transformation of Eigenstates:} Now we are faced with the question of which
eigenstate in the $\mu\to\infty$ limit transforms to which eigenstate in 
the $\mu\to 0$ limit as we change $\mu$. In general, this question is not as 
easy to answer for several Bethe ansatz variables as it is for a single 
variable. However, one key observation from our analysis of a single Bethe 
ansatz variable survives when we increase the number of Bethe ansatz variables: 
The highest energy eigenvalue of the Hamiltonian never becomes degenerate as we 
change $\mu$ from very large to very small values.

Let us first demonstrate this in the toy model with $10$ equally spaced
oscillation modes considered in Sec \ref{sec: one BAV} (See Eq. (\ref{toy
modes})) before giving a more general discussion about it.  Out of the expected
total of $1024$ eigenstates of this toy model, $1002$ have $m=\pm3, \pm2,
\pm1,0$ values which can be obtained with two, three, four, and five Bethe
ansatz variables, respectively.  We found all of the $1002$ solutions associated
with these $m$ values by numerically solving the corresponding Bethe ansatz
equations given in Eqs. (\ref{eq:general BA eq.}) and (\ref{eq:general BA eq.
-}). Our numerical solution utilizes the method introduced in Ref.
\cite{Faribault:2011rv}. Since each solution involves several complex
variables, it is impractical to present all of them here. In Fig.
\ref{fig:energies of BA states} we present the energy eigenvalues that we
calculate by substituting these solutions in Eqs.  (\ref{eq:general energies of
BA states}) and (\ref{eq:general energies of BA states -}). This figure also
includes the eigenvalues of $m=\pm4,\pm5$ eigenstates for completeness. 
Note that $m=\pm 4$ eigenvalues were already discussed in the previous section 
and $m=\pm5$ eigenvalues are taken from Eq. (\ref{eq:highest-lowest weight 
states}). Notice that for each $m$, the highest energy eigenvalue grows as 
$30\mu$ as expected from the fact that these states become $|5\; m\rangle$ in 
the $\mu\to 0$ limit.

In general, it is difficult to identify which eigenstate in the $\mu\to\infty$ 
limit is connected to which eigenstate in the $\mu\to0$ limit from Fig. 
\ref{fig:energies of BA states}. As can be seen in the insets, the eigenvalues 
cross each other at several points in the low $\mu$ region. The only exceptions 
are the highest energy eigenvalues. For each $m$, the highest energy eigenvalue 
is distinctly nondegenerate for any value of $\mu$. Using this observation, it 
is possible to identify which state in the $\mu\to0$ limit is connected to the 
state $|5,\; m\rangle$ in the $\mu\to\infty$ limit. All one needs to do is to 
identify the highest energy eigenstate at $\mu=0$ for a given value of $m$. As 
per Eq. (\ref{n1 and n2}) such a state should include $n^{(1)}=5+m$ neutrinos 
in the $\nu_1$ state and $n^{(2)}=5-m$ neutrinos in the $\nu_2$ state. Since 
having isospin-up (down) neutrinos at lower (higher) oscillation modes 
increases the energy, the highest energy state is found by placing all 
available $\nu_1$'s ($\nu_2$'s) in the lowest (highest) oscillation modes.  This 
way one concludes that the states
\begin{equation}
\label{transformation 5}
|5,\; m\rangle \longleftrightarrow
|\underbrace{\nu_1, \dots, \nu_1}_{5+m},
\underbrace{\nu_2, \dots, \nu_2}_{5-m}\rangle
\end{equation}
are analytically connected to each other through the running of $\mu$.

\begin{figure}[t!]
\includegraphics[width=1\columnwidth]{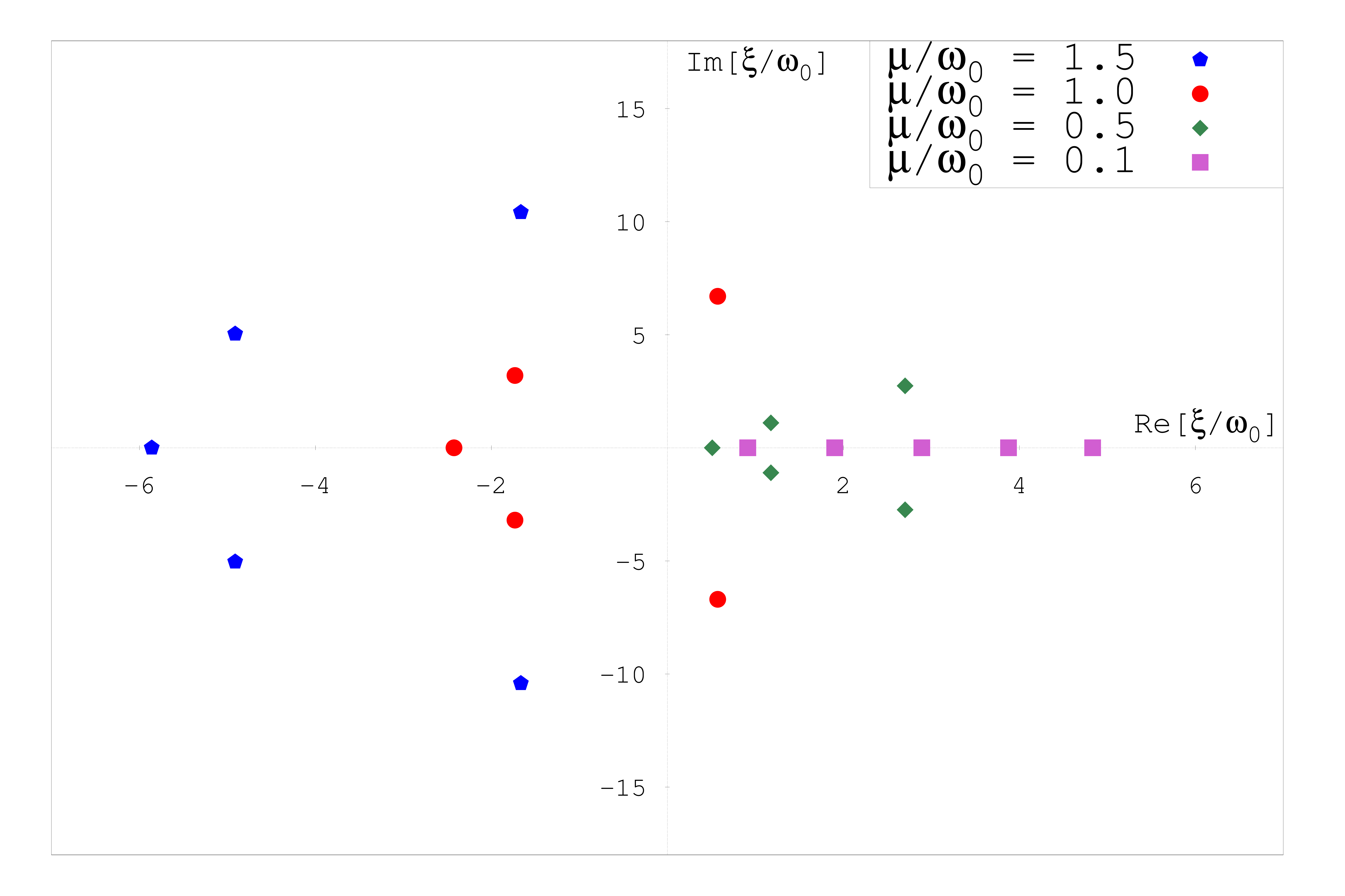}
\caption{\footnotesize \raggedright (Color online)
The behavior of the five Bethe ansatz variables corresponding to the $m=0$ case 
as $\mu$ decreases.  The free charges organize themselves into an arc in the
complex plane which is symmetric with respect to the $x$ axis.  As $\mu$
decreases, which increases the external electrostatic field, the arc of free
charges closes in on the fixed charges. As $\mu\to 0$, the Bethe ansatz 
variables settle on the lowest oscillation modes.
}
\label{arc}
\end{figure}

This can also be understood by examining the behavior of Bethe ansatz variables
corresponding to the highest energy eigenvalues. In Fig. \ref{fig:Solutions of
BAEs-highest level}, we show the solutions in Eq. (\ref{eq:general BA eq.}) with
two, three, four, and five Bethe ansatz variables corresponding to the highest
energy eigenstates with $m=-3,-2,-1,0$, respectively. (The solution with a
single Bethe ansatz variable corresponding to the highest energy eigenvalue with
$m=-4$ is already shown in the lowest line of Fig. \ref{one BAV numerical
solution}.) As expected from the discussion above, these solutions are such that
all variables start from the $x\to-\infty$ region when $\mu\to\infty$, yielding 
the maximum energy at this limit according to Eq.  (\ref{eq:general energies of 
BA states}). As $\mu$ decreases, the Bethe ansatz variables approach the 
finite region and settle on top of the lowest possible vacuum oscillation 
frequencies. This configuration yields the maximum energy in the $\mu\to 0$ 
limit. According to Eqs.  (\ref{mu to 0 limit of xi_alpha state}) and (\ref{mu 
to 0 limit of xi_alpha prime}), those neutrinos in the lowest oscillation modes 
are then converted to $\nu_1$, while those occupying the high oscillation modes 
remain $\nu_2$. This behavior is explicitly shown in Fig. \ref{arc} for five 
Bethe ansatz variables corresponding to the $m=0$ case.  As can be seen in this 
figure, the free charges form an arc in the complex plane which closes in on 
the fixed changes as $\mu$ decreases.

Although we obtained Eq. (\ref{transformation 5}) in the context of our simple
example, the rest of this paper is based on the assumption that it is always
true; i.e., the highest energy eigenvalues of the Hamiltonian for any value of
$m$ never become degenerate so the states
\begin{equation}
\label{transformation}
|\frac{n}{2},\; m\rangle
\longleftrightarrow |\underbrace{\nu_1, \dots, \nu_1}_{\frac{n}{2}+m },
\underbrace{\nu_2, \dots, \nu_2}_{\frac{n}{2}-m}\rangle.
\end{equation}
are analytically connected to each other by running $\mu$. We assume that
this is true even when we allow more than one neutrino in the same oscillation
mode. Whether or not the latter is allowed, the meaning of the right-hand side 
of Eq. (\ref{transformation}) is clear: $\nu_1$'s fill up all available states
starting from the lowest possible oscillation modes, and $\nu_2$'s fill up the
rest.  One oscillation mode in the middle can possibly contain both $\nu_1$ and
$\nu_2$. In that case, they should be symmetrized as per our discussion below
Eq. (\ref{energy for mu=0}).

As mentioned in Sect. \ref{section:Hamiltonian and Isospin Formulation of the
Problem}, the neutrino Hamiltonian given in Eq. (\ref{eq:H Total}) and the
pairing Hamiltonian given in Eq. (\ref{eq:pairing Hamiltonian}) differ by an
overall minus sign such that the highest energy eigenstates of the former (for
different values of $m$) correspond to the ground states of the latter (for
different numbers of pairs).  The nondegeneracy of these states is a well-known
phenomenon which is observed in numerical solutions of the Bethe ansatz
equations under a variety of conditions. Although these solutions are studied in
the context of fermion pairing in the literature, in what follows we discuss
them using the language of self-interacting neutrinos.  No general proof of Eq.
(\ref{transformation}) exists in the literature. However, a proof for large
values of $n$ and $m$ is originally given by Gaudin
\cite{Gaudin_Electrostatic_Analogy}, and later elaborated by Richardson
\cite{richardson:1802}. This proof is based on the observation that as the
number of Bethe ansatz variables ($\kappa$) increases, the solutions of the 
Bethe ansatz equations organize themselves into (piecewise) continuous arcs 
which are symmetric with respect to the $x$ axis in the electrostatic analogy. 
In particular, those solutions corresponding to the highest energy states on 
the left-hand side of Eq. (\ref{transformation}) form a single continuous arc. 
As the interaction constant decreases and the external electrostatic field 
becomes stronger, this arc of free charges closes itself onto the line of fixed 
charges on the $x$ axis in order to find a stable configuration.  During the 
transition the arc stays a single continuous structure.  As a result, when it 
closes in on the line of fixed charges, the Bethe ansatz variables approach the 
lowest oscillation frequencies, converting the $\nu_2$'s in these modes into 
$\nu_1$'s. The resulting state is the rigth-hand side of Eq. 
(\ref{transformation}). The analytical proofs of Gaudin and Richardson have 
been shown to agree with the numerical results for up to $n=1600$ and 
$\kappa=800$ in Ref. \cite{Roman:2002dh} for the case of equally spaced 
oscillation modes similar to the one introduced in Eq. (\ref{toy modes}).

One can also discuss Eq. (\ref{transformation}) from the point of view of
experiments involving cold atomic systems. In such experiments where one can
control the strength of the pairing interaction, the system is observed to move
from very weak (BCS) to very strong (BEC) interaction regimes smoothly. 
This so-called crossover behavior indicates that the ground state of the
system never undergoes a level crossing with one of the excited levels as it
moves between these two limits. A level crossing would show itself as abrupt
changes in the measurable quantities, which is not observed experimentally. This
is true for any number of particle pairs which correspond to different $m$
values in the neutrino case, whether or not these pairs occupy degenerate 
energy levels. For a review, see Refs. \cite{RevModPhys.80.885, 
RevModPhys.80.1215,
Randeria:2013kda}.

\section{Spectral Splits in Exact Many-Body Systems}
\label{section: Spectral Splits in Exact Many-Body States}

In this section, we show that, under the assumption of perfect adiabaticity, the
nondegeneracy of the highest energy eigenstates of the neutrino Hamiltonian for
different $m$ values [cf. Eq.  (\ref{transformation})] leads to a spectral
split in the energy spectrum of a neutrino ensemble which initially consists of
electron neutrinos only.

According to the adiabatic theorem, in a scenario in which $\mu$ changes
sufficiently slowly with time, the time evolution of the highest energy
eigenstates for each $m$ will be such that they will continue to occupy the same
instantaneous eigenstates. In particular, apart from a phase, the state on the 
left-hand side of Eq. (\ref{transformation}) will evolve into the state on the 
right-hand side as $\mu$ slowly decreases from very large to very small values 
under the adiabatic evolution conditions. Here we do not specify how slow is 
\emph{sufficiently} slow. But the conditions for perfect adiabaticity are 
typically satisfied in a core collapse supernova \cite{Raffelt:2007xt}.

In the experimental setups involving cold atom systems, one can control how
slowly the interaction constant changes with time, and ensure that the system
stays in its ground state instead of being excited to the next energy level. In
the case of neutrinos, one may intuitively think that, even if the perfectly
adiabatic conditions are satisfied, the system would not stay on the highest
energy eigenstate but would make a transition to a lower energy state.
However, one should keep in mind that the neutrino system that we consider in
Eq. (\ref{eq:H Total}) is dissipationless; i.e., the energy of the system
is conserved.

Since $J_z$ is a conserved quantity of the problem, the adiabatic theorem can be
applied to any combination of highest energy eigenstates for different $m$
values. In other words, based on Eq. (\ref{transformation}), if the initial 
state
at $\mu\to\infty$ is in the form
\begin{equation}
\label{initial state}
|\psi\rangle_{\mbox{\tiny initial}}=\sum_{m=-n/2}^{n/2} c_m |\frac{n}{2},
m\rangle,
\end{equation}
then, it will evolve into
\begin{align}
\label{final state}
|\psi\rangle_{\mbox{\tiny final}}
&=\mathcal{U}|\psi\rangle_{\mbox{\tiny initial}}\\
&=\sum_{m=-n/2}^{n/2} c_m \phi_m
|\underbrace{\nu_1, \dots, \nu_1}_{\frac{n}{2}+m },
\underbrace{\nu_2, \dots, \nu_2}_{\frac{n}{2}-m}\rangle \nonumber
\end{align}
in the $\mu\to 0$ limit. Here $\mathcal{U}$ denotes the evolution operator from
the $\mu\to\infty$ limit to the $\mu\to 0$ limit under the adiabatic 
approximation, and $\phi_m$ are some phases which contain both dynamical and 
geometrical components associated with the adiabatic evolution. Their actual 
values are irrelevant for our purposes because they do not affect the final 
energy distributions.

Equations (\ref{initial state}) and (\ref{final state}) are particularly 
useful when we consider an initial neutrino ensemble which consists entirely of
electron neutrinos. Regardless of how many neutrinos each oscillation mode
contains, such a state is the highest weight state in the flavor basis because 
all flavor isospins are up:
\begin{equation}
\label{eq: highest weigth flavor}
|\psi\rangle_{\mbox{\tiny initial}}=
|\nu_e,\nu_e,\dotsc,\nu_e\rangle=|n/2,n/2\rangle_{\mbox{\tiny flavor}}
\end{equation}
Using Eqs. (\ref{eq:global SU(2) transformation}) and (\ref{eq:transformation
operator}), this state can be written as
\begin{align}
\label{initial state electron}
&|\psi\rangle_{\mbox{\tiny initial}}= 
U^\dagger|\nu_1,\nu_1,\dotsc,\nu_1\rangle\\
&=\sum_{m=-n/2}^{n/2}
(\cos\theta)^{(\frac{n}{2}+m)}(\sin\theta)^{(\frac{n}{2}-m)}
 \sqrt{n\choose {\frac{n}{2}+m}}|n/2,m\rangle. \nonumber
\end{align}
Assuming that this state evolves adiabatically as described above, Eq.
(\ref{final state}) tells us that it will turn into
\begin{align}
\label{final state electron}
|\psi\rangle_{\mbox{\tiny final}} =&\sum_{m=-n/2}^{n/2}
(\cos\theta)^{(\frac{n}{2}+m)}(\sin\theta)^{(\frac{n}{2}-m)}
\nonumber \\
&\times \sqrt{n\choose {\frac{n}{2}+m}} \phi_m|\underbrace{\nu_1, \dots,
\nu_1}_{\frac{n}{2}+m }, \underbrace{\nu_2, \dots,
\nu_2}_{\frac{n}{2}-m}\rangle
\end{align}
as $\mu\to0$.

\begin{figure}
\begin{subfigure}[t]{1\columnwidth}
\includegraphics[width=1\textwidth]{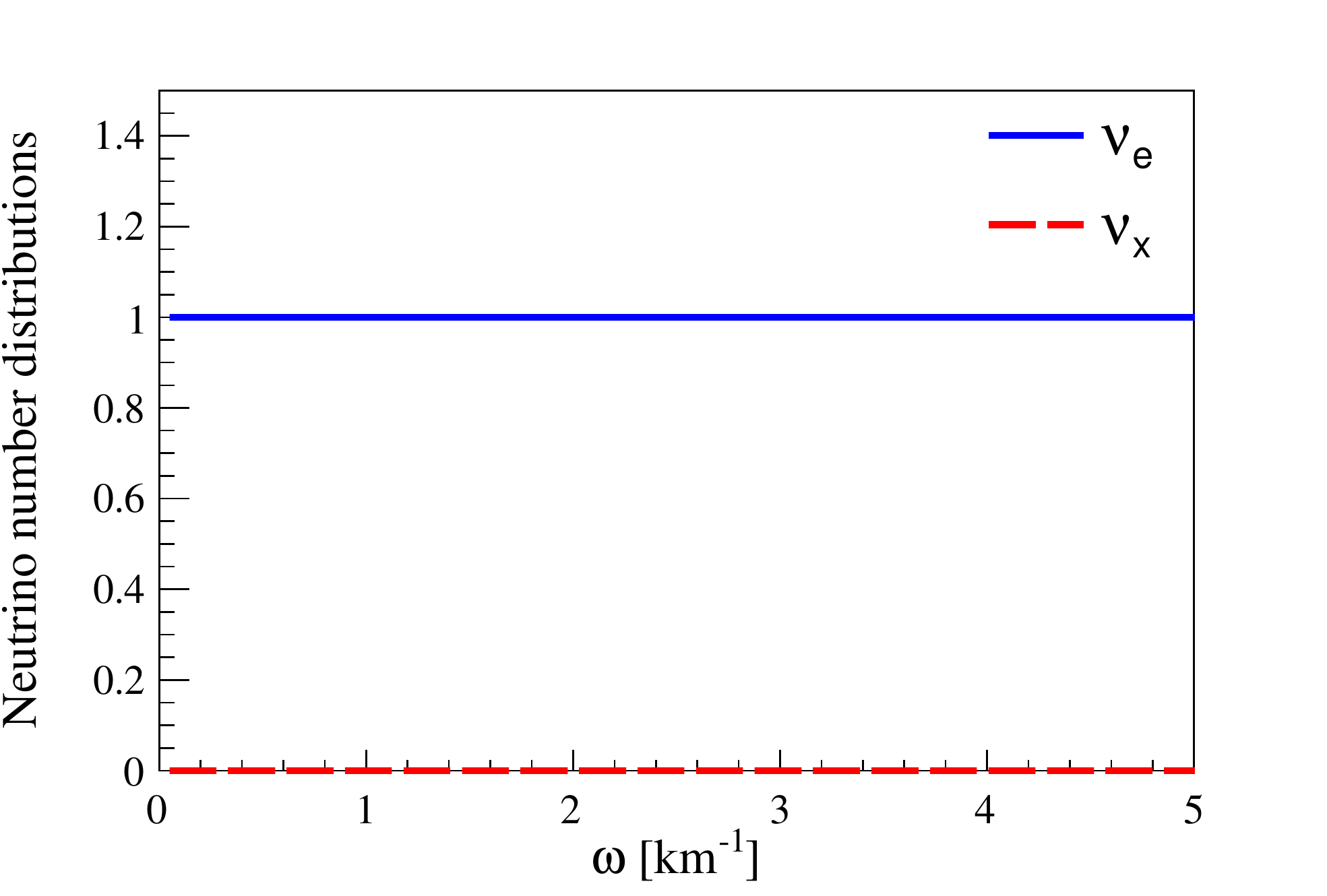}
\caption{\footnotesize \raggedright Initial distribution}
\label{fig:Box Distrubiton initial status}
\end{subfigure}
\begin{subfigure}[t]{1\columnwidth}
\includegraphics[width=1\textwidth]{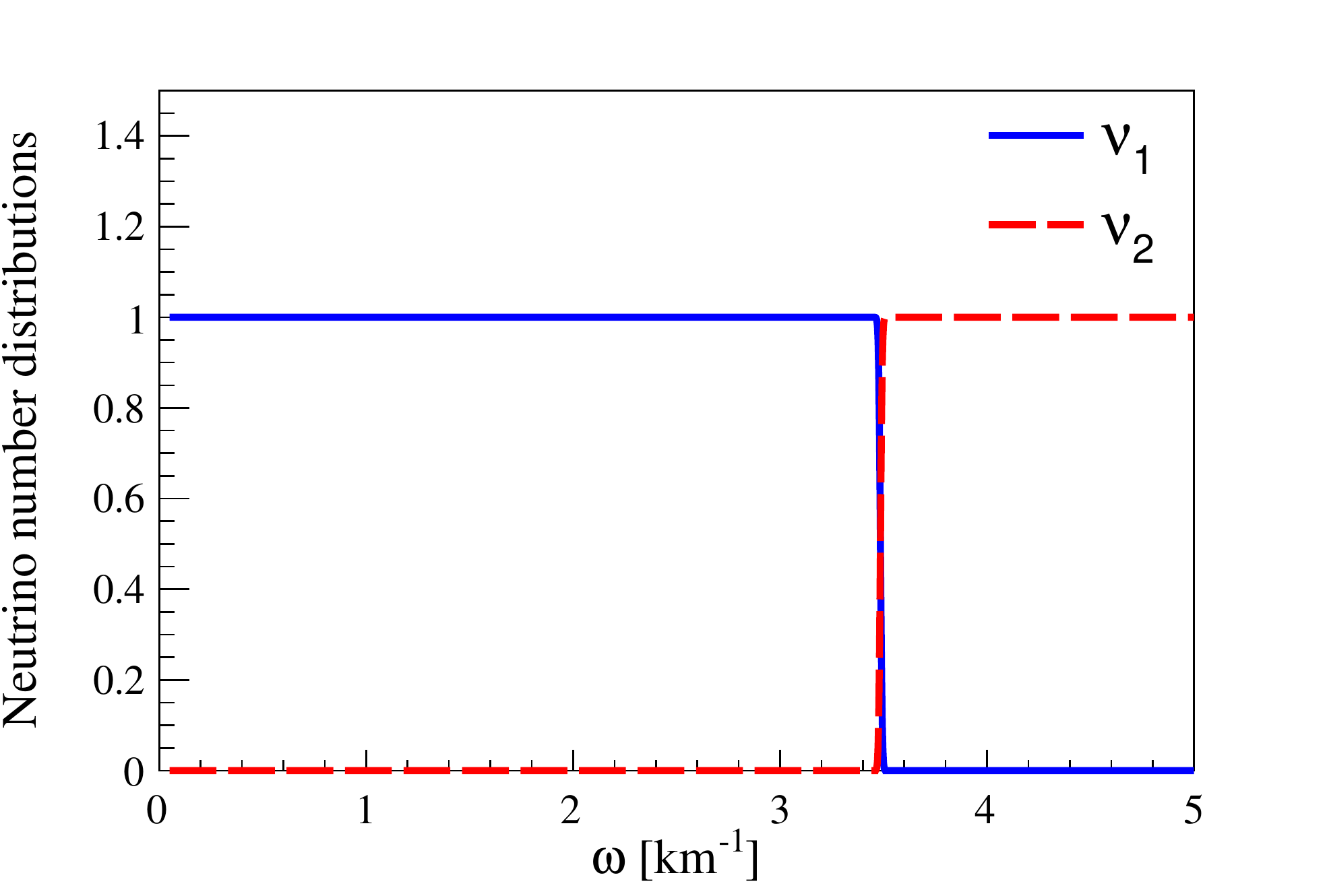}
\caption{\footnotesize \raggedright Final distributions}
\label{fig:Box Distrubiton final status}
\end{subfigure}
\caption{\footnotesize \raggedright (Color online) Adiabatic evolution of
initial box distribution of electron neutrinos. This is the extension of the toy
model that we introduced in Eq. (\ref{toy modes}). This time we consider
$\Omega=10^7$ equally spaced oscillation modes each containing a single
neutrino. We take $\omega_0=5 \times 10^{-7}$ km$^{-1}$ and adopt the normal
mass hierarchy with solar mixing parameters.}
\end{figure}
This final state is a superposition of $2n+1$ orthogonal components, each 
one with a split structure, i.e., filled by $\nu_1's$ up to a certain
point and by $\nu_2$'s after that. In fact, for each one of these components,
we can write
\begin{align}
\label{transformation right}
&|\underbrace{\nu_1, \dots, \nu_1}_{\frac{n}{2}+m },
\underbrace{\nu_2, \dots, \nu_2}_{\frac{n}{2}-m}\rangle=\\
&
(\prod^{s-1}_{k=1}|\frac{n_{\omega_k}}{2},\frac{n_{\omega_k}}{2}\rangle)
(|\frac{n_{\omega_s}}{2},m_{\omega_s}\rangle)
(\prod^\Omega_{k={s+1}}|\frac{n_{\omega_k}}{2},-\frac{n_{\omega_k}}{2}\rangle ).
\nonumber
\end{align}
This equation is very simple to understand. It tells us that lower oscillation 
modes containing only $\nu_1$'s live in the highest weight states, while the 
higher oscillation modes containing only $\nu_2$'s live in the lowest weight 
states. A single oscillation mode labeled by $s$ may contain both types of 
neutrinos in a completely symmetrized way [see the discussion following Eq.
(\ref{transformation})] and has $m_{\omega_s}=(n^{(1)}_{\omega_s}
-n^{(2)}_{\omega_s})/2$.  This particular mode can be found from the condition
that
\begin{align}
\label{eq:S(l)}
\sum_{k=1}^{s-1} n_{\omega_k}
\leq \frac{n}{2}+m\leq
\sum_{k=1}^{s} n_{\omega_k}
\end{align}
which simply tells us that the number of $\nu_1$'s on the left-hand side
of Eq. (\ref{transformation right}) is more than enough to fill the first $s-1$
oscillation modes, but not enough to fill the $s{\mbox{th}}$ oscillation
mode. We call $s$ the split index, and the corresponding frequency
$\omega_s$ the split frequency. Clearly they both depend on the value
of $m$. For this reason, in what follows we change the notation as
\begin{equation}
s\longrightarrow s(m)
\quad \mbox{and} \quad
\omega_s\longrightarrow \omega_{s(m)}
\end{equation}
In fact, since the  final state in Eq. (\ref{final state electron}) contains
components with all possible $m$ values, every allowed oscillation mode is a
split frequency for one or more of these components.  However, what we are
interested in, is the normalized energy distributions given by
\begin{equation}
\label{distribution}
\Phi^{(\alpha)}(\omega)=\frac{1}{n}\; \langle N^{(\alpha)}(\omega)\rangle.
\end{equation}
Here $\alpha$ can take values in $e,x,1,2$.  The initial neutrino energy
distributions can be easily written down in flavor basis as
\begin{equation}
\label{initial distribution}
\Phi^{(e)}_{\mbox{\tiny initial}}(\omega)=\frac{n_\omega}{n} \qquad
\Phi^{(x)}_{\mbox{\tiny initial}}(\omega)=0.
\end{equation}
This follows from the facts that the number of neutrinos in the oscillation mode
$\omega$ is $n_\omega$, and the initial state in Eq. (\ref{initial state
electron}) contains nothing but electron neutrinos.

Final neutrino energy distributions are easiest to calculate in the mass basis.
It is helpful to first note that Eq. (\ref{eq:isospin mass}) leads to
\begin{equation}
\label{N^a_omega expectation}
N^{(a)}_\omega=\frac{n_\omega}{2}\pm J^z_\omega
\end{equation}
where $a=1,2$. Here, and in what follows, we use the upper sign for $a=1$ and
lower sign for $a=2$. Substituting Eq. (\ref{transformation right}) into Eq.
(\ref{final state electron}) and calculating the expectation value of
$N^{(a)}_\omega$ using Eq. (\ref{N^a_omega expectation}) leads to
\begin{align}
\label{distribution in mass}
\Phi^{(a)}(\omega_k)=\frac{n_{\omega_k}}{2n}\pm
\frac{1}{n}\sum_{m=-n/2}^{n/2}
&(\cos\theta)^{(n+2m)}(\sin\theta)^{(n-2m)}\nonumber\\
&\times {n\choose {\frac{n}{2}+m}}
m_{\omega_k}
\end{align}
where
\begin{equation}
\label{distribution in mass m}
m_{\omega_k}=
\begin{cases}
\displaystyle{n_{\omega_k}/2}, & \mbox{for } k<s(m)\\
\displaystyle{\frac{n^{(1)}_{\omega_{s(m)}}-n^{(2)}_{\omega_{s(m)}}}{2}}, &
\mbox{for } k=s(m)\\
\displaystyle{-n_{\omega_k}/2}, & \mbox{for  } k>s(m)\\
\end{cases}
\end{equation}
in accordance with Eq. (\ref{transformation right}).

\begin{figure}[h]
\begin{subfigure}[b]{1\columnwidth}
\raggedright
\includegraphics[width=1\textwidth]{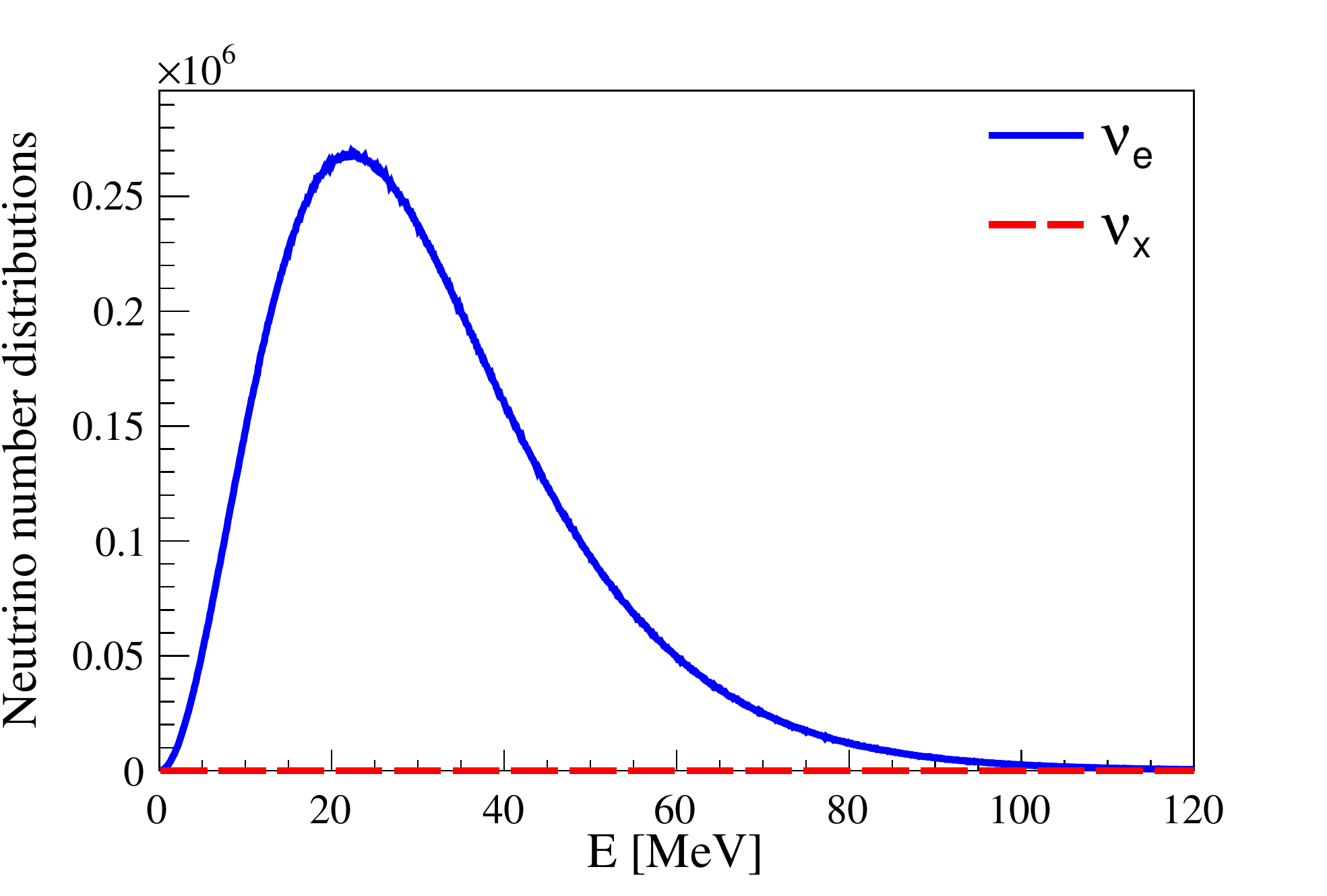}
\caption{\footnotesize \raggedright Initial Fermi-Dirac distribution of
electron neutrinos.}
\label{fig:expectation values of j_w for FD a}
\end{subfigure}
\begin{subfigure}[b]{1\columnwidth}
\raggedright
\includegraphics[width=1\textwidth]{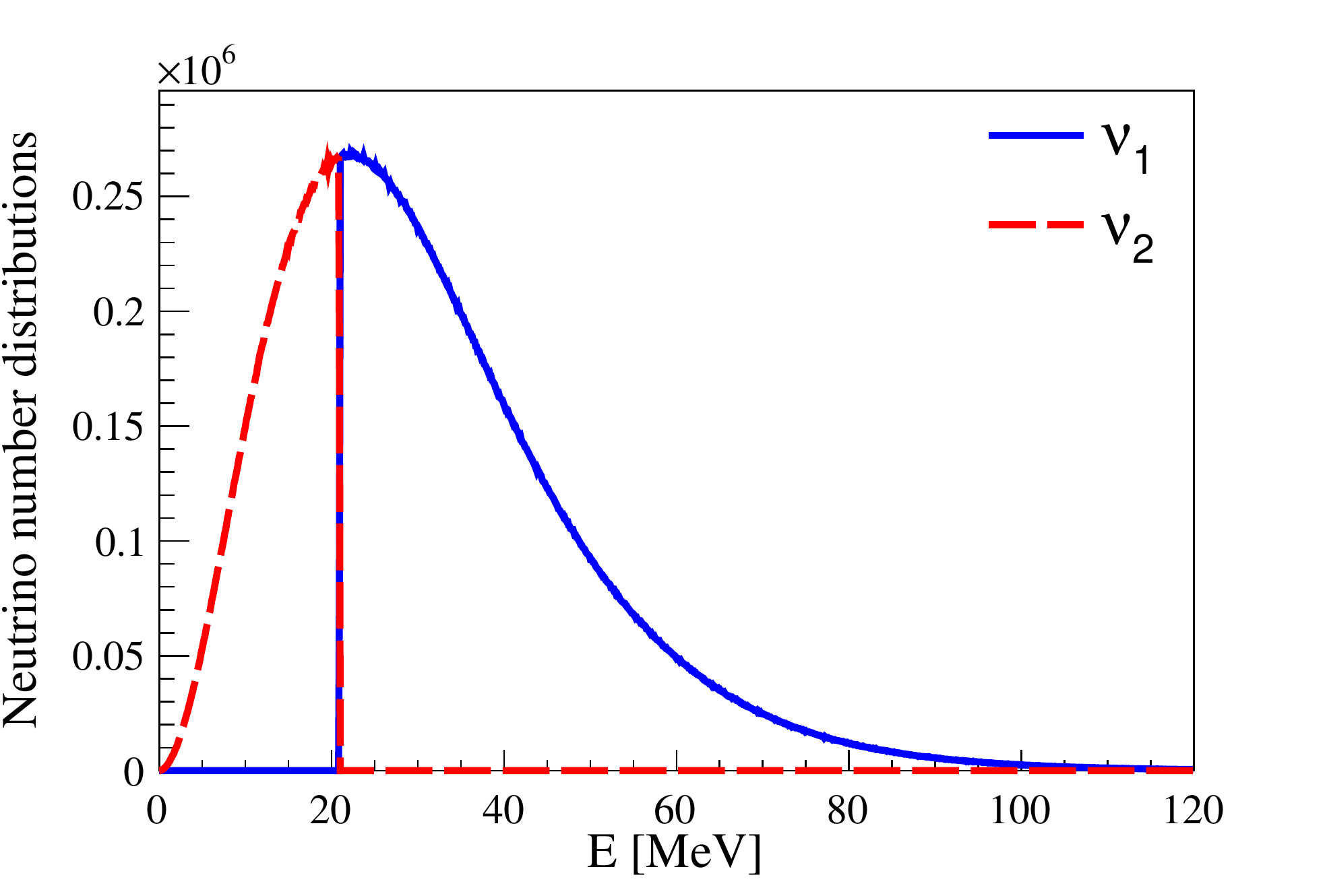}
\caption{\footnotesize \raggedright Final distributions in mass basis.
}
\label{fig:expectation values of j_w for FD b}
\end{subfigure}
\caption{\footnotesize \raggedright (Color online) Adiabatic evolution of an
initial thermal distribution of electron neutrinos. We take $n=10^8$ neutrinos
distributed over $\Omega=1200$ oscillation modes which are equally spaced in
energy [cf. Eq. (\ref{eq:FD distribution})]. We take the temperature
as $kT=10$ MeV, and adopt the normal mass hierarchy with solar mixing
parameters. Note that, we use a pseudo-random
number generator in order to distribute $n$ neutrinos into $\Omega$
energy bins according to Fermi-Dirac probability distribution. This guarantees
that each energy bin has an integer number of neutrinos and a corresponding
well-defined isospin quantum number.
The use of the pseudo-random number generator causes the wiggles seen in the
original distribution.
}
\label{fig:FD distributions}
\end{figure}

The calculation of the final energy distribution of neutrinos involves the
numerical computation of Eqs. (\ref{distribution in mass}) and
(\ref{distribution in mass m}). However, for the typical number of neutrinos
that we work with, which extends up to $n=10^8$, it becomes impractical to
directly calculate the factorials involved in the ${n\choose {\frac{n}{2}+m}}$
term. For this reason,  we use the fact that
\begin{align}
\label{eq:Gaussian approximation}
(\cos\theta)^{(n+2m)}&(\sin\theta)^{(n-2m)}
{n\choose {\frac{n}{2}+m}}\nonumber \\
&\approx
\frac{1}{\sqrt{2\sigma^2\pi}}\exp\left({-\frac{\left(m-\bar{m}\right)^2}{
2\sigma^2}}\right)
\end{align}
where
\begin{align}
\label{gaussian parameters}
\bar{m}=n(\cos^2\theta-\frac{1}{2}) \qquad \sigma=\sqrt{n}\cos\theta\sin\theta
\end{align}
The left-hand side of Eq. (\ref{eq:Gaussian approximation}) is nothing more than
the binomial distribution, while the right-hand side is the Gaussian 
distribution with mean value $\bar{m}$ and standard deviation $\sigma$.
These two distributions approximate very well to each other at the large $n$
values that we work with. Substituting Eq. (\ref{eq:Gaussian approximation})
into Eq. (\ref{distribution in mass}) leads to the formula for the final
neutrino energy distributions that we use in our numerical computations:
\begin{align}
\label{final}
&\Phi^{(a)}(\omega_k)=\\ \nonumber
&\frac{n_{\omega_k}}{2n}\pm
\frac{1}{n}\sum_{m=-n/2}^{n/2}
\frac{1}{\sqrt{2\sigma^2\pi}}\exp\left({-\frac{\left(m-\bar{m}\right)^2}{
2\sigma^2}}\right)
m_{\omega_k}
\end{align}
One can now calculate the final energy distribution by numerically computing
$m_{\omega_k}$ from Eqs. (\ref{eq:S(l)}) and (\ref{distribution in mass m}) for
a given initial distribution and by substituting them in Eq. (\ref{final}).

The simplest case that one can consider is an extension of the toy model that we
introduced in Eq. (\ref{toy modes}).  This time we divide the interval into
$\Omega=10^7$ equally spaced oscillation modes and allow each mode to contain
only a single neutrino (i.e., $n_{\omega_k}=1$) so that the total number of
neutrinos is $n=10^7$. This brings us to the continuum limit of this example.
[See Fig. \ref{fig:Box Distrubiton initial status}.] In this particular case,
Eq. (\ref{distribution in mass m}) is simply reduced to
\begin{equation}
m_{\omega_k}= \left\{
	 \begin{array}{l l}
	  -1/2 & \quad \text{if $k>\frac{n}{2}+m$} \\
	  +1/2 & \quad \text{otherwise}
	 \end{array}
\right.
\end{equation}
whose substitution in Eq. (\ref{final}) yields the final energy distribution
shown in Fig. \ref{fig:Box Distrubiton final status}. In this example we take
$\omega_0=5\times 10^{-7}$ km$^{-1}$. We adopt solar mixing parameters for
demonstration purposes because a smaller mixing angle brings the split point in
Fig. \ref{fig:expectation values of j_w for FD b} too close to the edge of the
distribution.

Next, we consider another example in which the electron neutrinos are initially
in a thermal energy distribution, i.e., the fraction of neutrinos in the
$(\omega,\omega+d\omega)$ interval is given by
\begin{equation}
\label{eq:FD distribution}
\Phi^{(e)}_{\mbox{\tiny initial}}(\omega)d\omega
=\frac{1}{(kT)^3\Gamma(3)F_2(0)}\frac{E^2dE}{e^{E/kT}+1}
\end{equation}
Here $k$, $\Gamma$, and $F_2$ denote the Boltzmann constant, the Gamma function 
and the complete Fermi-Dirac integral of rank 2, respectively. The variables
$\omega$ and $E$ are related by Eq. (\ref{w}). This initial distribution is
shown in Fig. \ref{fig:expectation values of j_w for FD a} as a function of
energy. In this example, we take $n=10^8$ neutrinos distributed over
$\Omega=1200$ oscillation modes which are equally spaced in energy. We take
the temperature as $kT=10$ MeV. Numerical calculation of the final
energy distribution involves finding $s(m)$ for each $m$ from Eq.
(\ref{eq:S(l)}) and substituting them in Eqs.  (\ref{distribution in mass m})
and (\ref{final}). The result is shown in Fig.  \ref{fig:expectation values of
j_w for FD b}.  The final distribution involves a single sharp swap of neutrino
energy distributions at the low energy region.


In spite of the fact that the final state given in Eq. (\ref{final state
electron}) is a superposition of $2n+1$ states with different split frequencies,
the final energy spectra involves only a single split. This is due to the fact
that the Gaussian distribution in Eq. (\ref{eq:Gaussian approximation}) has a
very small fractional width. In a distribution, the concept of fractional
width describes the ratio between the terms which are considerably different
from zero and the total number of terms. In the context of our problem, it is 
given by
$\sigma/(2n+1)$ and its significance can be seen from the following discussion:
Since $\sigma$ is proportional to $\sqrt{n}$ as shown in Eq. (\ref{gaussian
parameters}), we see that only those states in the interval
\begin{equation}
\label{m interval}
m\sim \bar{m}\pm {\cal O}(\sqrt{n})
\end{equation}
significantly contribute to the sum in Eq. (\ref{final}). As we increase the 
number of neutrinos, the number of such states increases. However, since the 
total number of the terms in Eq. (\ref{final}) is equal to $2n+1$, the 
\emph{fraction} of the states which significantly contribute to the sum 
decreases as $\sigma/(2n+1)\sim n^{-1/2}$.  Given that the number of 
oscillation modes in the system is constant, one can intuitively infer that the 
split frequencies of these significantly contributing states should
approach each other. In fact, for the states in the interval given in Eq. 
(\ref{m interval}) the split frequencies fall into the interval
\begin{equation}
\label{omega interval}
\omega_s\sim \bar{\omega}_s \left[1 \pm {\cal O}(\frac{1}{\sqrt{n}})\right].
\end{equation}
Here $\bar{\omega}_s$ is the split frequency corresponding to $\bar{m}$.  This
can be easily seen by taking the continuum limit of Eq. (\ref{eq:S(l)}).  Let
$\rho_\omega$ be the density of oscillation modes so that $\rho_\omega d\omega$
is the number of modes between $\omega$ and $\omega+d\omega$. In this limit, the
difference between the upper and lower bounds of Eq. (\ref{eq:S(l)}) is
infinitesimally small and one can write
\begin{equation}
\label{eq:split equation continuum}
\int_0^{\omega_s}n_{\omega}\rho_{\omega}d\omega=\frac{n}{2}+m
\end{equation}
In particular, substituting the value of $\bar{m}$ from Eq. (\ref{gaussian
parameters}) leads to the equation for the corresponding split frequency 
$\bar{\omega}_s$:
\begin{equation}
\label{omega bar}
\int_0^{\bar{\omega}_s}n_{\omega}\rho_{\omega}d\omega=n\cos^2\theta
\end{equation}
Equation (\ref{eq:split equation continuum}) also tells us that a small spread
$dm\sim{\cal O}(\sqrt{n})$ around $\bar{m}$ leads to a spread
\begin{equation}
\label{dm s} d\omega_s =\frac{dm}{n_{\bar{\omega}_s}\rho_{\bar{\omega}_s}}
\sim \frac{\sqrt{n}}{n_{\bar{\omega}_s}\rho_{\bar{\omega}_s}}
\end{equation}
around $\bar{\omega}_s$. Assuming that the neutrino distribution is practically
nonzero in a finite region around $\omega_s$, we can write
$n_{\bar{\omega}_s}\rho_{\bar{\omega}_s} \sim {\cal O}(n/\omega_s)$ from Eq.
(\ref{omega bar}).  When substituted in Eq. (\ref{dm s}), this directly leads to
Eq. (\ref{omega interval}).

For the first example presented in Fig. \ref{fig:Box Distrubiton final status},
Eq. (\ref{omega bar}) yields $\bar{\omega}_s= 3.435$ km$^{-1}$.  For the second
example presented in  Fig.  \ref{fig:expectation values of j_w for FD b} it
leads to $\bar{\omega}_s= 9.03\times 10^{-3}$ km$^{-1}$ which corresponds to 
$E=21.0$ MeV.
Both results agree very well with the values seen in the respective figures.

The results obtained in this section agree very well with the
established understanding of the mean field behavior of the system. In recent
years, systematic numerical studies of the mean field flavor evolution equations
have shown that spectral splits develop around some instabilities which
are located at particular spectral crossing\footnote{These are different 
than the level crossings between different many-body energy eigenstates. 
Spectral crossings are defined as the energies (or frequencies) at which the 
initial spectra of different neutrino flavors become equal.} points 
\cite{Dasgupta:2009mg, Dasgupta:2010cd}.  However, those initial spectra with 
no crossings can still display such behavior (see, e.g., Refs. 
\cite{Raffelt:2007cb,Pehlivan:2016lxx}).  In general an elegant way to
analytically understand the appearance of spectral splits in the mean field case
is through the linearized stability analysis which reformulates the small
amplitude solutions of the mean field equations as linearized
eigenvalue-eigenfunction relations \cite{Banerjee:2011fj}. In this analysis,
some spectral crossings are associated with complex eigenvalues  which lead to
instabilities and induce spectral splits. However, spectra with no crossings are
associated with purely real eigenvalues  for which the linearized analysis
breaks down. This special case is considered in   Ref. \cite{Raffelt:2007cb} and
also applies to the examples that we consider in this paper because initially we
have only a single flavor.  As was discussed in Ref. \cite{Raffelt:2007cb}, one
expects only a single spectral split to develop in this particular case.  For
the normal mass hierarchy, the spectral swap occurs for the frequencies which
are higher than the split frequency (or for energies lower than the split
energy). Given these considerations, the single split frequency in the final
spectra can be directly calculated from a simple conservation law in the mean
field picture as explained below. 

The $z$ component of total mass isospin, which is an exact invariant of the
Hamiltonian in Eq. (\ref{eq:H Total}), is still conserved on average under
the mean field approximation. In other words, the mean field flavor evolution
equations leave the expectation value
\begin{equation}
\label{conservation mean field}
\langle J_z\rangle
=\cos2\theta\langle J^z_{\mbox{\tiny flavor}}\rangle
-\sin2\theta\langle J^x_{\mbox{\tiny flavor}}\rangle
\end{equation}
unchanged.  Here the right-hand side of the equality follows from the inversion
of Eq. (\ref{BCH formula}).  The conservation of this quantity was first
discussed in  Ref. \cite{Raffelt:2007cb} and is generally known as lepton
number conservation in the literature due to the typically small effective
mixing angles employed close to the proto-neutron star.  As was shown in Ref.
\cite{Raffelt:2007cb} this conservation law can be used to calculate a single
split frequency. The value of the conserved quantity in Eq. (\ref{conservation 
mean field}) is most easily computed using the right-hand side for the initial 
state and the left-hand side for the final state. This gives
\begin{equation}
\label{conserved quantity value}
\frac{1}{2}\left(\int_0^{\bar{\omega}_s} n_\omega \rho_\omega d\omega
-\int_{\bar{\omega}_s}^\infty n_\omega \rho_\omega d\omega\right)
=\frac{n}{2}\cos2\theta
\end{equation}
where we used the definitions of isospin operators given in Eqs.
(\ref{eq:isospin mass}) and (\ref{eq:isospin flavor})  together with the 
summation conventions introduced in Sec. \ref{subsection: summation 
convention}. The left-hand side of Eq. (\ref{conserved quantity value})  simply 
reflects the fact that in the final state the $\nu_1$ neutrinos are assumed to 
fill up the levels up to $\bar{\omega}_s$, while $\nu_2$ neutrinos fill up the 
rest; the right-hand side follows from the fact that all neutrinos are 
$\nu_e$ in the initial state.  On the other hand, the total number of neutrinos 
$n$ is equal to
\begin{equation}
\label{n}
\int_0^\infty n_\omega \rho_\omega d\omega=n.
\end{equation}
Substitution of Eq. (\ref{n}) in Eq. (\ref{conserved quantity value})
immediately leads to
\begin{equation}
\label{omega bar mean field}
\int_0^{\bar{\omega}_s}n_{\omega}\rho_{\omega}d\omega=n\cos^2\theta
\end{equation}
which gives the same spectral split frequency as
Eq. (\ref{omega bar}).
However, note that while
Eq. (\ref{omega bar}) is derived using the exact conservation of
$J^z$ in the original many-body formalism,
Eq. (\ref{omega bar mean field}) is derived from the conservation of
the average value $\langle J^z\rangle$.
The derivation of the split frequency using the conservation of
$\langle J^z\rangle$ in the mean field formalism was first carried out in
Ref. \cite{Raffelt:2007cb}.

\section{Inverted hierarchy}

\begin{figure}[h]
\raggedright
\includegraphics[width=1\columnwidth]{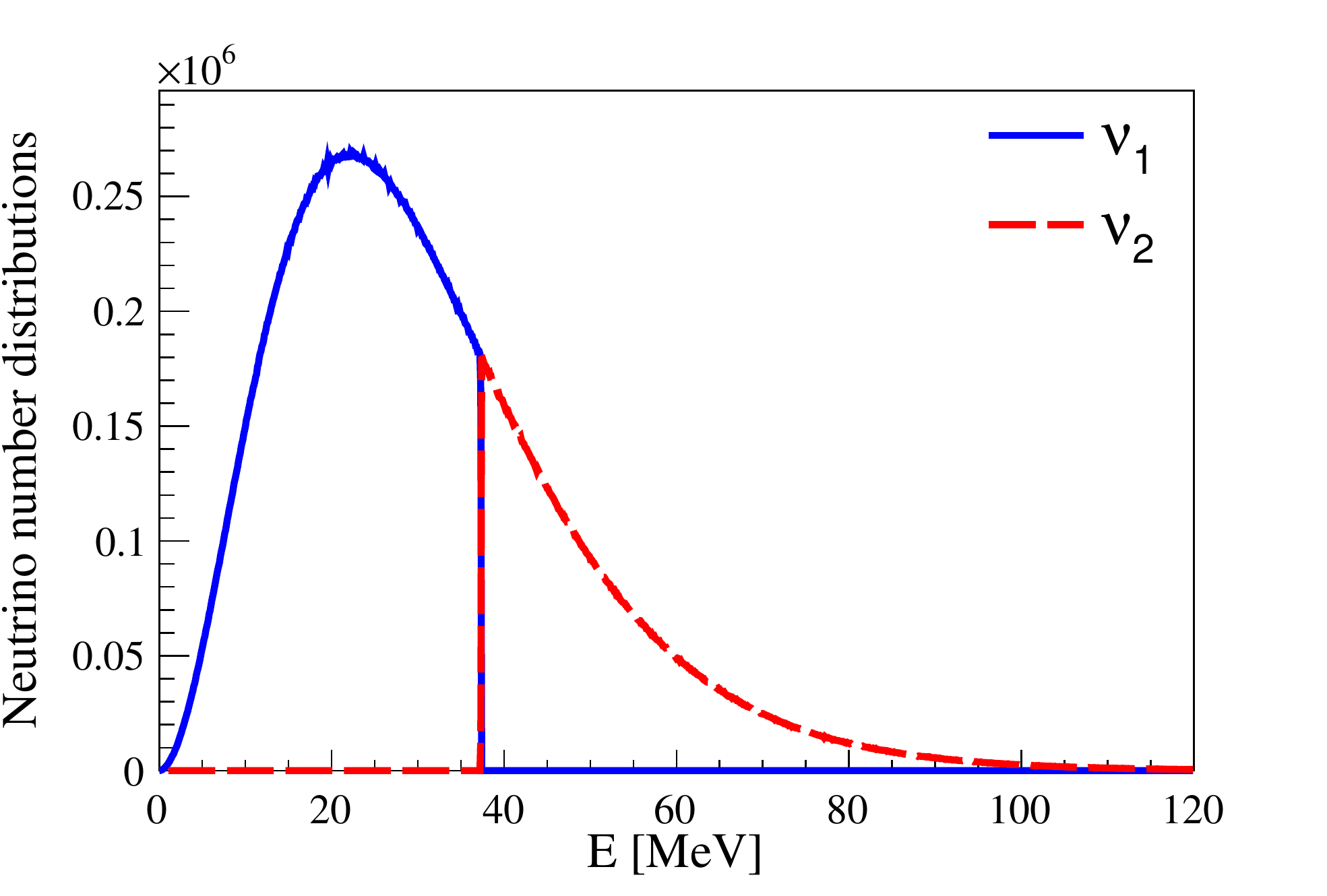}
\caption{\footnotesize \raggedright (Color online) Final distributions of
neutrinos in the mass basis in the case of inverted mass hierarchy. All other
parameters are the same as those in Fig. \ref{fig:FD distributions}.}
\label{fig:FD distributions_inverted}
\end{figure}

In previous sections we worked in the normal mass hierarchy by setting
$m_1<m_2$.  In this section, we convert our results into inverted mass
hierarchy.  For this purpose, we introduce the operator
\begin{equation}
\label{R}
R=e^{-i\pi J^x}
\end{equation}
which converts $\nu_1$ and $\nu_2$ into each other:
\begin{equation}
\label{R transforms a}
R^\dagger a_1(\mathbf{p})R= -ia_2(\mathbf{p})
\qquad
R^\dagger a_2(\mathbf{p})R= -ia_1(\mathbf{p})
\end{equation}
As a result, it transforms the vacuum oscillation Hamiltonian in Eq.
(\ref{eq:vacuum mixing mb H}) into
\begin{equation}
\label{eq:rotation for vacuum term}
R^\dagger H_\nu R =
\sum_{\mathbf{p}}
\left(E_1(p)N^{(2)}_\mathbf{p}+E_2(p)N^{(1)}_\mathbf{p}\right)
\end{equation}
so that now the heavier mass belongs to $\nu_1$ and the lighter mass belongs to
$\nu_2$. The operator $R$ transforms the  isospin operators as
\begin{equation}
\label{eq:rotation for operators}
R^\dagger J_\omega^z R =-J_\omega^z
\qquad
R^\dagger J_\omega^\pm R =J_\omega^\mp
\end{equation}
and leaves the self-interaction term $\vec{J}\cdot\vec{J}$ invariant.  As a
result, it converts the Hamiltonian $H$ given in Eq. (\ref{eq:H Total}) into
\begin{equation}
\label{eq:rotation for Hamiltonian}
R^\dagger H R
=-\sum_\omega\omega\hat{B}\cdot\vec{J}_\omega+\mu(r){\vec{J}}\cdot{\vec{J}}
\end{equation}
which describes the vacuum oscillations and self-interaction of neutrinos in the
inverted mass hierarchy.  An initial state in the form of Eq. (\ref{initial
state}) evolves into Eq.  (\ref{final state}) under the adiabatic conditions in
the normal mass hierarchy. In the case of inverted mass hierarchy, the same
initial state would evolve into a different final state given by
\begin{equation}
\label{inverted evolution}
|\tilde{\psi}\rangle_{\mbox{\tiny final}}
= R ^\dagger \mathcal{U}  R |\psi\rangle_{\mbox{\tiny initial}}
\end{equation}
This new final state can be easily found by first noting that
Eq. (\ref{eq:rotation for operators}) implies
\begin{align}
\label{eq:rotation for states}
 R |j,m\rangle=(-1)^{j}|j,-m\rangle
\end{align}
which leads to
\begin{align}
\label{rotation initial 1}
R |\psi\rangle_{\mbox{\tiny initial}}
=\mkern-10mu\sum_{m=-n/2}^{n/2}\mkern-10mu c_{-m}(-1)^{\frac{n}{2}} 
|\frac{n}{2},
m\rangle,
\end{align}
where we substituted Eq. (\ref{eq:rotation for states}) into
Eq. (\ref{initial state}) and changed $m\to-m$ in the summation.
According to Eq. (\ref{final state}), this
state would evolve into
\begin{align}
\mathcal{U} R |\psi\rangle_{\mbox{\tiny initial}}
=\mkern-15mu\sum_{m=-n/2}^{n/2} \mkern-15mu c_{-m} \phi_{m}
(-1)^{\frac{n}{2}}|\underbrace{\nu_1, \dots, \nu_1}_{\frac{n}{2}+m },
\underbrace{\nu_2, \dots, \nu_2}_{\frac{n}{2}-m}\rangle
\end{align}
Then, another application of $R^\dagger$ leads to
\begin{align}
\label{rotation initial 2}
|\tilde{\psi}\rangle_{\mbox{\tiny final}}
=\mkern-10mu\sum_{m=-n/2}^{n/2} \mkern-10mu c_{-m} \phi_{m}
|\underbrace{\nu_2, \dots, \nu_2}_{\frac{n}{2}+m },
\underbrace{\nu_1, \dots, \nu_1}_{\frac{n}{2}-m}\rangle
\end{align}
where we used Eq. (\ref{R transforms a}). A comparison of the final states
given in Eqs. (\ref{final state}) and (\ref{rotation initial 2})  reveals that
the final states in normal and inverted mass hierarchies are related by
$c_m\leftrightarrow c_{-m}$ and $\nu_1\leftrightarrow\nu_2$.
The rest of the analysis follows the same lines as in the case of the normal
mass hierarchy. For an initial state in the form of Eq. (\ref{eq: highest weigth
flavor}), this leads to the final energy distributions given by
\begin{align}
\label{final inverted}
&\Phi^{(a)}(\omega_k)=\\ \nonumber
&\frac{n_{\omega_k}}{2n}\pm
\frac{1}{n}\sum_{m=-n/2}^{n/2}
\frac{1}{\sqrt{2\sigma^2\pi}}\exp\left({-\frac{\left(m+\bar{m}\right)^2}{
2\sigma^2}}\right)
(-m_{\omega_k})
\end{align}
where $\sigma$, $\bar{m}$ and $m_{\omega_k}$ have the same values as in the case
of normal mass hierarchy given in Eqs. (\ref{distribution in mass m}) and
(\ref{gaussian parameters}).  Once again, the upper sign is for $a=1$ and the
lower sign for $a=2$.  For the thermal initial distribution of electron
neutrinos given in Fig. \ref{fig:expectation values of j_w for FD a}, Eq.
(\ref{final inverted}) yields the final distribution in the mass basis given in 
Fig.
\ref{fig:FD distributions_inverted}. The single split frequency $\bar{\omega}_s$
in this figure can be calculated using an analysis similar to the one which led 
to Eq.
(\ref{omega bar}). In the case of inverted mass hierarchy,
this analysis leads to the analogous equation
\begin{equation}
\label{omega bar inverted}
\int_{\bar{\omega}_s}^\infty n_{\omega}\rho_{\omega}d\omega=n\cos^2\theta
\end{equation}
which agrees with what one would calculate from the mean field approximation
using the conservation of the average value $\langle J^z\rangle$.  Equation
(\ref{omega bar inverted}) yields $\bar{\omega}_s=5.04\times10^{-3}$ km$^{-1}$ 
which
corresponds to $E=37.4$ MeV for the parameters that we use in this example. This
value agrees very well with Fig.  \ref{fig:FD distributions_inverted}.

\section{Antineutrinos}
\label{section:Antineutrinos}
The most convenient way to include antineutrinos into this formalism is by using
the doublets
\begin{equation}
\label{doublets anti}
\begin{pmatrix}
-|\bar{\nu}_2,\mathbf{p}\rangle \\ |\bar{\nu}_1,\mathbf{p}\rangle
\end{pmatrix}
\quad
\mbox{and}
\quad
\begin{pmatrix}
-|\bar{\nu}_x,\mathbf{p}\rangle \\ |\bar{\nu}_e,\mathbf{p}\rangle
\end{pmatrix}
\end{equation}
Here $|\bar{\nu}_a,\mathbf{p}\rangle$ denotes an antineutrino in a mass
($a=1,2$) or flavor ($a=e,x$) eigenstate with momentum $\mathbf{p}$. If
we denote the corresponding annihilation operator by $b_a(\mathbf{p})$, the
isospin operators for this doublet structure are given by
\begin{subequations}
\begin{align}
\label{eq:isospin mass antineutrinos}
&\bar{J}^+_{{\tiny \mathbf{p}},{\mbox{\tiny 
mass}}}=-b^\dagger_2(\mathbf{p})b_1(\mathbf{p}),\quad
\bar{J}^{-}_{{\tiny \mathbf{p}},{\mbox{\tiny 
mass}}}=-b^\dagger_1(\mathbf{p})b_2(\mathbf{p}) \nonumber \\
&\bar{J}^z_{{\tiny \mathbf{p}},{\mbox{\tiny 
mass}}}={\frac{1}{2}}{\big(b^\dagger_2(\mathbf{p})b_2(\mathbf{p}
)-b^\dagger_1(\mathbf{p})b_1(\mathbf{p})\big)}
\end{align}
\begin{align}
\label{eq:isospin flavor antineutrinos}
&\bar{J}^+_{{\tiny \mathbf{p}},{\mbox{\tiny 
flavor}}}=-b^\dagger_x(\mathbf{p})b_e(\mathbf{p}),\quad
\bar{J}^{-}_{{\tiny \mathbf{p}},{\mbox{\tiny 
flavor}}}=-b^\dagger_e(\mathbf{p})b_x(\mathbf{p}) \nonumber \\
&\bar{J}^z_{{\tiny \mathbf{p}},{\mbox{\tiny 
flavor}}}={\frac{1}{2}}{\big(b^\dagger_x(\mathbf{p})b_x(\mathbf{p}
)-b^\dagger_e(\mathbf{p})b_e(\mathbf{p})\big)}
\end{align}
\end{subequations}
in mass and flavor bases respectively. As we did for neutrinos, we drop the
``mass'' index from the antineutrino mass isospin operators:
\begin{equation}
\bar{J}^{\pm,0}_{{\tiny \mathbf{p}},{\mbox{\tiny mass}}}\to
\bar{J}^{\pm,0}_{{\tiny \mathbf{p}}}.
\end{equation}
Following the established practice in this field, we define the ``energy'' as
$E=-|\mathbf{p}|$ for antineutrinos. Accordingly their ``vacuum oscillation
frequencies'' $\omega$ defined in Eq. (\ref{w}) are also allowed to take
negative values. Characterizing antineutrinos with the doublet in Eq.
(\ref{doublets anti}) and using negative energies helps one to seamlessly
integrate them into the formulation (see Refs.  \cite{Duan:2007fw, Duan:2007bt},
for example). One only needs to keep in mind that the physical values of
energy and vacuum oscillation frequency for antineutrinos are equal to $-E$ and
$-\omega$, respectively. The summation formula in Eq. (\ref{eq:total isospin
operators for each mode}) generalizes to antineutrinos as
\begin{equation}
\label{eq:total isospin operators for each mode antineutrino}
{\vec{J}_\omega}=\sum_{|\mathbf{p}|=-E}{\vec{\bar{J}}_{\mathbf{p}}}
\end{equation}
This formula tells us that $\vec{J}_\omega$ for $\omega<0$ represents the total
isospin of all antineutrinos with energy $E<0$. The formula for the summation
over all modes given in Eq. (\ref{eq:global isospin operator}) is now
generalized to include both positive and negative oscillation frequencies so
that $\vec{J}$ represents the total isospin of both neutrinos and antineutrinos.
In this case the operator $U$ defined in Eq. (\ref{eq:transformation operator})
transforms both neutrinos and antineutrinos between flavor and mass bases.
Finally, the summation convention for other quantities described in Sec.
\ref{subsection: summation convention} also generalizes in a similar fashion:
$\bar{Q}_{\mathbf{p}}$ and $Q_{\omega<0}$ refer to the antineutrino analogs of
the corresponding neutrino quantities, while $Q$ refers to the same quantity
summed over all neutrinos and antineutrinos.  In particular, $n$ now denotes the
total number of neutrinos and antineutrinos in the ensemble.

With these definitions, the many-body Hamiltonian describing neutrinos and
antineutrinos which undergo vacuum oscillations and self-interactions is given
by
\begin{align}
\label{eq:H including antineutrinos}
 H=-\sum_\omega \omega J_\omega^z+\mu \vec{J}\cdot\vec{J}\msp,
\end{align}
This Hamiltonian has the same form as the one given in Eq. (\ref{eq:H Total})
except that now the range of $\omega$ extends to include negative values
corresponding to antineutrinos.
Because of the reversed definition of isospin doublets for
antineutrinos, Eq. (\ref{analogy 1}), describing its analogy with the other
many-body systems, is now generalized to
\begin{equation}
\label{analogy anti}
\begin{split}
&
|\f\rangle\leftrightarrow
|\uparrow\rangle\leftrightarrow
\begin{cases}
-|\bar{\nu}_2\rangle & \mbox{ for } \omega<0, \\
|\nu_1\rangle & \mbox{ for } \omega>0, \\
\end{cases}\\
&
|\e\rangle\leftrightarrow
|\downarrow\rangle\leftrightarrow
\begin{cases}
|\bar{\nu}_1\rangle & \mbox{ for } \omega<0, \\
|\nu_2\rangle & \mbox{ for } \omega>0, \\
\end{cases}
\end{split}
\end{equation}
Since the form of the Hamiltonian does not change when we include antineutrinos,
and since all calculations that we carried out so far
depend only on the isospin structure and the corresponding $SU(2)$ commutators,
our results can be extended to include antineutrinos in a trivial way.
The following procedure achieves this goal:
\\
{\bf 1.} One first shifts the origin of the range of $\omega$ to allow for 
negative
frequencies.
\\
{\bf 2.} Then, for the negative frequencies one needs to substitute
\begin{equation}
\begin{split}
\label{substitution}
\nu_1\to-\bar{\nu}_2, \qquad \nu_2\to\bar{\nu}_1,\\
\end{split}
\end{equation}

In connection with the Bethe ansatz formalism, the first step would
correspond to shifting the origin of the coordinate system in the electrostatic
analogy shown in Fig. \ref{fig:Electrostatic analogy}.  Clearly the
electrostatic system is invariant under a translation along the $x$ axis,
reflecting the invariance of Bethe ansatz equations under a transformation which
takes $\omega\to\omega+a$ and $\xi_\alpha\to\xi_\alpha+a$ where $a$ is a real
parameter. Therefore, all the conclusions that we draw from the Bethe ansatz
formalism about the many-body eigenstates of the Hamiltonian and how they change
as the neutrino self-interactions decrease are still valid.

The substitution in Eq. (\ref{substitution})
also leads to
\begin{equation}
\begin{split}
\label{substitution flavor}
\nu_e\to-\bar{\nu}_x, \qquad \nu_x\to\bar{\nu}_e
\end{split}
\end{equation}
in agreement with the mixing formula in Eq. (\ref{eq:fermion relations}).  In
particular, the initial state that we started with in Eq. (\ref{eq:
highest weigth flavor}) becomes
\begin{equation}
\label{eq: highest weigth flavor anti}
|\psi\rangle_{\mbox{\tiny initial}}=
|\underbrace{\bar{\nu_x}, \dots, \bar{\nu}_x}_{\omega<0 },
\underbrace{\nu_e, \dots, \nu_e}_{\omega>0}\rangle
=|\frac{n}{2}, \frac{n}{2}\rangle_{\mbox{\tiny flavor}}
\end{equation}
As $\mu$ decreases from very large to very small values under the assumption of
perfect adiabaticity, this state evolves to
\begin{align}
\label{final state antineutrino}
|\psi\rangle_{\mbox{\tiny final}} =&\sum_{m=-n/2}^{n/2}
(\cos\theta)^{(\frac{n}{2}+m)}(\sin\theta)^{(\frac{n}{2}-m)}
\nonumber \\
&\times \sqrt{n\choose {\frac{n}{2}+m}} \phi_m|\underbrace{\Uparrow, \dots,
\Uparrow}_{\frac{n}{2}+m }, \underbrace{\Downarrow, \dots,
\Downarrow}_{\frac{n}{2}-m}\rangle.
\end{align}
This result is obtained by making the substitution given in Eq.
(\ref{substitution}) for negative frequencies in
Eq. (\ref{final state electron}). The states $|\Uparrow\rangle$ and
$|\Downarrow\rangle$ refer to the neutrino mass isospin: They are defined
as
\begin{equation}
\label{analogy anti isospins}
\begin{split}
&
|\Uparrow\rangle=
\begin{cases}
-|\bar{\nu}_2\rangle & \mbox{ for } \omega<0, \\
|\nu_1\rangle & \mbox{ for } \omega>0, \\
\end{cases}\\
&
|\Downarrow\rangle=
\begin{cases}
|\bar{\nu}_1\rangle & \mbox{ for } \omega<0, \\
|\nu_2\rangle & \mbox{ for } \omega>0. \\
\end{cases}
\end{split}
\end{equation}
For the final state in Eq. (\ref{final state antineutrino}), the normalized
energy distribution functions are given by the same formula as in
Eq. (\ref{final}) except that $\omega_k$ can now be both positive and negative, 
and
$a$ takes values in $\{\Uparrow, \Downarrow\}$. The $+$ sign in Eq. 
(\ref{final}) is for
$\Uparrow$ and the $-$ sign is for $\Downarrow$.
The value of $m_{\omega_k}$ in Eq. (\ref{final}) can now be found from
\begin{equation}
\label{distribution in mass m with antineutrinos}
m_{\omega_k}=
\begin{cases}
\displaystyle{n_{\omega_k}/2}, & \mbox{for } k<s(m)\\
\displaystyle{\frac{n^{(\Uparrow)}_{\omega_{s(m)}}-n^{(\Downarrow)}_{\omega_{
s(m)}}}{2}}, &
\mbox{for } k=s(m)\\
\displaystyle{-n_{\omega_k}/2}, & \mbox{for  } k>s(m)\\
\end{cases}
\end{equation}
with $s(m)$ defined by the same formula as in Eq. (\ref{eq:S(l)}).
With these definitions, Eq. (\ref{final}) now
gives us both neutrino and antineutrino distributions as
\begin{equation}
\label{distribution anti isospins}
\begin{split}
&
\Phi^{(\Uparrow)}(\omega) =
\begin{cases}
\Phi^{(\bar{\nu}_2)}(\omega) & \mbox{ for } \omega<0, \\
\Phi^{(\nu_1)}(\omega) & \mbox{ for } \omega>0,
\end{cases}\\
&
\Phi^{(\Downarrow)}(\omega) =
\begin{cases}
\Phi^{(\bar{\nu}_1)}(\omega) & \mbox{ for } \omega<0,\\
\Phi^{(\nu_2)}(\omega) & \mbox{ for } \omega>0,
\end{cases}
\end{split}
\end{equation}
in accordance with Eq. (\ref{analogy anti isospins}). Note that the minus sign
in Eq. (\ref{analogy anti isospins}) has no consequences in
the final energy spectra given in Eq. (\ref{distribution anti isospins}) because
they cancel each other when we calculate the expectation values.

\begin{figure}[t]
\begin{subfigure}[t]{1\columnwidth}
\includegraphics[width=\textwidth]{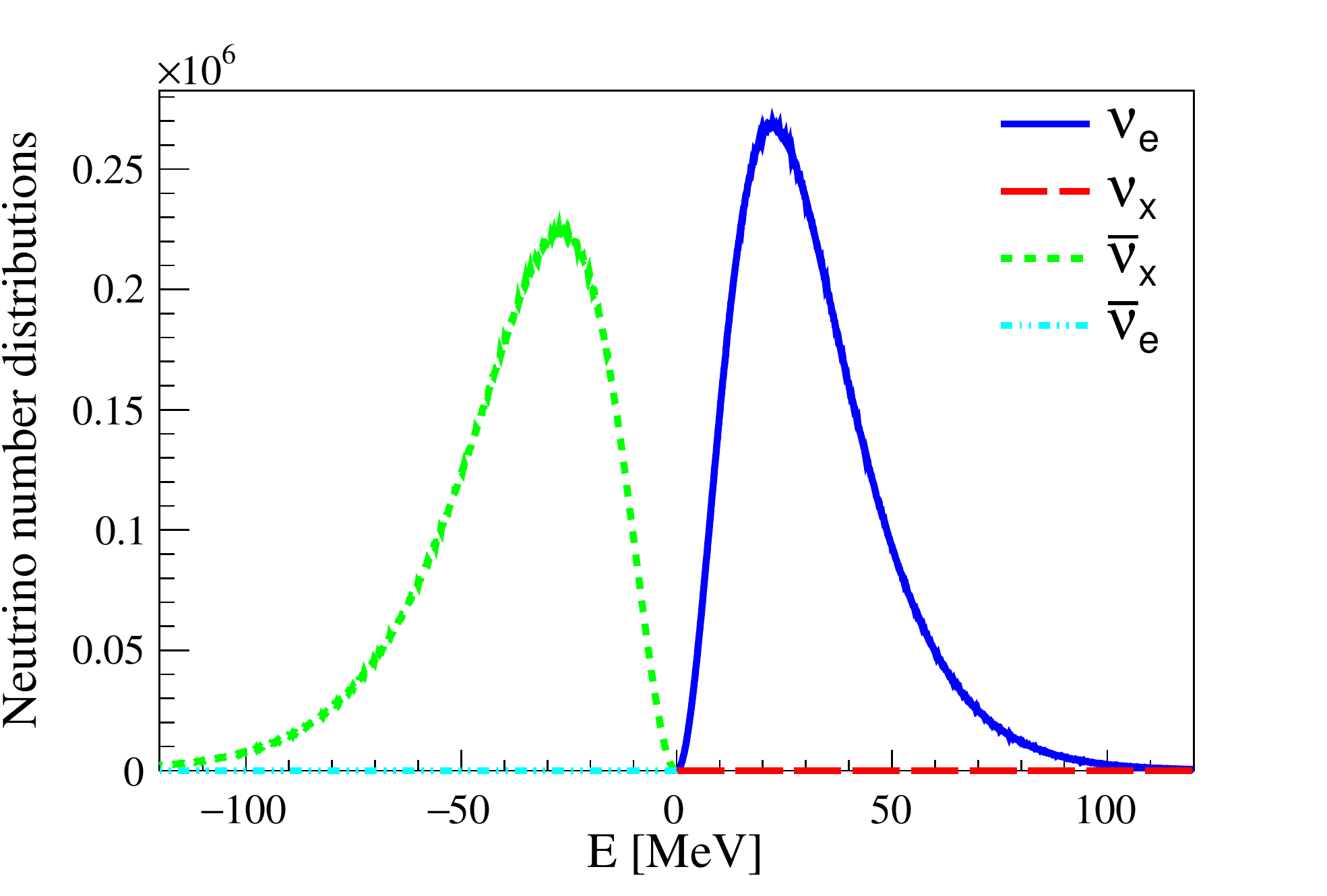}
\caption{\footnotesize \raggedright
Initial thermal distributions of $\nu_e$ and $\bar{\nu}_x$.
}
\label{fig:initial thermal distribution with antineutrinos}
\end{subfigure}
\begin{subfigure}[t]{\columnwidth}
\includegraphics[width=1\textwidth]{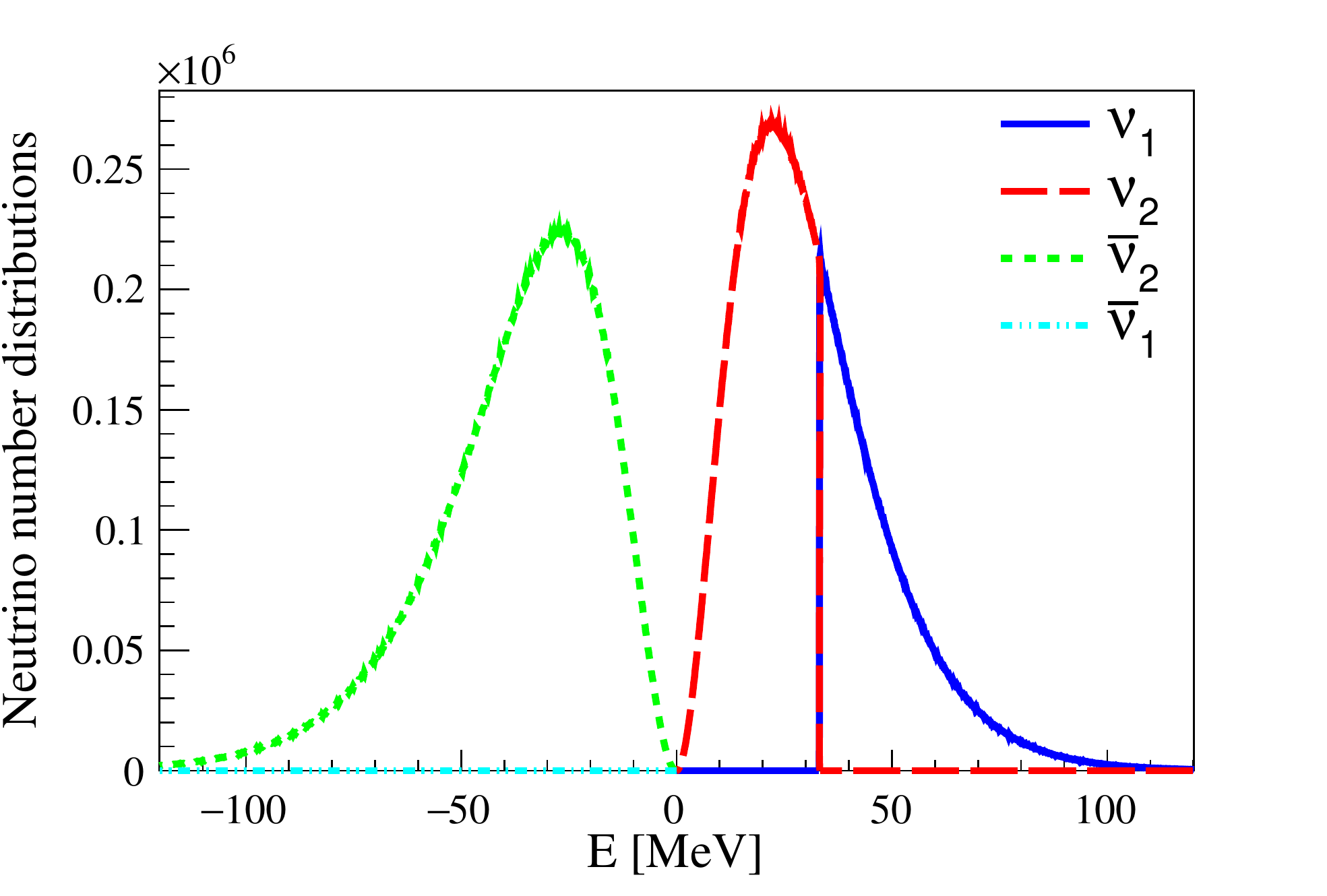}
\caption{\footnotesize \raggedright
Final distributions in mass basis.}
\label{fig:final distribution with antineutrinos}
\end{subfigure}
\caption{\footnotesize \raggedright (Color online) Adiabatic evolution of an
initial distribution of thermal $\nu_e$ and $\bar{\nu}_x$. We take the
temperature $kT=10$ MeV for neutrinos and $kT=12$ MeV for antineutrinos.  We
have $10^8$ neutrinos occupying $1200$ oscillation modes which are equally
spaced in energy. The same is also true for antineutrinos. We adopt the normal
mass hierarchy with solar mixing parameters.}
\label{figure antineutrinos}
\end{figure}

In Fig. \ref{figure antineutrinos}, we show the numerical results for an initial
thermal distribution of $\nu_e$ and $\bar{\nu}_x$ with the respective
temperatures of $10$ MeV and $12$ MeV [Fig.  \ref{fig:initial thermal
distribution with antineutrinos}]. The final energy distributions shown in Fig.
\ref{fig:final distribution with antineutrinos} exhibit a single spectral split
in the neutrino sector.  This can be easily understood with an analysis similar
to the one provided between Eqs. (\ref{m interval}) and (\ref{dm s}) which leads
to the same formula as in Eq. (\ref{omega interval}) with the mean split
frequency $\bar{\omega}_s$ given by the formula
\begin{equation}
\label{omega bar anti}
\int_{-\infty}^{\bar{\omega}_s}n_{\omega}\rho_{\omega}d\omega=n\cos^2\theta
\end{equation}
This formula is the same as what one would derive from the conservation of
$\langle J_z \rangle$ within the mean field approximation.  For the thermal
distributions adopted in Eq. (\ref{eq:FD distribution}), it yields
$\bar{\omega}_s=5.70\times10^{-3}$ km$^{-1}$, which corresponds to $E=33.3$ MeV 
and agrees
with Fig. \ref{figure antineutrinos}.

\section{Conclusions}
\label{section:Conclusion}

We considered the many-body system formed by neutrinos undergoing vacuum
oscillations and self-interactions through neutral current weak force.  As is
standard in the literature, we represented neutrinos as plane
waves in a box so that they all interact with each other at the same time. In
the effective two flavor mixing scenario that we work with, this
many-body system is analogous to a system of spins with long-range interactions,
and to a system of fermions with pairing.  We study this neutrino many-body 
system in the context of a core collapse supernova where it is believed to play 
an important role with such emergent effects as the spectral splits.  However, 
this current study is not meant to be a comprehensive analysis of many-body 
effects under the complicated setting of a real supernova. We retain only some 
of the simple aspects of supernova such as a decreasing self-interaction rate 
as the neutrinos radiate from the proto-neutron star at the center. A constant 
matter background can also be incorporated by using matter effective mixing 
parameters instead of vacuum mixing parameters, although we did not explicitly 
do this in our numerical calculations.

The focus of our study is the exact many-body behavior of self-interacting
neutrinos in comparison to their behavior under the commonly used mean field
approximation.  The latter formulation reduces the $2^n$-dimensional Hilbert 
space to $n$ individual two-dimensional
Hilbert spaces for $n$ neutrinos by omitting the entangled neutrino
states.  Our technique also involves the reduction of the
dimensionality of the Hilbert space, but in a different way. Rather than
omitting part of the Hilbert space,  we first determined and classified the
exact many-body eigenstates of the neutrino Hamiltonian using the
Richardson-Gaudin formalism and then identified those eigenstates which project
onto our initial state. This strategy allows one to work with a smaller part of
the full Hilbert space without any omissions.

Our choice of the initial state in this paper is the simplest one that can be
studied with this prescription. It consists of only electron neutrinos (to which
antineutrinos of the orthogonal flavor can also be added). It projects only on
the highest energy eigenstates of the Hamiltonian for all possible values of the
conserved $z$ component of the total mass isospin operator. Those eigenstates do
not undergo any level crossings with the other eigenstates, allowing one to
easily follow their evolution under the assumption of perfect adiabaticity.  We
have shown that our initial state adiabatically develops into a superposition of
some states, each with a different split point. However, we have shown
that those states which significantly contribute to the sum have their split
frequencies within $(100/\sqrt{n})\%$ of a mean split frequency. This mean split
frequency dominates the final energy distribution, so it is the only apparent
pattern in it.  The formula for this mean split frequency is the same as what
one would obtain with the mean field approximation based on average 
conservation of the $z$ component of total mass isospin. Therefore, our study 
demonstrates the validity of the mean field approximation in this particular 
context in an analytical way.

Although previous studies of the exact many-body dynamics of self-interacting
neutrinos have demonstrated the validity of the mean field approximation in
their specific settings, this is the first study to include the effects of the
vacuum oscillations and the first one to demonstrate the formation of a spectral
split which results from an interplay between vacuum oscillation and 
self-interaction terms. The spectral splits that we obtain in the exact 
many-body 
picture are the same as those in the mean field case. In particular, they 
develop at the same frequency or energy as indicated by Eqs. (\ref{omega bar}) 
and (\ref{omega bar mean field}). As is the case in the mean field formulation, 
the swap appears in the region which has lower (higher) energy than the split 
energy in the normal (inverted) mass hierarchy scenarios.

Two important caveats of our study are the lack of a real dynamical evolution
and the choice of a particularly simple initial state.  Our initial
conditions involve no spectral crossings between different neutrino flavors. In
that sense, they correspond to the special case examined in Ref.
\cite{Raffelt:2007cb} in the mean field approximation. At present, our
formulation does not apply to the splits which arise from instabilities which
develop around spectral crossing points, i.e., those discussed in Refs.
\cite{Dasgupta:2009mg, Dasgupta:2010cd, Banerjee:2011fj}.  It may be possible to
overcome this limitation and apply the formalism to a more general initial
condition. This would require a careful examination of the behavior of the
system around the points where many-body energy eigenvalues overlap with one
another. Self-interacting neutrinos have several dynamical symmetries, and our
preliminary studies indicate that the corresponding conservation laws may help
us to identify the evolution of a particular many-body energy eigenstate at an
energy crossing point.  This will be the subject of a future study. 


\vspace*{3mm} \noindent
This work was supported
in part by T{\"{U}}B{\.{I}}TAK under Project No. 115F214,
in part by the US National Science
Foundation Grant No. PHY-1514695,
and in part by Grants-in-Aid for Scientific Research of JSPS (No. 15H03665 and 
No. 17K05459).

\bibliography{bibyamac}

\section*{Appendix: Energy eigenvalues of bethe ansatz states in strongly 
interacting regime}
\label{appendix}

In the text we have argued that the Bethe ansatz states with $\kappa$ variables
given in Eq. (\ref{eq:trial states from particle vacuum}) go to the
$|j,m\rangle$ states of the total isospin operator given in Eq. (\ref{jm states
for kappa variables in infty}) in $\mu\to\infty$ limit. In this Appendix, we
show that in particular the solution of Bethe ansatz equations in which $k$ of
the Bethe ansatz variables remain finite in $\mu\to\infty$ limit while
$\kappa-k$ of them go to $-\infty$, produces the eigenstate $|\frac{n}{2}-k,\;
-\frac{n}{2}+\kappa\rangle$.

Since the eigenvalue of $J_z$ is fixed to $-n/2+\kappa$ by the number of Bethe
ansatz variables, we only only need to establish the value of $j$. This can be
determined by calculating the energy of the Bethe ansatz state in $\mu\to\infty$
limit and comparing it with Eq. (\ref{eq:eigenstates at mu=infinite}).

Since the ordering of the Bethe ansatz variables is not important, we can begin
by reordering our variables in such a way that those that approach to $-\infty$
for large $\mu$ are $\xi_1,\xi_2,\dotsc,\xi_{\kappa-k}$ and those that stay
finite are $\xi_{\kappa-k+1},\dots,\xi_{\kappa}$. Then, the Bethe ansatz
equations for the first $p$ variables can be written as follows in
$\mu\to\infty$ limit:
\begin{equation}
\label{eq:BA at mu infinite}
-\frac{n}{\xi_\alpha} = \frac{1}{\mu} - {\sum_{\beta\neq\alpha}^\kappa} 
\frac{2}{\xi_\alpha-\xi_\beta}
\end{equation}
for $\alpha=1,2,\dots,\kappa-k$. Here we ignored the $\omega$ values which are
finite valued so that the sum on the left-hand side of Eq. (\ref{eq:general BA
eq.}) is performed to yield the total number of neutrinos, $n$. The sum on the
right-hand side of Eq. (\ref{eq:BA at mu infinite}) can separated into two
parts, one running over first $\kappa-k$ (infinite valued) $\xi_\beta$'s and the
other running over the last $k$ (finite valued)  $\xi_\beta$'s:
\begin{eqnarray}
\label{eq:finete and infine sums of terms in BA eq}
\sum_{\beta\neq\alpha}^\kappa \frac{2}{\xi_\alpha-\xi_\beta}&=&
\sum_{\beta(\neq\alpha)=1}^{\kappa-k} \frac{2}{\xi_\alpha-\xi_\beta} +
\sum_{\beta=\kappa-k+1}^{\kappa}\frac{2}{\xi_\alpha-\xi_\beta}\nonumber \\
&=&\frac{2}{\xi_\alpha} (k+L_\alpha)
\end{eqnarray}
Here $L_\alpha$ is given by
\begin{equation}
\label{L_alpha}
L_\alpha=\sum_{\beta(\neq\alpha)=1}^p\frac{\xi_\alpha}{\xi_\alpha-{\xi_\beta}}\msp.
\end{equation}
Substituting it in Eq.\s(\ref{eq:BA at mu infinite}) we find
\begin{equation}
\label{eq: ksi_alpha at mu infinite}
\xi_\alpha=\mu(2L_\alpha+2k-n)
\end{equation}
Note that, $L_\alpha$ depends on $\xi_\alpha$ so that Eq. (\ref{eq: ksi_alpha at
mu infinite}) is not an explicit solution. However, this expression is very
useful in calculating the energy. By substituting
Eq. (\ref{eq: ksi_alpha at mu infinite}) in Eq. (\ref{eq:general energies of BA
states}), we find the energy of the eigenstate
$|\xi_1,\xi_2,\dotsc,\xi_{\kappa-k},\xi_{\kappa-k+1},\dots,\xi_{\kappa}\rangle$ as
\begin{eqnarray}
\label{eq:Energy eigenvalues at mu infinite}
E
&=E_{-n/2}-\kappa\mu(n-\kappa+1)-\sum_{\alpha=1}^{\kappa-k}\xi_\alpha \nonumber \\
&
=E_{-n/2}-\mu p(2k-n)+2\mu\sum_{\alpha=1}^{\kappa-k}L_\alpha
\end{eqnarray}
where we ignored the contribution of finite valued Bethe ansatz variables in the
energy. By summing both sides of Eq. (\ref{L_alpha}) over $\alpha$ and antisymmetrizing
the result, one finds that
\begin{equation}
\label{sum of L_alpha}
\sum_{\alpha=1}^{\kappa-k} L_\alpha=(\kappa-k)(\kappa-k-1)
\end{equation}
Therefore, the energy in Eq. (\ref{eq:Energy eigenvalues at mu infinite})
is equal to
\begin{equation}
\label{eq:Energy eigenvalues at mu infinite compact form}
E=\mu(n/2-k)(n/2-k+1)
\end{equation}
This energy is consistent with Eq. (\ref{eq:eigenstates at mu=infinite}) with
$j=n/2-k$. However, we know that all representations with $j<n/2$ come in
multiplicities Therefore Eq. (\ref{eq:Energy eigenvalues at mu infinite compact
form}) can only tell us that, in $\mu\to\infty$ limit, the state
$|\xi_1,\xi_2,\dotsc,\xi_{\kappa-k},\xi_{\kappa-k+1},\dots,\xi_{\kappa}\rangle$
approaches to a linear combination of multiple $|n/2-k, -n/2+\kappa\rangle$
states belonging to different representations with the same $j=n/2-k$. On the
other hand, those representations with the same $j$ can always be linearly
combined to produce another such representation because any linear combination of
$|j,m\rangle$ states gives $j(j+1)$ and
$m$ under the actions of $\vec{J}\cdot\vec{J}$ and $J_z$, respectively. For
this reason, we have a degree of freedom in choosing those representations
which come with multiplicities when we add several spins or isospins, as long as
they are orthogonal to each other. Since the eigenstates of a Hermitian
Hamiltonian are also orthogonal to each other, the $j=n/2-k$ representations can
be chosen in such a way that the limit of our Bethe ansatz state
$|\xi_1,\xi_2,\dotsc,\xi_{\kappa-k},\xi_{\kappa-k+1},\dots,\xi_{\kappa}\rangle$
as $\mu\to\infty$ coincides with a particular $|n/2-k, -n/2+\kappa\rangle$
state.

\end{document}